\documentclass[aps,prd,manuscript,showpacs,amsfonts,amssymb,amsmath]{revtex4-1}
\usepackage[dvips]{graphicx}
\usepackage{float}

\begin{document}

\title{Partial wave analysis of the Dirac fermions scattered from Schwarzschild  black holes}

\author{Ion I.  Cot\u aescu}
 \email{cota@physics.uvt.ro}
\author{Cosmin Crucean}%
 \email{crucean@physics.uvt.ro}
\author{Ciprian A. Sporea}%
 \email{ciprian.sporea89@e-uvt.ro}
\affiliation{\small \it
 West University of Timi\c soara,\\
V.  P\^ arvan Ave.  4, RO-300223 Timi\c soara, Romania
}

\date{\today}

\begin{abstract}
Asymptotic analytic solutions of the Dirac equation, giving the
scattering modes  (of the continuous energy spectrum, $E>mc^2$) in
Schwarzschild's chart and Cartesian gauge,  are used for building
the partial wave analysis of Dirac fermions scattered by black
holes. The contribution of the bound states to absorption and
possible resonant scattering is neglected because of some technical
difficulties related to the discrete spectrum that is less studied
so far. In this framework, the analytic expressions of the
differential cross section and induced polarization degree are
derived in terms of scattering angle, mass of the black hole, energy
and mass of the fermion. Moreover, the closed form of the absorption
cross section due to the scattering modes  is derived showing  that
in the high-energy limit this tends to the event horizon area
regardless of the fermion mass (including zero). A  graphical study
presents the differential cross section analyzing the
forward/backward scattering (known also as glory scattering) and the
polarization degree as functions of scattering angle. The graphical
analysis shows the presence of oscillations in scattering intensity
around forward/backward directions, phenomena known as spiral
scattering. The energy dependence of the differential cross section
is also established by using analytical and graphical methods.
\end{abstract}


\maketitle

\section{introduction}

The complex problem  of the quantum particles scattered from  black
holes was studied extensively mainly considering massless scalar
\cite{sc1}-\cite{no} or other massless boson fields
\cite{n1}-\cite{n3} since the equations  of massive fields cannot be
solved analytically  in the Schwarzschild geometry. This is one of
the reasons why the scattering of massive Dirac fermions by black
holes was studied either in particular cases \cite{FHM} or by using
combined  analytical and numerical methods  \cite{S1}-\cite{S3}.
Thus in Refs. \cite{bh1,S3} such methods  were applied for
investigating the absorption cross section, the scattering intensity
(pointing out  the glory and spiral scattering) and the polarization
degree of these scattering processes. However, in the case of the
interactions of  quantum particles with black holes the numerical
methods may represent a difficult task since one must  combine
quantities at quantum scale (mass, energy of the fermion) with
quantities at galactic scale (as  the black hole mass). Therefore,
it is obvious that any new analytical study may improve this
investigation helping us to understand the quantum mechanisms
governing the scattering process.

For this reason we would like to propose in this paper the partial
wave analysis of the Dirac fermions scattered from  Schwarzschild
black holes constructed applying exclusively analytical methods. In
order to do this, we exploit the analytical properties of the
(approximative) asymptotic solutions of the Dirac equation we have
found some time ago \cite{C4}. These solutions were obtained in the
chart with Schwarzschild coordinates where we considered the
Cartesian gauge that preserves the global central symmetry of the
field equations \cite{ES}, allowing the separation of spherical
variables just as in the central problems of special relativity
\cite{TH}. We note that thanks to this gauge one of us (IIC)
succeeded to solve analytically  the Dirac equation on the central
charts of the de Sitter and anti-de Sitter spacetimes
\cite{C1}-\cite{C1C}. In the case of the  Schwarzschild chart and
Cartesian gauge, after the separation of angular variables, we
remain with a pair of simple radial equations depending only on the
gauge field components that can be  approximatively solved by using
the Novikov radial coordinate \cite{Nov,GRAV}. We obtained thus the
asymptotic radial solutions that are either of scattering type or
describing bound states   resulted from a quantization condition of
the third order in energy \cite{C4}. However, these last mentioned
solutions are less studied such that  we restrict ourselves   to
consider only the scattering solutions (with $E>mc^2$) for
performing the partial wave analysis of the Dirac fermions scattered
from black holes.

In general, when one considers the analytic expressions of the
scattering quantum modes, one must impose suitable boundary
conditions (in origin or at event horizon) in order to fix the
integration constants determining the asymptotic behavior and
implicitly the phase shifts of the partial wave analysis. Having
here only the  asymptotic form of the Dirac spinors in the black
hole field we must replace the boundary conditions with suitable
{\em asymptotic} conditions playing the same role in determining the
integration constants of our asymptotic solutions.  Fortunately,
this can be done assuming that  for large values of angular momentum
the collision becomes elastic  approaching to the Newtonian limit
\cite{FHM}. This hypothesis is enough for determining completely the
integration constants  and deriving the analytical forms  of the
phase shifts, scattering amplitudes, differential cross section and
polarization degree. The nice surprise is that this approach
emphasis the absorption of the fermions by black hole in a natural
manner laying out the absorption cross section in an analytical
closed form with some associated selection rules. However, we must
specify that this cross section encapsulates only the contribution
of the scattering modes since the influence of the bound states  is
neglected here.

The paper is organized as follows. In the second section we briefly
present our method of separating variables of the Dirac equation on
central backgrounds and Cartesian gauge focusing on  the mentioned
approximative solutions of the Dirac equation in Schwarzschild's
charts, corresponding to the continuous energy spectrum \cite{C4}.
The third section is devoted to our partial wave analysis  based on
the asymptotic condition (discussed in the Appendix C) that fixes
the integration constants giving the analytical form of the phase
shifts encapsulating both the cases studied here,  the elastic
scattering of fermions and their absorption by black hole. We verify
that our scattering amplitudes have a correct Newtonian limit and we
derive the scattering intensity and polarization degree. A special
attention is paid to the absorption cross section for which we give
the analytic expression of the partial cross sections and the
selection rules indicating in which partial wave we can find
absorption.   The fourth section is devoted to the graphical
analysis and discussion of the physical consequences of our results.
Here we show that our analytical results concerning the elastic
scattering are very similar to those obtained by using
analytical-numerical methods  \cite{bh1,S3} but there are some
differences in what concerns the absorption. We suggest that this
could be a consequence of fact that we ignore the bound states that
might give rise to a resonant scattering.   Other conclusions are
summarized in the last section.

In what follows we use natural units with $c=\hbar=G=1$.

\section{Approximating  Dirac spinors in Schwarzschild's geometry}

The Dirac equation in curved spacetimes  is defined in frames
$\{x;e\}$  formed by a local chart of coordinates $x^{\mu}$, labeled
by natural indices, $\alpha,..,\mu, \nu,...=0,1,2,3$, and an
orthogonal  local frame and coframe  defined by the gauge  fields
(or tetrads), $e_{\hat\alpha}$ and  respectively $\hat
e^{\hat\alpha}$, labeled by the local indices
$\hat\alpha,..,\hat\mu,...$ with the same range.

In local-Minkowskian manifolds $(M,g)$, having as flat model the
Minkowski spacetime $(M_0,\eta)$ of metric $\eta={\rm
diag}(1,-1,-1,-1)$, the gauge fields satisfy the  usual duality
conditions, $\hat e^{\hat\mu}_{\alpha}\,
e_{\hat\nu}^{\alpha}=\delta^{\hat\mu}_{\hat\nu} \hat
e^{\hat\mu}_{\alpha}\, e_{\hat\mu}^{\beta}=\delta^{\beta}_{\alpha}$
and the orthogonality relations, $e_{\hat\mu}\cdot
e_{\hat\nu}=\eta_{\hat\mu \hat\nu}\,,\, \hat e^{\hat\mu}\cdot \hat
e^{\hat\nu}=\eta^{\hat\mu \hat\nu}$. The gauge fields define the
local derivatives
$\hat\partial_{\hat\mu}=e_{\hat\mu}^{\nu}\partial_{\nu}$ and the
1-forms $\omega^{\hat\mu}=\hat e^{\hat\mu}_{\nu}dx^{\nu}$ giving the
line element
$ds^2=\eta_{\hat\alpha\hat\beta}\omega^{\alpha}\omega^{\beta}=g_{\mu\nu}dx^{\mu}dx^{\nu}$
(with $g_{\mu\nu}=\eta_{\hat\alpha\hat\beta}\hat
e^{\hat\alpha}_{\mu}\hat e^{\hat\beta}_{\nu}$).

\subsection{The Dirac equation in central charts and Cartesian gauge}

In a given frame $\{x;e\}$, the Dirac equation of a free spinor
field $\psi$  of  mass $m$ has the form
\begin{equation}\label{Dir1}
 i\gamma^{\hat\alpha}D_{\hat\alpha}\psi - m\psi=0\,,
\end{equation}
where $\gamma^{\hat\alpha}$ are the point-independent Dirac matrices
that  satisfy  $\{ \gamma^{\hat\alpha}, \gamma^{\hat\beta}
\}=2\eta^{\hat\alpha \hat\beta}$ and define the generators of the
spinor representation of the $SL(2,C)$ group, $S^{\hat\alpha
\hat\beta}=\frac{i}{4}[\gamma^{\hat\alpha}, \gamma^{\hat\beta} ]$,
which give the spin connections of the covariant derivatives,
\begin{equation}
D_{\hat\alpha}=e_{\hat\alpha}^{\mu}D_{\mu}=\hat\partial_{\hat\alpha}+\frac{i}{2}S^{\hat\beta \cdot}_{\cdot
\hat\gamma}\hat\Gamma^{\hat\gamma}_{\hat\alpha \hat\beta}\,,
\end{equation}
depending on the connection  components in local frames
$\hat\Gamma^{\hat\sigma}_{\hat\mu \hat\nu}=e_{\hat\mu}^{\alpha}
e_{\hat\nu}^{\beta}
(\hat e_{\gamma}^{\hat\sigma}\Gamma^{\gamma}_{\alpha \beta}
-\hat e^{\hat\sigma}_{\beta, \alpha})$
where the notation  $\Gamma^{\gamma}_{\alpha \beta}$ stands for the usual Christoffel symbols.

In this approach the Dirac equation (\ref{Dir1}) takes the explicit form
\begin{equation}\label{Dir2}
i\gamma^{\hat\alpha}e_{\hat\alpha}^{\mu}\partial_{\mu}\psi - m\psi
+ \frac{i}{2} \frac{1}{\sqrt{-g}}\partial_{\mu}(\sqrt{-g}e_{\hat\alpha}^{\mu})
\gamma^{\hat\alpha}\psi
-\frac{1}{4}
\{\gamma^{\hat\alpha}, S^{\hat\beta \cdot}_{\cdot \hat\gamma} \}
\hat\Gamma^{\hat\gamma}_{\hat\alpha \hat\beta}\psi =0\,,
\end{equation}
where $g={\rm det}(g_{\mu\nu})$. Moreover,  from the conservation of
the  electric charge, one  deduces that the time-independent
relativistic scalar product of two spinors  \cite{C1},
\begin{equation}\label{sp}
(\psi,\psi')=\int_{D}d^{3}x\, \sqrt{-g(x)}\,e_{\hat\mu}^{0}(x)\bar\psi(x)\gamma^{\hat\mu}\psi'(x)\,,
\end{equation}
given by the integral over the space domain $D$ of the local chart
under consideration.

In general, a manifold with central symmetry has a static central
chart with spherical coordinates $(t, r, \theta, \phi)$, associated
to the Cartesian ones $(t,\vec{x})$, with $r=|\vec{x}|$, covering
the space domain $D=D_{r}\times S^{2}$, i. e. $r\in D_{r}$ while
$\theta$ and $\phi$ cover the sphere $S^{2}$. The form of the Dirac
equation in any chart is strongly dependent on the choice of the
tetrad gauge. For this reason, our approach is based on the virtues
of the mentioned Cartesian gauge which is defined by the 1-forms
\cite{C1,ES},
\begin{eqnarray}
\omega^0&=&w(r)dt \,,\\
\omega^1&=&\frac{w(r)}{u(r)}\sin\theta\cos\phi \,dr+ \frac{r w(r)}{v(r)}\cos\theta\cos\phi \,d\theta-\frac{r w(r)}{v(r)}\sin\theta\sin\phi \,d\phi\,, \\
\omega^2&=&\frac{w(r)}{u(r)}\sin\theta\sin\phi \,dr+ \frac{r w(r)}{v(r)}\cos\theta\sin\phi \,d\theta+\frac{r w(r)}{v(r)}\sin\theta\cos\phi \,d\phi\,, \\
\omega^3&=&\frac{w(r)}{u(r)}\cos\theta \,dr- \frac{r w(r)}{v(r)}\sin\theta \,d\theta\,,
\end{eqnarray}
expressed in terms of three arbitrary functions of $r$, denoted by
$u$, $v$ and $w$,  which  allow us to write  the general  line
element,
\begin{equation}\label{(muvw)}
ds^{2}=\eta_{\hat\alpha\hat\beta}\omega^{\hat\alpha}\omega^{\hat\beta}
=w(r)^{2}\left[dt^{2}-\frac{dr^{2}}{u(r)^2}-
\frac{r^2}{v(r)^2}(d\theta^{2}+\sin^{2}\theta d\phi^{2})\right]\,.
\end{equation}

We have shown that in this gauge the  last term of Eq. (\ref{Dir2})
does not  contribute and, moreover, there is a simple
transformation, $\psi\to vw^{-\frac{3}{2}}\psi$,  able to eliminate
the terms containing the derivatives of the functions $u$, $v$ and
$w$, leading thus  to a simpler {\em reduced} Dirac equation
\cite{C1}. The advantage of this equation is that its spherical
variables can be separated just as in the case of the central
problems in Minkowski  spacetime \cite{TH}. Consequently,   the
Dirac field can be written as a linear combination of particular
solutions of  given energy, $E$. Those of positive frequency,
\begin{eqnarray}
U_{E,\kappa,m_{j}}({x})&=&
U_{E,\kappa,m_{j}}(t,r,\theta,\phi)\label{(u)}\\
&=&\frac{v(r)}{rw(r)^{3/2}}[f^{+}_{E,\kappa}(r)\Phi^{+}_{m_{j},\kappa}(\theta,\phi)
+f^{-}_{E,\kappa}(r)\Phi^{-}_{m_{j},\kappa}(\theta,\phi)]e^{-iEt}\,,\nonumber
\end{eqnarray}
are particle-like energy eigenspinors expressed in terms of radial
wave functions,  $f^{\pm}_{E,\kappa}$, and  usual four-component
angular spinors $\Phi^{\pm}_{m_{j}, \kappa}$ \cite{TH}.  It is known
that these spinors are orthogonal to each other being  labelled by
the angular quantum numbers $m_{j}$ and
\begin{equation}\label{kjl}
\kappa=\left\{\begin{array}{lcc}
~~~~\,j+\frac{1}{2}=l&{\rm for}& j=l-\frac{1}{2}\\
-(j+\frac{1}{2})=-l-1&{\rm for}& j=l+\frac{1}{2}
\end{array}\right.
\end{equation}
which encapsulates the information about the quantum numbers $l$ and
$j=l\pm\frac{1}{2}$ as defined in Refs. \cite{TH,LL} (while  in Ref.
\cite{S3}  $\kappa$ is of opposite sign).   The spherical spinors
are normalized to unity with respect to their own angular scalar
product. We note that the antiparticle-like energy eigenspinors  can
be obtained directly using the charge conjugation  as in the flat
case \cite{C3}.

Thus the problem of the angular motion is completely solved for any
central background.  We remain with  a pair of  radial wave
functions, $f^{\pm}$, (denoted from now without indices)  which
satisfy two radial equations  that can be written in compact form as
the eigenvalue problem $H_r{\cal F}=E{\cal F}$ of the radial
Hamiltonian \cite{C1},
\begin{equation}\label{HR}
H_r=\left(\begin{array}{cc}
    m\,w(r)& -u(r)\frac{\textstyle d}{\textstyle dr}+\kappa\frac{\textstyle v(r)}
{\textstyle r}\\
&\\
  u(r)\frac{\textstyle d}{\textstyle dr}+\kappa\frac{\textstyle v(r)}
{\textstyle r}& -m\,w(r)
\end{array}\right)\,\,,
\end{equation}
in the space of  two-component vectors (or doublets),  ${\cal
F}=(f^{+}, f^{-})^{T}$,  equipped with the radial scalar product
\cite{C1}
\begin{equation}\label{(spf)}
({\cal F},{\cal F}')=\langle U,U'\rangle=\int_{D_{r}}\frac{dr}{u(r)}\, {\cal
F}^{\dagger}{\cal F}'\,,
\end{equation}
resulted from the general formula (\ref{sp}) where  we have to  take
$\sqrt{-g(x)}=\frac{w(r)^4}{u(r)v(r)^2}\,r^2\sin\theta$, the spinors
$U$ and $U'$ of the form (\ref{(u)}) and
$e_{\hat\mu}^0\gamma^{\hat\mu}=\frac{1}{w(r)}\gamma^0$. This scalar
product selects the 'good' radial wave functions, i.e. square
integrable functions or tempered distributions, which enter in the
structure of the particle-like energy eigenspinors.

In the central charts each particular solution  (\ref{(u)}) gives
rise to a  {\em partial} radial current defined as
\begin{equation}
J_{rad}=\int_{S^2} d\theta\,d\phi \sqrt{-g(x)}\, \overline{U}(x) \gamma_{rad} U(x)
\end{equation}
where  we take into account that in our Cartesian gauge we have
\begin{equation}
\gamma_{rad}
=e_{\hat\mu}^r\gamma^{\hat\mu}=\frac{u(r)}{w(r)}\gamma_x \,, \quad
\gamma_x=\frac{1}{r}\,\vec{x}\cdot \vec{\gamma}\,.
\end{equation}
Exploiting then the property $\gamma^0\gamma_x \Phi^{\pm}=\pm
i\Phi^{\mp}$  and the orthogonality of the spherical spinors
\cite{TH} we obtain the final result
\begin{equation}\label{Jrad}
J_{rad}=i\left( f^+ {f^-}^*-{f^+}^*f^-\right)\,.
\end{equation}
Now it is obvious that these currents are conserved and independent
on $r$  since $\partial_r J_{rad}=0$ whenever the functions
$f^{\pm}$ satisfy the above radial equations. These will represent
useful  significant constants of motion in the scattering process.

\subsection{Analytic asymptotic solutions in Scwarzschild's charts with Cartesian gauge}

Let us assume that a Dirac particle of mass $m$ is  moving freely (as a perturbation) in
the central gravitational field of a black hole of mass $M$ with  the Scwarzschild line element
\begin{equation}\label{(le)}
ds^{2}=\left(1-\frac{r_{0}}{r}\right)dt^{2}-\frac{dr^{2}}{1-
\frac{\textstyle r_{0}}
{\textstyle r}}- r^{2} (d\theta^{2}+\sin^{2}\theta~d\phi^{2})\,,
\end{equation}
defined on the radial domain $D_{r}=(r_{0}, \infty)$ where $r_0=2M$. Hereby we identify the functions
\begin{equation}
u(r)=1-\frac{r_0}{r}\,, \quad v(r)=w(r)=\sqrt{1-\frac{r_0}{r}}\,,
\end{equation}
that give the radial Hamiltonian (\ref{HR}). The resulting radial
problem   cannot be solved analytically as it stays forcing one to
resort to  numerical methods \cite{S3} or to some approximations.

In Ref. \cite{C4} we proposed an effective method of approximating
this radial   problem expanding the radial equations in terms of the
Novikov  dimensionless coordinate \cite{Nov,GRAV},
\begin{equation}\label{x}
x=\sqrt{\frac{r}{r_{0}}-1}\,\in\,(0,\infty)\,.
\end{equation}
Using this new variable and introducing the notations
\begin{equation}
\mu=r_{0}m\,,\quad \epsilon=r_{0}E\,,
\end{equation}
we rewrite the exact radial problem as
\begin{equation}\label{rrr}
\left(\begin{array}{cc}
 \mu\sqrt{1+x^{2}}-\epsilon\left(x+\frac{\textstyle 1}{\textstyle x}\right)
& -\frac{\textstyle 1}{\textstyle 2}\frac{\textstyle d}{\textstyle
dx}+
\frac{\textstyle \kappa}{\textstyle \sqrt{1+x^2}}\\
&\\
    \frac{\textstyle 1}{\textstyle 2}   \frac{\textstyle d}{\textstyle dx}+
\frac{\textstyle \kappa}{\textstyle \sqrt{1+x^2}} & -
    \mu\sqrt{1+x^{2}}-\epsilon\left(x+\frac{\textstyle 1}{\textstyle
x}\right)
\end{array}\right)\,\left(
\begin{array}{c}
f^{+}(x)\\
\\
f^{-}(x)
\end{array}\right)=0\,.
\end{equation}
Moreover, from Eq.(\ref{(spf)}) we find that the radial scalar
product  takes now the form
\begin{equation}\label{(norm)}
({\cal F}_1,{\cal F}_2)=
2r_{0}\int_{0}^{\infty}dx\,\left(x+\frac{1}{x}\right) {\cal
F}^{\dagger}_1{\cal F}_2\,.
\end{equation}
We note that the singular form above,  due to our special
parametrization and  use of the Novikov variable, is merely apparent
since  this comes from the usual regular scalar product (\ref{sp}).
Moreover, we have shown that there exist square integrable spinors
with respect to this scalar product \cite{C4}.

For very large values of $x$, we can use the Taylor expansion with respect to $\frac{1}{x}$ of the operator  (\ref{rrr}) neglecting  the terms of the order $O(1/x^2)$. We obtain thus  the
{\em asymptotic} radial problem \cite{C4} which can be rewritten as
\begin{equation}\label{RA}
\left(\begin{array}{cc}
 \frac{\textstyle 1}{\textstyle 2} \frac{\textstyle d}{\textstyle dx}
 +\frac{\textstyle\kappa}{\textstyle x}
& -\mu\left( x+\frac{\textstyle 1}{\textstyle 2x}\right)
-\epsilon \left(x+\frac{\textstyle 1}{\textstyle x}\right)\\
&\\
 -\mu \left(x+\frac{\textstyle 1}{\textstyle 2x}\right)
 +\epsilon\left(x+\frac{\textstyle 1}{\textstyle x}\right)
&
   \frac{\textstyle 1}{\textstyle 2} \frac{\textstyle d}{\textstyle dx}
   -\frac{\textstyle \kappa}{\textstyle x}
\end{array}\right)\,\left(
\begin{array}{c}
f^{+}(x)\\
\\
f^{-}(x)
\end{array}\right)=0\,,
\end{equation}
after reversing between themselves the lines of the matrix operator.
This is  necessary for diagonalizing simultaneously the term
containing derivatives and the one proportional to $x$, as in the
Dirac-Coulomb case \cite{LL}. This can be done by using the matrix
\begin{equation}
T=\left|
\begin{array}{cc}
-i\sqrt{\mu+\epsilon}&i\sqrt{\mu+\epsilon}\\
\sqrt{\epsilon-\mu}&\sqrt{\epsilon-\mu}\\
\end{array}\right|\,,
\end{equation}
for transforming the radial doublet as ${\cal F}\to \hat{\cal F}=T^{-1}{\cal F}=(\hat
f^{+},\,\hat f^{-})^T$,  obtaining the new system of radial equations
\begin{equation}\label{TE}
\left[\frac{1}{2} x
\frac{d}{dx}\pm i \left(\frac{\mu^2- 2\epsilon^2}{2\nu}-\nu x^2
\right)\right]\hat f^{\pm}
=\left(\kappa\mp\frac{i\epsilon\mu}{2\nu}\right)\hat
f^{\mp}\,,
\end{equation}
where  $\nu=\sqrt{\epsilon^2-\mu^2}$. These equations can be  solved
analytically  for any values of $\epsilon$.  In Ref. \cite{C4} we
derived the spinors  of the discrete spectrum and the quantization
condition, in the domain $\epsilon<\mu$. Moreover,  we outlined the
scattering modes  corresponding  to  the continuous spectrum
$\epsilon\in [\mu, \infty)$, but without  investigating scattering
effects.

Here we would like to complete this study by calculating the fermion
scattering amplitudes starting with the asymptotic solutions of Ref.
\cite{C4} we briefly present below.  The radial equations (\ref{TE})
can be analytically solved for  the continuous energy spectrum,
$\epsilon>\mu$,  in terms of Whittaker functions as \cite{C4}
\begin{eqnarray}
\hat f^+(x)&=&C_1^+\frac{1}{x}M_{r_+,s}(2i\nu x^2)
+C_2^+\frac{1}{x}W_{r_+,s}(2i\nu x^2)\,,\label{E11}\\
\hat f^-(x)&=&C_1^-\frac{1}{x}M_{r_-,s}(2i\nu x^2)
+C_2^-\frac{1}{x}W_{r_-,s}(2i\nu x^2)\,,\label{E22}
\end{eqnarray}
where  we denote
\begin{equation}
s=\sqrt{\kappa^2+\frac{\mu^2}{4}-\epsilon^2},\quad
r_{\pm}=\mp\frac{1}{2}-i q,\quad q=\nu+ \frac{\mu^2}{2\nu} \,.
\end{equation}
The  integration constants must satisfy \cite{C4}
\begin{equation}\label{C1C1}
\frac{C_1^-}{C_1^+}=\frac{s-i q}{\kappa-i\lambda}\,,\quad
\frac{C_2^-}{C_2^+}=-\frac{1}{\kappa-i\lambda}\,,\quad \lambda=\frac{\epsilon\mu}{2\nu}\,.
\end{equation}

We observe that these solutions are similar to those of the
relativistic  Dirac-Coulomb problem. The functions
$M_{r_{\pm},s}(2i\nu x^2)=(2i\nu x^2)^{s+\frac{1}{2}}[1+O(x^2)]$ are
regular in $x= 0$, where the functions $W_{r_{\pm},s}(2i\nu x^2)$
diverge as $x^{1-2s}$ if $s>\frac{1}{2}$ \cite{NIST}. These
solutions will help us to find the scattering amplitudes of the
Dirac particles by black holes, after fixing the integration
constants.

\section{Partial wave analysis}

We consider now the scattering of Dirac fermions on a black hole.
This   is described by the energy eigenspinor $U$ whose asymptotic
form,
\begin{equation}
U\to U_{plane}(\vec{p})+A(\vec{p},\vec{n}) U_{sph}\,,
\end{equation}
for $r\to \infty$  (where the gravitational field vanishes)   is
given by the plane wave spinor of momentum $\vec{p}$ and the free
spherical spinors of the flat case behaving as
\begin{equation}\label{Up}
U_{sph}\propto \frac{1}{r}\,e^{ipr-iEt}\,,\quad p=\sqrt{E^2-m^2}=\frac{\nu}{r_0}\,,
\end{equation}
Here we fix the geometry such that $\vec{p}=p\vec{e_3}$ while the
direction  of the scattered fermion is given by the scattering
angles $\theta$ and $\phi$ which are just the spheric angles of the
unit vector $\vec{n}$.  Then the scattering amplitude
\begin{equation}\label{Ampl}
A(\vec{p},\vec{n})=f(\theta)+ig(\theta)\frac{ \vec{p}\land \vec{n}}{|\vec{p}\land \vec{n}|}\cdot\vec{\sigma}
\end{equation}
depend on  two scalar amplitudes, $f(\theta)$ and $g(\theta)$, that
can be  studied by using the partial wave analysis.

\subsection{Asymptotic conditions and phase shifts}

The partial wave analysis exploits the asymptotic form of the exact
analytic  solutions which satisfy suitable boundary conditions that
in our case might be fixed at the event horizon (where $x=0$).
Unfortunately, here we have here only the asymptotic solutions
(\ref{E11}) and (\ref{E22}) whose integration constants cannot be
related to those of the approximative solutions near event horizon
\cite{C4} without resorting to numerical methods
\cite{FHM}-\cite{S3}. Therefore, as long as we restrict ourselves
only to an analytical study, we cannot match  boundary conditions
near event horizon being forced to find suitable  asymptotic
conditions for determining the integration constants. Then, the
behaviour near horizon can be discussed, a posteriori, focusing on
the partial radial currents and especially on the absorption cross
section that can be easily derived in our approach.

When the Dirac fermions are scattered from central potentials, the
particular  spinors of parameters $(p,\kappa)$ represent partial
waves having the asymptotic form
\begin{equation}\label{AsiF}
{\cal F}\propto \begin{array}{c}
\sqrt{E+m}\sin\\
\sqrt{E-m}\cos
\end{array}\left( pr-\frac{\pi l}{2} +\delta_{\kappa}\right)\,,
\end{equation}
where the phase shifts $\delta_{\kappa}$ can be, in general, complex
numbers  which allow one to write down  the amplitudes in terms of
Lagrange polynomials of $\cos \theta$. Each partial wave gives rise
to the radial current (\ref{Jrad}) that can be put now in the form
\begin{equation}\label{Jrad1}
J_{rad}(p,\kappa)\propto -p \sinh(2 \Im \delta_{\kappa})\,.
\end{equation}
The scattering is elastic when the phase shifts $\delta_{\kappa}$
are real  numbers such that the partial radial currents vanish.
Whenever there are inelastic processes the phase shifts become
complex numbers, whose imaginary parts are related to the inelastic
behavior, producing the non-vanishing radial currents indicating
absorption.

In the Appendix C we show that in our approach it is necessary to
adopt the  general  asymptotic condition  $C_2^+=C_2^-=0$  in order
to have {\em elastic} collisions with a correct Newtonian limit for
large angular momentum.   It is remarkable that  this asymptotic
condition selects the asymptotic spinors that are regular in $x=0$
but this cannot be interpreted as a boundary condition since the
exact solutions have a different structure near the event horizon
\cite{C4}.

Under such circumstances we have to use the asymptotic forms of the
radial  functions for large $x$ resulted from  Eq. (\ref{A3}). We
observe that the first term from (\ref{A3}) is dominant for $\hat
f^+$, while the second term in (\ref{A3}) is dominant for  $\hat
f^-$, such that for $x\to \infty$  the radial functions behave as:
\begin{eqnarray}
\hat f^+(x)&\to& C_1^+e^{-\frac{1}{2}\pi q}\frac{\Gamma(2s+1)}{\Gamma(1+s+iq)}\,e^{i[\nu x^2+q\ln(2\nu x^2)]}\,,\\
\hat f^-(x)&\to& C_1^-e^{-\frac{1}{2}\pi q}\frac{\Gamma(2s+1)}{\Gamma(1+s-iq)}\,e^{-i[\nu x^2+\pi s+q\ln(2\nu x^2)]}\,.
\end{eqnarray}
Hereby  we  derive the asymptotic form (\ref{AsiF}) of the doublet
${\cal F} = T\hat{\cal F}$ observing that the argument of the
trigonometric functions can be deduced as $\frac{1}{2} {\arg}
\left(\frac{\hat f^+}{\hat f^-}\right)$. Then by using Eq.
(\ref{C1C1}a) and taking into account that Eqs. (\ref{x}) and
(\ref{Up}b) allow us to replace $\nu x^2=p(r-r_0)$, we obtain  the
definitive asymptotic form of the radial functions for the
scattering by black holes,
\begin{equation}
{\cal F}=\left(
\begin{array}{c}
i\sqrt{\epsilon+\mu}\,(\hat f^- -\hat f^+)\\
\sqrt{\epsilon-\mu}\,(\hat f^+ +\hat f^-)
\end{array}\right)
\propto \begin{array}{c}
\sqrt{E+m}\,\sin\\
\sqrt{E-m}\,\cos
\end{array}\left( pr-\frac{\pi l}{2} +\delta_{\kappa}+\vartheta(r)\right)\,,
\end{equation}
whose point-independent phase shifts $\delta_{\kappa}$  give the quantities
\begin{equation}\label{final}
S_{\kappa}=e^{2i\delta_{\kappa}}=\left(\frac{\kappa-i\lambda}{s-iq}\right)\,\frac{\Gamma(1+s-iq)}{\Gamma(1+s+iq)} e^{i\pi(l-s)}\,.
\end{equation}
Notice that the values of  $\kappa$ and $l$ are related as in Eq. (\ref{kjl}), i. e. $l=|\kappa|-\frac{1}{2}(1-{\rm sign}\,\kappa)$. The remaining point-dependent phase,
\begin{equation}
\vartheta(r)= -p r_0+q \ln [2p(r-r_0)]\,,
\end{equation}
which does not depend on angular quantum numbers, may be ignored as
in the  Dirac-Coulomb case \cite{LL,S3}.

We arrived thus at the final result (\ref{final}) depending on the
parameters  introduced above that can be expressed in terms of
physical quantities by using Eq. (\ref{Up}b) as,
\begin{eqnarray}
s&=&\sqrt{\kappa^2-k^2}\,,\quad k=M\sqrt{4p^2+3m^2}\,,\\
q&=&\frac{M}{p}\,(2p^2+m^2)\,,\label{q}\\
\lambda&=&\frac{M}{p} m\sqrt{m^2+p^2}\,. \label{lam}
\end{eqnarray}
The parameters $k,q,\lambda \in {\Bbb R}^+$ are  positively defined
and satisfy the identity
\begin{equation}\label{kqlam}
k^2=q^2-\lambda^2\,.
\end{equation}
In the particular case of the massless fermions ($m=0$) we remain
with  the unique parameter $k=q=2pM$ since $\lambda=0$.

The parameter $s$ has a special position since this can take either
real  values  or pure imaginary ones. Let us briefly discuss these
two cases:

{\em The elastic scattering} arises for the values of $\kappa$  (at
given $p$) that satisfy the condition
\begin{equation}\label{cond}
|\kappa|\geq n+1\,,
\end{equation}
where $n={\rm floor} (k)$ is the largest integer less than $k$.
Then $s=\sqrt{\kappa^2-k^2}\in{\Bbb R}$ and the identity
(\ref{kqlam})  guarantees that  the phase shifts of Eq.
$(\ref{final})$ are real numbers  such that $|S_{\kappa}|=1$ .

{\em The absorbtion} is present in the partial waves for which  we have
\begin{equation}\label{cond1}
1\le |\kappa|\leq n\,.
\end{equation}
Here we meet a branch point in $s=0$ and two solutions $s=\pm
i|s|=\pm i\sqrt{k^2-\kappa^2}$  among them we must chose $s=-i|s|$
since only in this manner we select the physical case of
$|S_{\kappa}|<1$. More specific, by substituting $s=-i|s|$ in Eq.
(\ref{final})  we obtain the simple closed form
\begin{equation}\label{SSk}
|S_{\kappa}|=|S_{-\kappa}|=e^{-2\Im \delta_{\kappa}}=e^{-\pi |s|}\sqrt{\frac{\sinh \pi (q-|s|)}{\sinh \pi (q+|s|)}}
\end{equation}
showing that $0< |S_{\kappa}|< 1$ since $|s|<q$ for any $(p,\kappa)$
obeying the condition (\ref{cond1}).  Moreover, we can verify that
in the limit of the large momentum (or energy) the absorption tends
to become maximal since
\begin{equation}\label{limS}
\lim_{p\to \infty} |S_{\kappa}|=0\,,
\end{equation}
regardless of the fermion mass.

Finally, we note that for any values of the above parameters the
phase  shifts $\delta_l$ and $\delta_{-l}$ are related as
\begin{equation}
e^{2i(\delta_l-\delta_{-l})}=\frac{pl-imM \sqrt{p^2+m^2}}{pl+imM \sqrt{p^2+m^2}}
\end{equation}
which means that for massless Dirac fermions we have $\delta_l=\delta_{-l}$.

This is the basic framework of the relativistic partial wave
analysis of the Dirac fermions scattered by Schwarzschild black
holes in which we consider exclusively the contribution of the
scattering modes. Our results are in accordance with the Newtonian
limit since in the large-$l$ limits and for very small momentum we
can take $s\sim |\kappa|\sim l$ and  $\lambda\sim q$   such that our
phase shifts (\ref{final}) becomes just the Newtonian ones
(\ref{shN}) \cite{FHM}.

On the other hand, we observe that our phase shifts are given by a
formula which has the same form as that deduced for the
Dirac-Coulomb scattering \cite{LL}. Indeed, by replacing in Eq.
(\ref{final}) the parameters
\begin{equation}\label{Coul}
s\to \sqrt{\kappa^2-Z^2\alpha^2}\,,\quad q\to \frac{Z\alpha}{p}E\,,\quad \lambda \to \frac{Z\alpha}{p}m\,.
\end{equation}
 we recover the phase shifts of the Dirac-Coulomb scattering. However, there are significant
  differences between these  parameterizations showing that these two systems are of different natures.

\subsection{Partial amplitudes and cross sections}

Now we have all the elements for calculating the analytic
expressions  of the  amplitudes and cross sections in terms of the
phase shifts defined by Eq. (\ref{final}). In the relativistic
theory, the scalar amplitudes of Eq. (\ref{Ampl}),
\begin{eqnarray}
f(\theta)&=&\sum_{l=0}^{\infty}a_l\,P_l(\cos \theta)\,,\label{f}\\
g(\theta)&=&\sum_{l=1}^{\infty}b_l\,P_l^1(\cos\theta)\,, \label{g}
\end{eqnarray}
depend on the following {\em partial} amplitudes \cite{LL,S3},
\begin{eqnarray}
&&a_l=(2l+1)f_l=\frac{1}{2ip}\left[(l+1)(S_{-l-1}-1)+l(S_l-1)\right]\,,\label{fl}\\
&&b_l=(2l+1)g_l=\frac{1}{2ip}\left(S_{-l-1}-S_l\right)\,.\label{gl}
\end{eqnarray}
In our problem the quantities $S_{\kappa}$ are given by the analytic
expression  (\ref{final}) such that we can apply analytic methods
for studying the above amplitudes that give  the {\em elastic}
scattering intensity or  differential cross section,
\begin{equation}\label{int}
\frac{d\sigma}{d\Omega}=|f(\theta)|^2+|g(\theta)|^2\,,
\end{equation}
and the polarization degree,
\begin{equation}\label{pol}
{\cal P}(\theta)=-i\frac{f(\theta)^*g(\theta)-f(\theta) g(\theta)^*}{|f(\theta)|^2+|g(\theta)|^2}.
\end{equation}
This last quantity is interesting for the scattering of massive
fermions  representing the induced polarization for an unpolarized
initial beam.

Let us verify first that our approach recovers the correct Newtonian
limit in  weak gravitational fields (with small $M$) where the
partial amplitudes can be expanded as,
\begin{eqnarray}
f_l&=&M f_l^{(1)}+M^2 f_l^{(2)}+\cdots \quad l=0,1,2,...\,,\\
g_l&=&M g_l^{(1)}+M^2 g_l^{(2)}+\cdots \quad l=1,2,...\,.
\end{eqnarray}
Our algebraic codes indicate that  the cases $l=0$ and $l>0$ must be
studied separately  since  the expansion  of $f_l$ does not commute
with its limit for $l\to 0$.   However, this is not surprising since
a similar phenomenon can be meet in the Dirac-Coulomb problem.
Then, according to Eqs. (\ref{fl}), (\ref{gl}) and (\ref{final}), we
obtain first the expansion for $l=0$,
\begin{eqnarray}
f_0^{(1)}&=&\gamma\frac{2p^2+m^2}{p^2}-\frac{2p^2+m^2-m\sqrt{p^2+m^2}}{2p^2}\,,\\
f_0^{(2)}&=&i\gamma^2\frac{(2p^2+m^2)^2}{p^3}-i\gamma\frac{2p^2+m^2}{p^3}(2p^2+m^2-m\sqrt{p^2+m^2})\nonumber\\
&-&i\frac{m(2p^2+m^2)}{2p^3}\sqrt{p^2+m^2}+i\frac{4p^4+5p^2m^2+2m^4}{4p^3}+\pi\frac{4p^2+3m^2}{4p}\,,
\end{eqnarray}
where $\gamma$ is the Euler constant. Furthermore,  for any $l>0$ we find the terms  of first order,
\begin{eqnarray}
f_l^{(1)}&=&-\psi(l+1)\frac{2p^2+m^2}{p^2}\,,\hspace*{21mm}.\\
g_l^{(1)}&=&-\frac{1}{2l(l+1)}\frac{2p^2+m^2 -m\sqrt{p^2+m^2}}{p^2}\,,
\end{eqnarray}
depending  on the digamma function $\psi$, and the more complicated
ones of second order,
\begin{eqnarray}
f_l^{(2)}&=&i\psi(l+1)^2\frac{(2p^2+m^2)^2}{p^3}-\frac{i}{2l(l+1)}\frac{m(2p^2+m^2)}{p^3}\sqrt{p^2+m^2}\nonumber\\
&+&\frac{i}{4l(l+1)}\frac{2m^4+5p^2m^2+4p^4}{p^3}+\frac{\pi}{2 (2l+1)}\frac{4p^2+3m^2}{p}\,,\\
g_l^{(2)}&=&\frac{i\psi(l+1)}{l(l+1)}\frac{2p^2+m^2}{p^3}(2p^2+m^2 -m\sqrt{p^2+m^2})+\frac{i}{l^2(l+1)^2}\frac{m(2p^2+m^2)}{p^3}\sqrt{p^2+m^2}\nonumber\\
&-&\frac{i}{4l^2(l+1)^2}\frac{4p^4+5m^2p^2+2m^4}{p^3}-\frac{\pi}{4l(l+1)(2l+1)} \frac{4p^2+3m^2}{p}\,,
\end{eqnarray}
which lay out the dependence on $l$ and $p$. Then, from Eq.
(\ref{fN}) we  observe that $f_l^{(1)}$ and the first term of
$f_l^{(2)}$ are of Newtonian form, increasing with $l$ because of
the digamma function. The other terms decrease when $l$ is
increasing so that we verify again  that our partial wave analysis
has a correct Newtonian limit for large $l$, as it happens  in the
well-known case of the scalar particles \cite{FHM}.

In other respects, we observe that the partial amplitudes of the
massless  fermions are regular in $p=0$ while those of the massive
particles  diverge in any order. Thus  the problem of removing the
infrared catastrophe in the massive case seems to remain a serious
challenge.

Important global quantities, independent on $\theta$,  are the total
cross sections. The {\em elastic}  cross section,
\begin{eqnarray}
\sigma_e&=&2\pi\int_{-1}^{1}d\cos\theta \, \left[|f(\theta)|^2+|g(\theta)|^2\right]\nonumber\\
&=&4\pi\sum_{\kappa} (2l+1)\left[|f_l|^2+l(l+1)|g_l|^2\right]\nonumber\\
&=&\frac{\pi}{p^2}\sum_{\kappa}\left\{2l+1+(l+1)|S_{-l-1}|^2+l|S_l|^2-2\left[(l+1)\Re S_{-l-1}+l\Re S_l\right]\right\}\,,\label{sigmae}
\label{ts}
\end{eqnarray}
is obtained by integrating the scattering intensity (\ref{int}) over
$\theta$ and $\phi$, according to the normalization integral
\begin{equation}
\int_{-1}^{1}dx P_l^m(x)P_{l'}^m(x)=\frac{2\delta_{ll'}}{2l+1}\frac{(l+m)!}{(l-m)!}\,.
\end{equation}
The general expression  (\ref{sigmae}) gives the elastic cross
section  even in the presence of absorption  when $|S_{\kappa}|<1$.
For the genuine elastic scattering  (without  absorption and
$|S_{\kappa}|=1$),  the elastic cross section $\sigma_e$ represents
just the total cross section $\sigma_t$ such that we can write
\begin{equation}
\sigma_t=\frac{2\pi}{p^2}\sum_{\kappa}\left[2l+1-(l+1)\Re S_{-l-1}-l\Re S_l\right]\,,
\end{equation}
deducing that the absorption cross section,
$\sigma_a=\sigma_t-\sigma_e$,   reads \cite{S3}
\begin{equation}\label{sigmaaa}
\sigma_a=\frac{\pi}{p^2}\sum_{\kappa}\left[(l+1)(1-|S_{-l-1}|^2)+l(1-|S_l|^2)\right]
=\frac{2\pi}{p^2}\sum_{l=1}^{n}l (1-|S_l|^2)\,,
\end{equation}
since for $s=-i|s|$ we have $|S_{-\kappa}|=|S_{\kappa}|$ as in Eq. (\ref{SSk}).

This last cross section deserves a special attention since this can
be calculated at any time as a finite sum indicating how the
fermions can be absorbed by black holes. This is a function on the
momentum $p$ since all our parameters, including $n$, depend on it.
For this reason it is convenient to represent the absorption cross
section as,
\begin{equation}\label{abss}
\sigma_a=\sum_{l\ge 1}\sigma_a^l (p)
\end{equation}
in terms of the partial cross sections whose  definitive closed form,
\begin{equation}\label{sigmaa}
\sigma_a^l(p)=\theta(k-l)\frac{2\pi l}{p^2}
\left[1-e^{-2\pi\sqrt{k^2-l^2}}\, \frac{\sinh
\pi(q-\sqrt{k^2-l^2})}{\sinh \pi(q+\sqrt{k^2-l^2})}\right]\,,
\end{equation}
is derived according to Eqs. (\ref{SSk}) and (\ref{sigmaaa}) while
the condition (\ref{cond1})  introduces the Heaviside step function
$\theta(k-l)$. Hereby we understand that the absorption arises in
the partial wave $l$ for the values of $p$ satisfying the condition
$k>l$. This means that for large values of $l$  there may appear
non-vanishing thresholds,
\begin{equation}
p_l =\left\{
\begin{array}{clr}
0&{\rm if}& 1\le l\le \sqrt{3}mM\\
\frac{\sqrt{l^2-3m^2M^2}}{2M}&{\rm if}& l>\sqrt{3}mM
\end{array}\right. \,,
\end{equation}
indicating that the fermions with $|\kappa|=l$ can be absorbed by
black hole only if $p\ge p_l$.  Obviously, when $\sqrt{3}mM<1$ we
have $p_l\not=0$  for any $l$ such that we meet a  non-vanishing
threshold in every partial wave.  This happens for massless fermions
too when we have the equidistant thresholds $p_l=\frac{l}{2M}$ for
any $l\ge 1$.

The existence of these thresholds is important since these keep
under control the effect of the singularities in $p=0$. Thus for any
partial wave with $p_l>0$ the Heaviside function prevents the
partial cross section to be singular but when $p_l=0$ then $p$
reaches the singularity point $p=0$ where the partial cross section
diverges. Obviously, for $\sqrt{3}mM<1$ all the partial sections are
finite.

Finally, we must specify that in the high-energy  limit  all these
absorption cross sections tend to the event horizon (apparent) area,
indifferent on the fermion mass  $m\ge 0$, as it results from  Eq.
(\ref{limS}) that yields
\begin{equation}\label{asy}
\lim_{p\to \infty}\sigma_a =\lim_{p\to
\infty}\frac{2\pi}{p^2}\,\sum_{l=1}^{n}  1=\lim_{p\to
\infty}\frac{\pi}{p^2}\,n(n+1)=4\pi M^2
\end{equation}
since  $n(n+1)\sim k^2\sim 4M^2 p^2$. This asymptotic value is less
than the geometrical optics value of $27\pi M^2$ \cite{Un} which is
the high-energy limit of the absorption cross sections obtained
applying analytical-numerical methods \cite{bh1,S3}. The explanation
could be that here we neglected the effects of the bound states that
may give rise to a resonant scattering or even to a supplemental
absorption mechanism able to increase the cross section.

\section{Numerical examples}

Our purpose now is to use the graphical analysis for understanding
the  physical consequences of our analytical results encapsulated in
quite complicated formulas and the infinite series (\ref{f}) and
(\ref{g}) which are  poorly convergent or even divergent since their
coefficients $ a_l$ and $b_l$ are increasing with $l$  faster than
$l \ln l$ and respectively $ l^{-1}$.

This is an unwanted effect of the singularity at $\theta=0$ but
which can be attenuated  adopting  the method of Ref. \cite{Yeni}
that resides in replacing the series  (\ref{f}) and (\ref{g}) by the
$m$th reduced ones,
\begin{eqnarray}
f(\theta)&=&\frac{1}{(1-\cos\theta)^{m_1}} \sum_{l\ge 0} a_l^{(m_1)} P_l(cos\theta)\,,\label{gf}\\
g(\theta)&=&\frac{1}{(1-\cos\theta)^{m_2}} \sum_{l\ge 1} b_l^{(m_2)} P_l^1(\cos\theta)\,.\label{gf1}
\end{eqnarray}
The recurrence relations satisfied by the Legendre polynomials
$P_{l}(x)\,,P_{l}^{1}(x)$ lead  to    the iterative rules giving the
reduced coefficients in any order
\begin{eqnarray}
a_l^{(i+1)}&=&a_l^{(i)}-\frac{l+1}{2l+3}a_{l+1}^{(i)}-\frac{l}{2l-1}a_{l-1}^{(i)}\,,\\
b_l^{(i+1)}&=&b_l^{(i)}-\frac{l+2}{2l+3}b_{l+1}^{(i)}-\frac{l-1}{2l-1}b_{l-1}^{(i)}\,,
\end{eqnarray}
if we start with $a_l^{(0)}=a_l$ and $b_l^{(0)}=b_l$ as defined by
Eqs. (\ref{fl})  and (\ref{gl}) and (\ref{final}). Note that in this
last equation we replace $s\to \Re s-i |\Im s|$ in order to cover
automatically both the cases of interest here,  elastic collision
and absorption.  Then, we will see that this method is very
effective assuring the convergence of the reduced series  for any
value of $\theta$ apart from the singularity in $\theta=0$. Here we
present the numerical results obtained by using the second iteration
for $f$  ($m_1=2$) and the first one for $g$  ($m_2=1$) that seem to
be satisfactory without distorting the analytical results.

In this approach the elastic and total cross sections cannot be
calculated  since their series remain divergent in any conditions
because of the mentioned singularity.  Even if we replace  the
amplitudes $f$ and $g$  with their reduced series we cannot assure
the convergence since then we  introduce additional factors of the
type $(1-\cos\theta)^{-n},\, n\geq2$, leading to divergent integrals
over $\theta$ when we calculate $\sigma_e$ as in Eq. (\ref{sigmae}).
Therefore, we are able to study only the absorption cross section
(\ref{sigmaa}) , which is given by a finite sum,  following  then
to focus on the scattering amplitudes, scattering intensity and
polarization degree.

All these quantities depend on three free parameters: the fermion
mass  $m$ and momentum $p$ and the black hole mass $M$. Since we
work in the asymptotic zone where the fermion energy is
$E=\sqrt{m^2+p^2}$, we can use  the fermion velocity $v=p/E$ as an
auxiliary parameter. With these parameters one can construct two
relevant dimensionless quantities that in usual units read
$GME/(\hbar c^3)$ and $mGM/(\hbar c)$. The last one  can be seen as
(proportional to) the ratio of  the black hole horizon to the
fermion Compton wavelength. In our natural units (with
$c=\hbar=G=1$) these quantities appear as $ME$ and respectively $mM$
being used for labeling our graphs.

We should mention that our graphical analysis is performed in what
follow only for the case of small or micro black holes, for which ME
and $mM$ take relatively small values, since in this manner we can
compare our results with those obtained by using
analytical-numerical methods \cite{S1}-\cite{S3}. However, our
analytical results presented above are valid for any values of these
parameters.

\subsection{Forward and backward scattering}

\begin{figure}[h!t]
\includegraphics[scale=0.4]{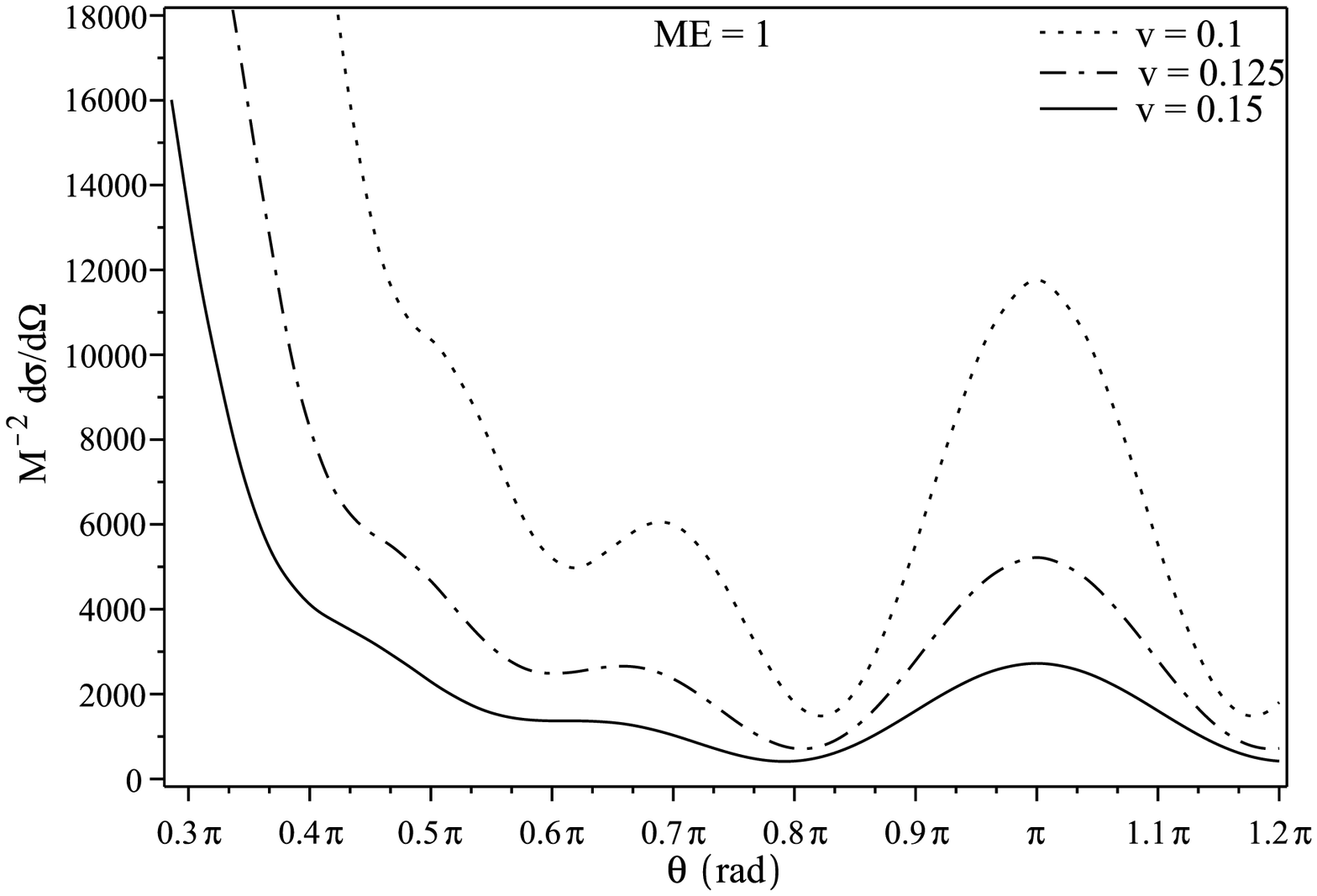}
\quad
\includegraphics[scale=0.4]{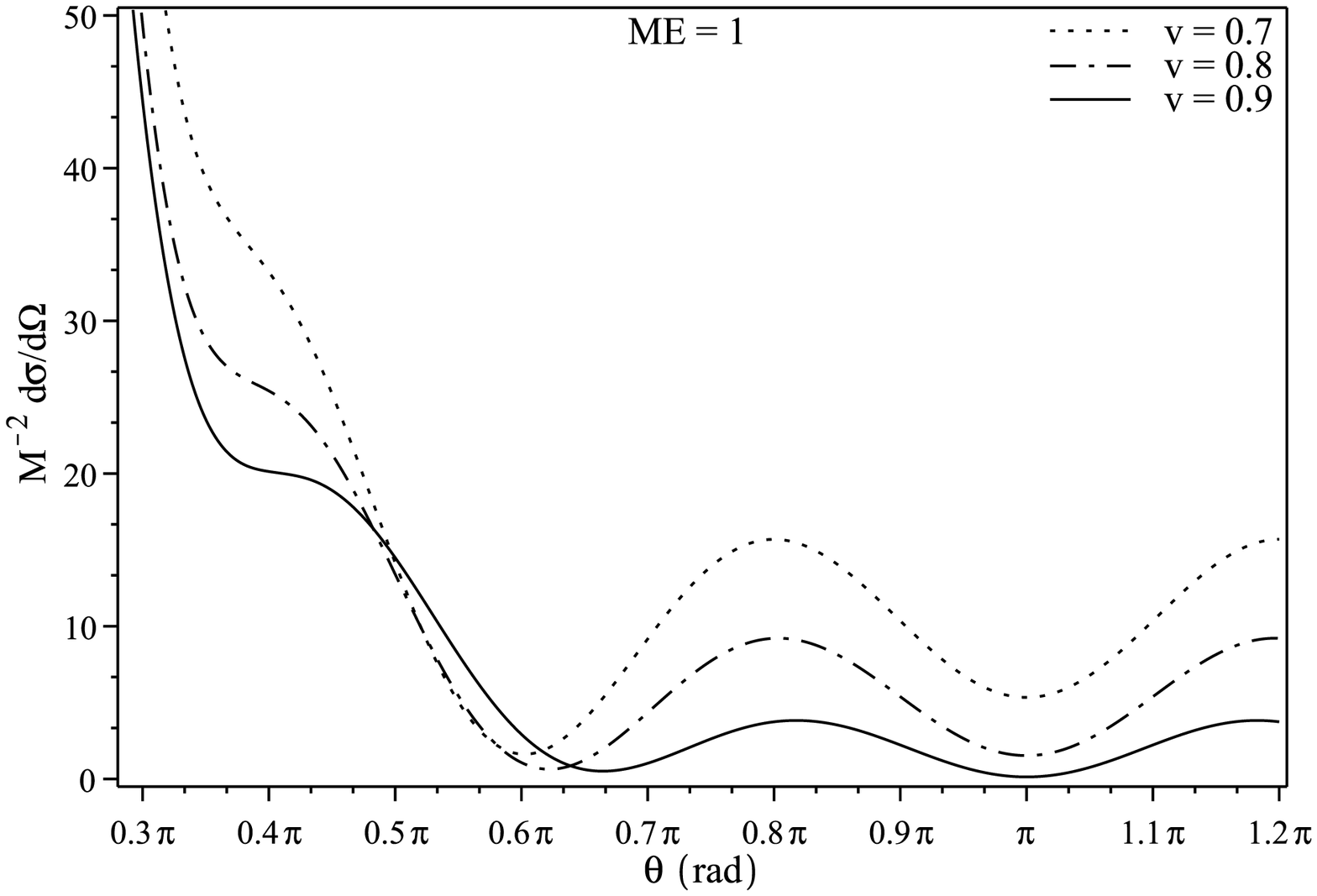}
\quad
\includegraphics[scale=0.4]{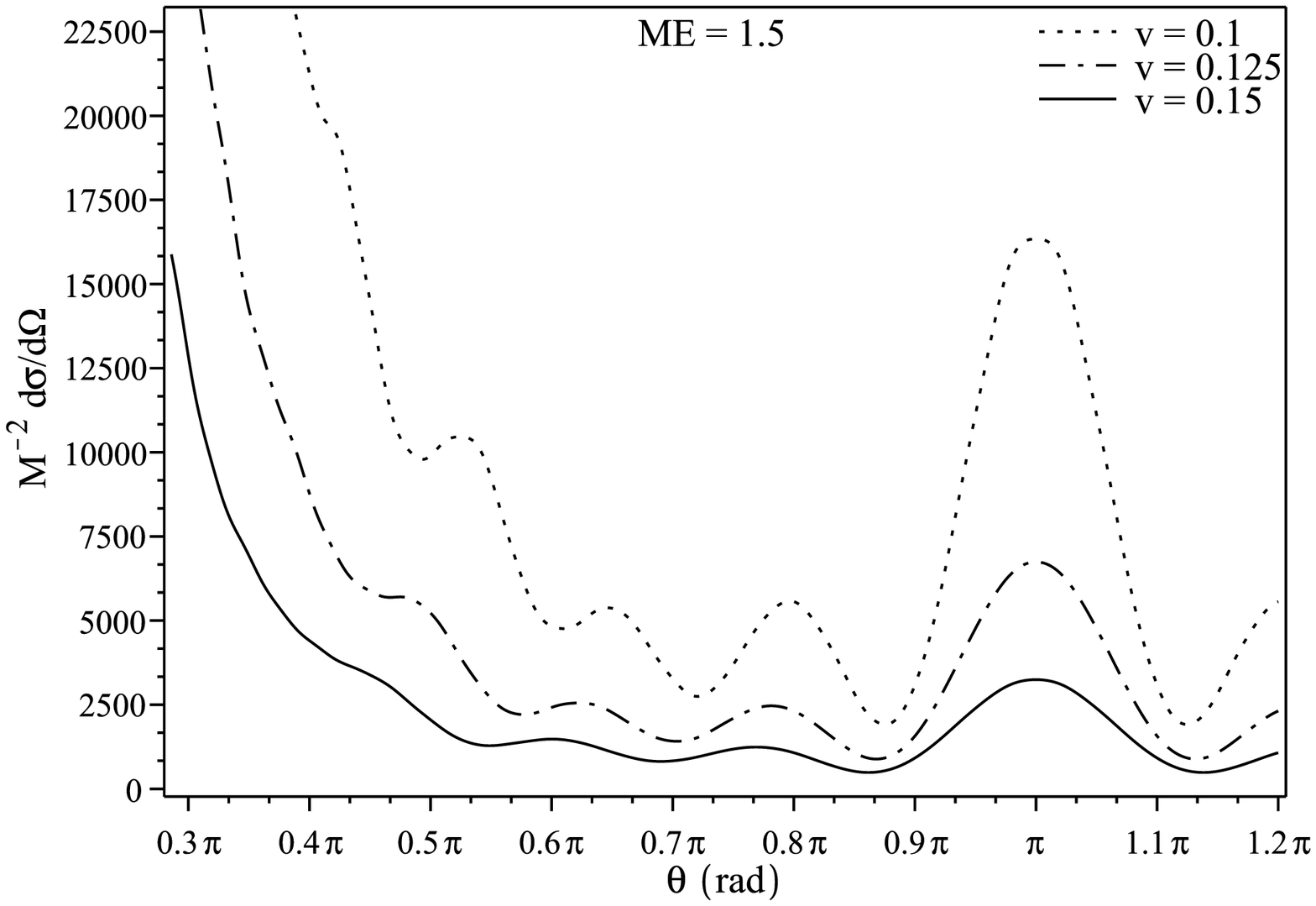}
\quad
\includegraphics[scale=0.4]{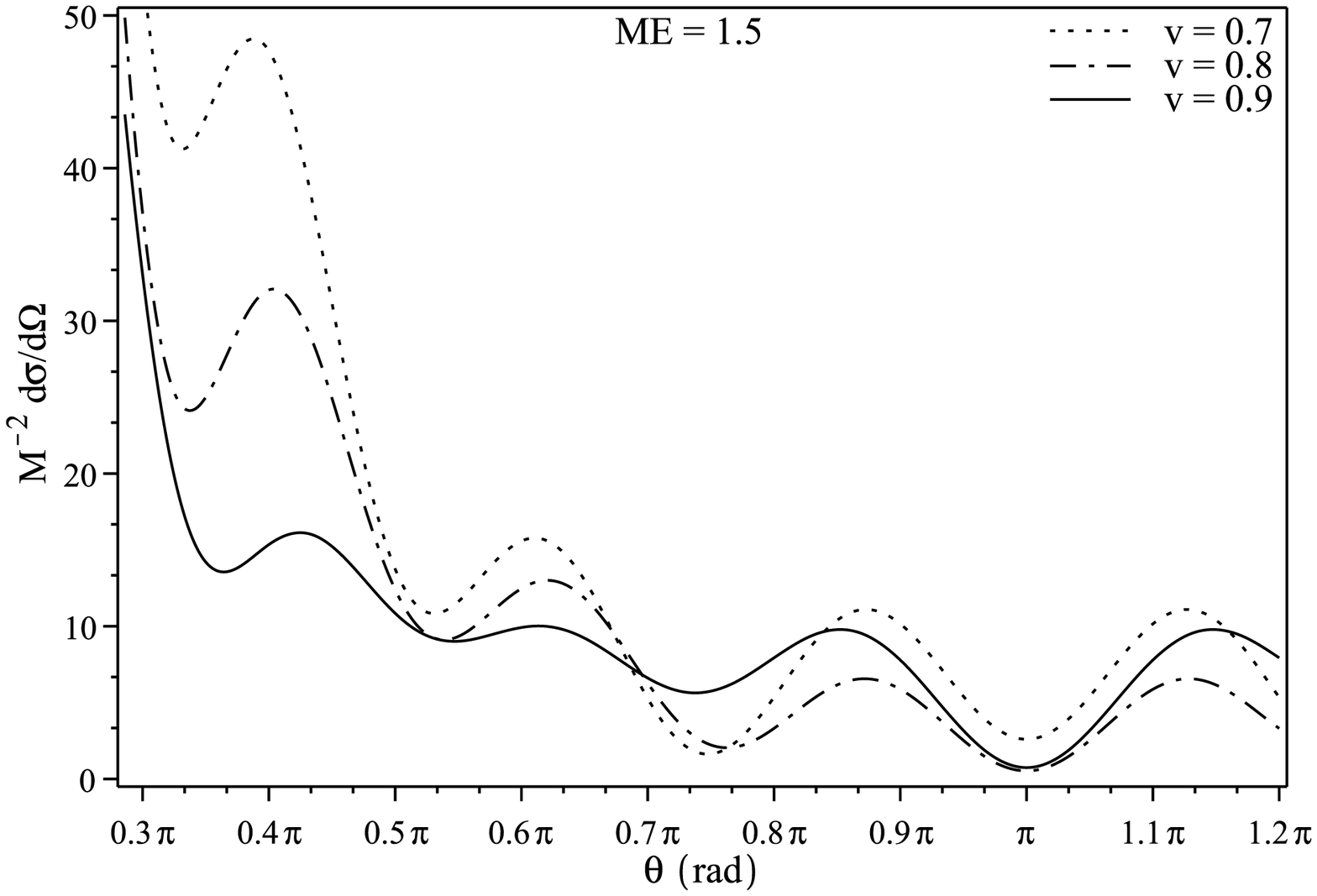}
\caption{The differential cross section as function  of $\theta$
for different values  of the fermion speed.} \label{f1}
\end{figure}

\begin{figure}[h!t]
\includegraphics[scale=0.4]{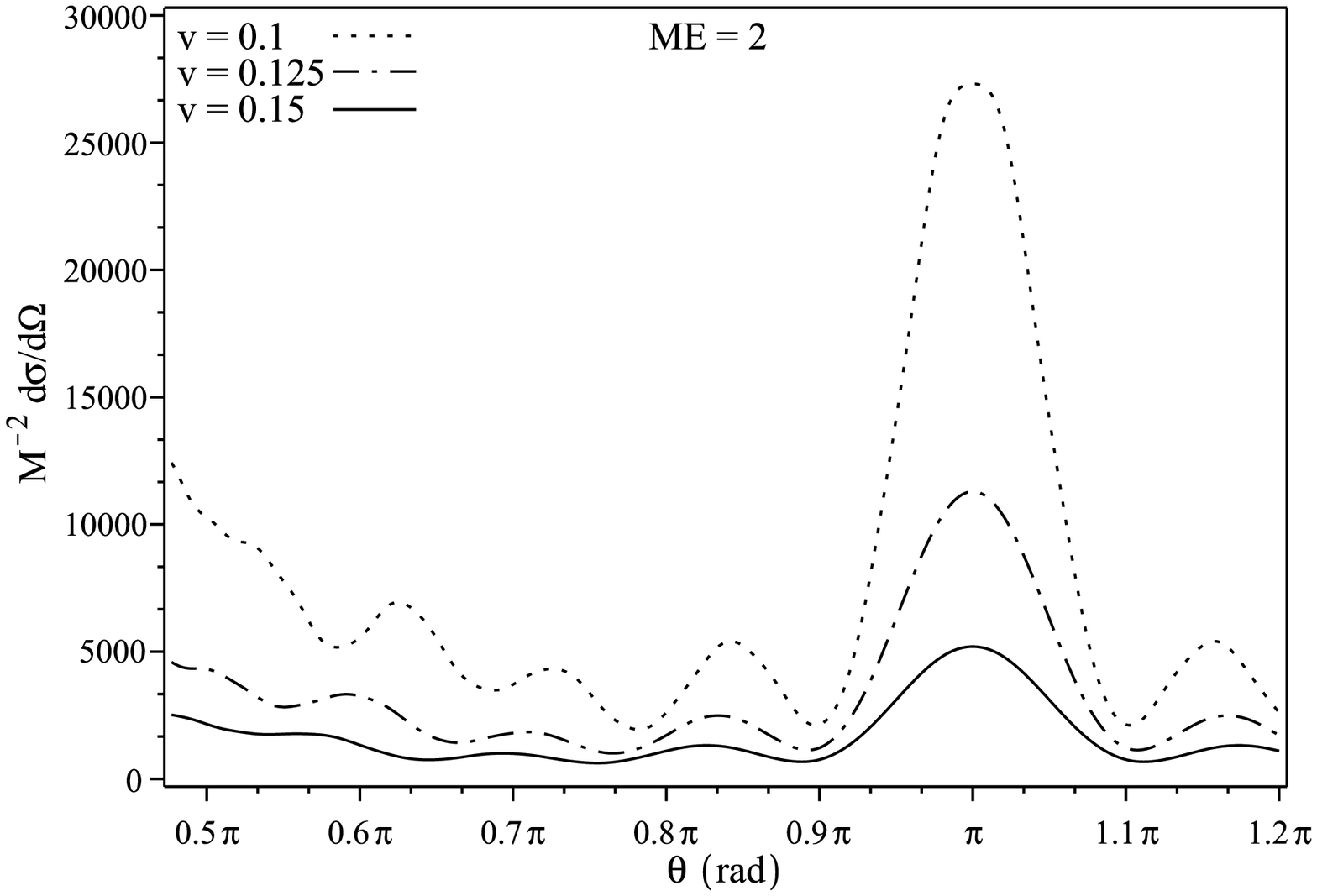}
\quad
\includegraphics[scale=0.4]{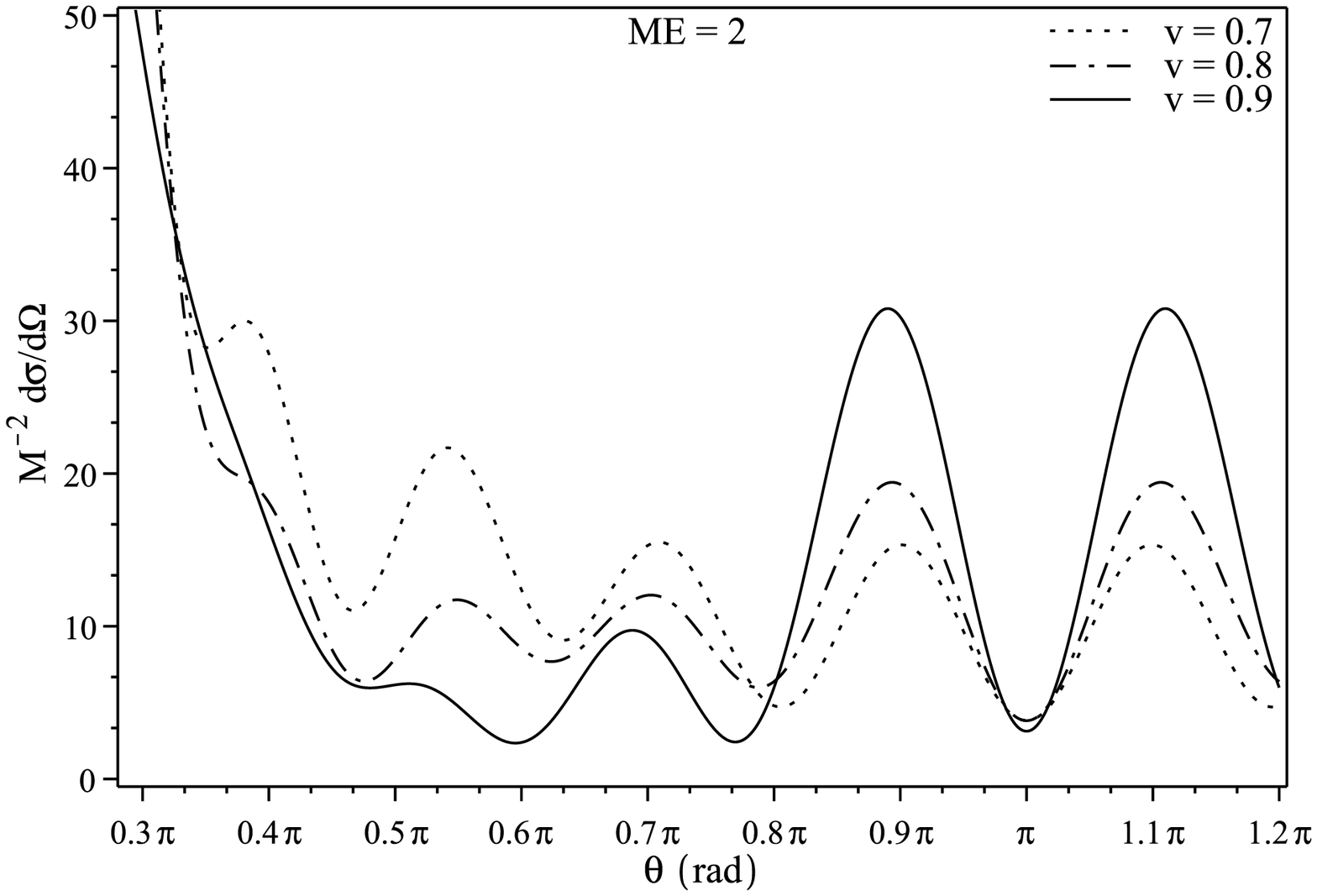}
\quad
\includegraphics[scale=0.4]{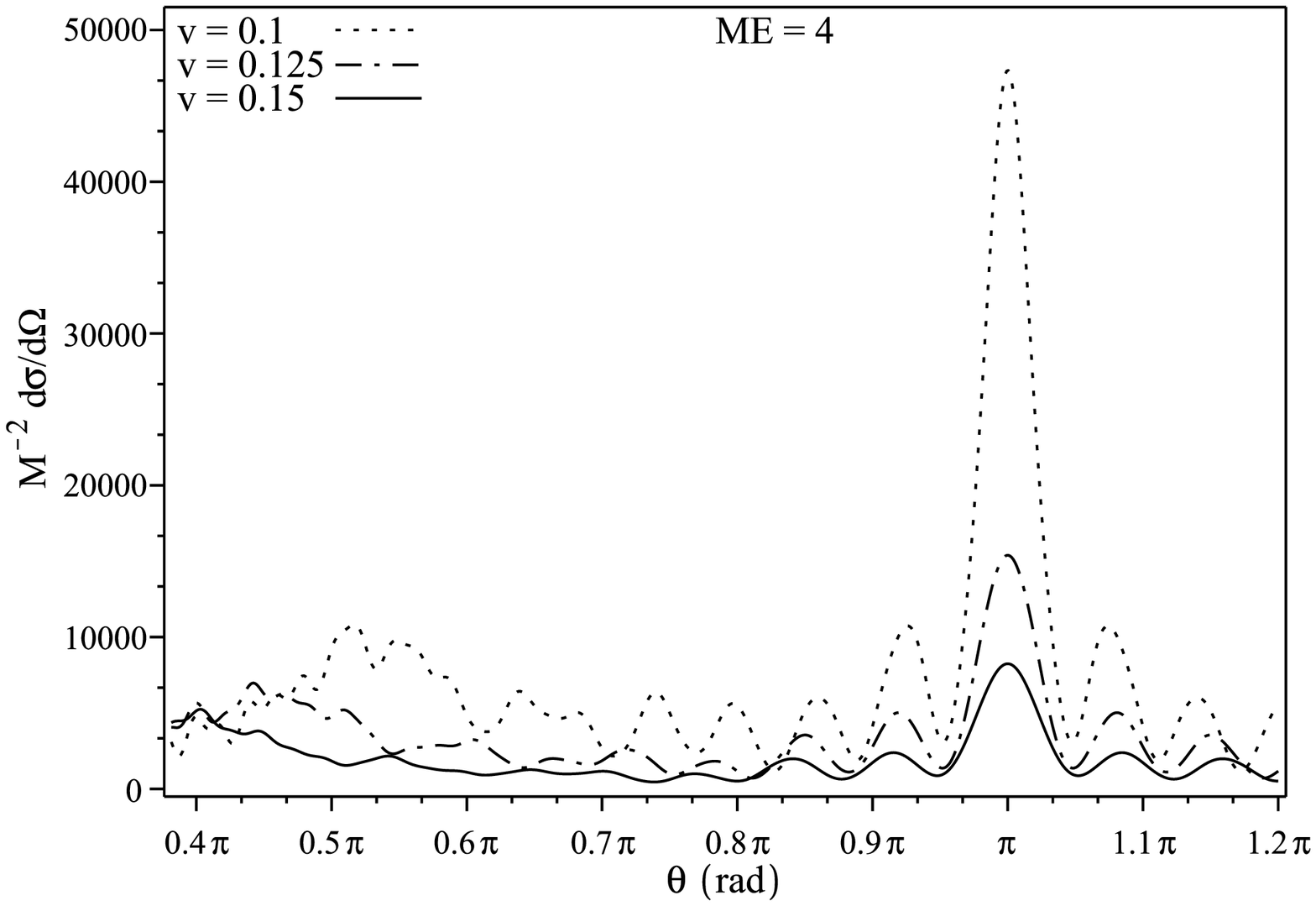}
\quad
\includegraphics[scale=0.4]{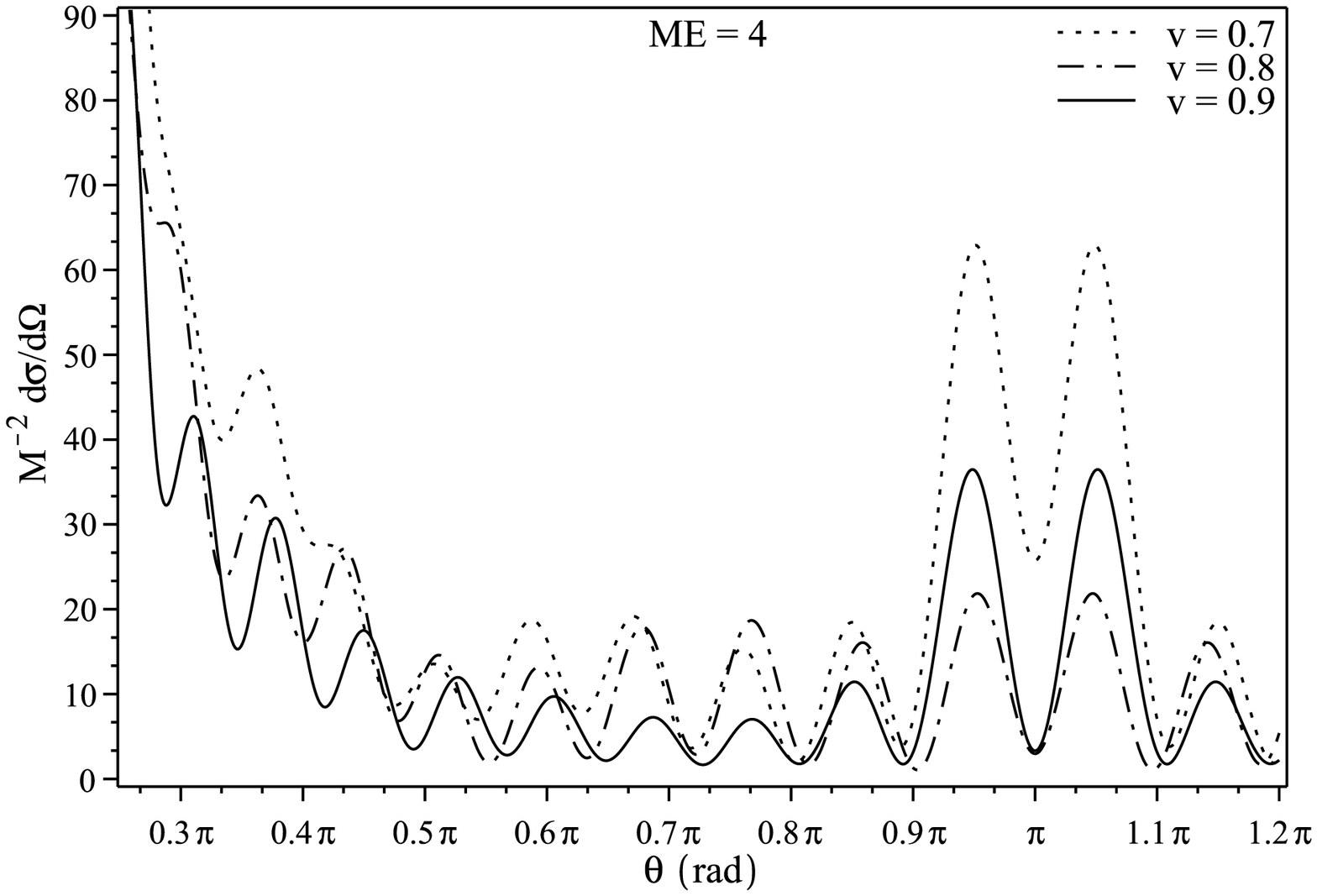}
\caption{The differential cross section as function  of $\theta$
for different values  of the fermion speed.} \label{f1a}
\end{figure}

We begin the graphical analysis by plotting the differential cross
section (\ref{int}), as  function of the angle $\theta$ for
different numerical values of $mM$ and $ME$. Since
$E=\sqrt{m^2+p^2}$, the condition $ME\geq mM$, must be always
satisfied. In addition, multiplying by $M$ the expression of energy
we obtain $mM=ME\sqrt{1-v^2}$ which gives the connection between the
pair of parameters that define our analytical formulas. The graphs
in Figs.(\ref{f1}-\ref{f1a}) show how the scattering intensity
depends on the scattering angle for small/large  fermion velocities.
In order to observe the oscillations in the scattering intensity
around $\theta=\pi$, corresponding to backward scattering, we
shorten the axis of $\theta$  because   the cross section is
divergent in $\theta=0$.

In Figs.(\ref{f1}-\ref{f1a}) we observe the presence of  a maximum
in scattering intensity in the backward direction.  This is known as
the glory scattering \cite{WF}, while the oscillations in the
scattering intensity at intermediate scattering angles, around
$\theta=\pi$, are known as orbiting or spiral scattering \cite{WF}.
Another important observation that emerge from these graphs, is that
the scattering intensity has large values only for small fermion
velocities while at large ones the scattering intensity is sensibly
smaller. The conclusion is that the glory and orbiting scattering
are significant only for non-relativistic fermions.

At $\theta=0$ the scattering intensity becomes divergent for both
small  or large values of the fermion speed. For this reason, in
order to obtain the behavior of scattering intensity at small
scattering angles for different fermion velocities, we restrict
ourselves only to values of $\theta$ close to $0$.

\begin{figure}[h!t]
\includegraphics[scale=0.4]{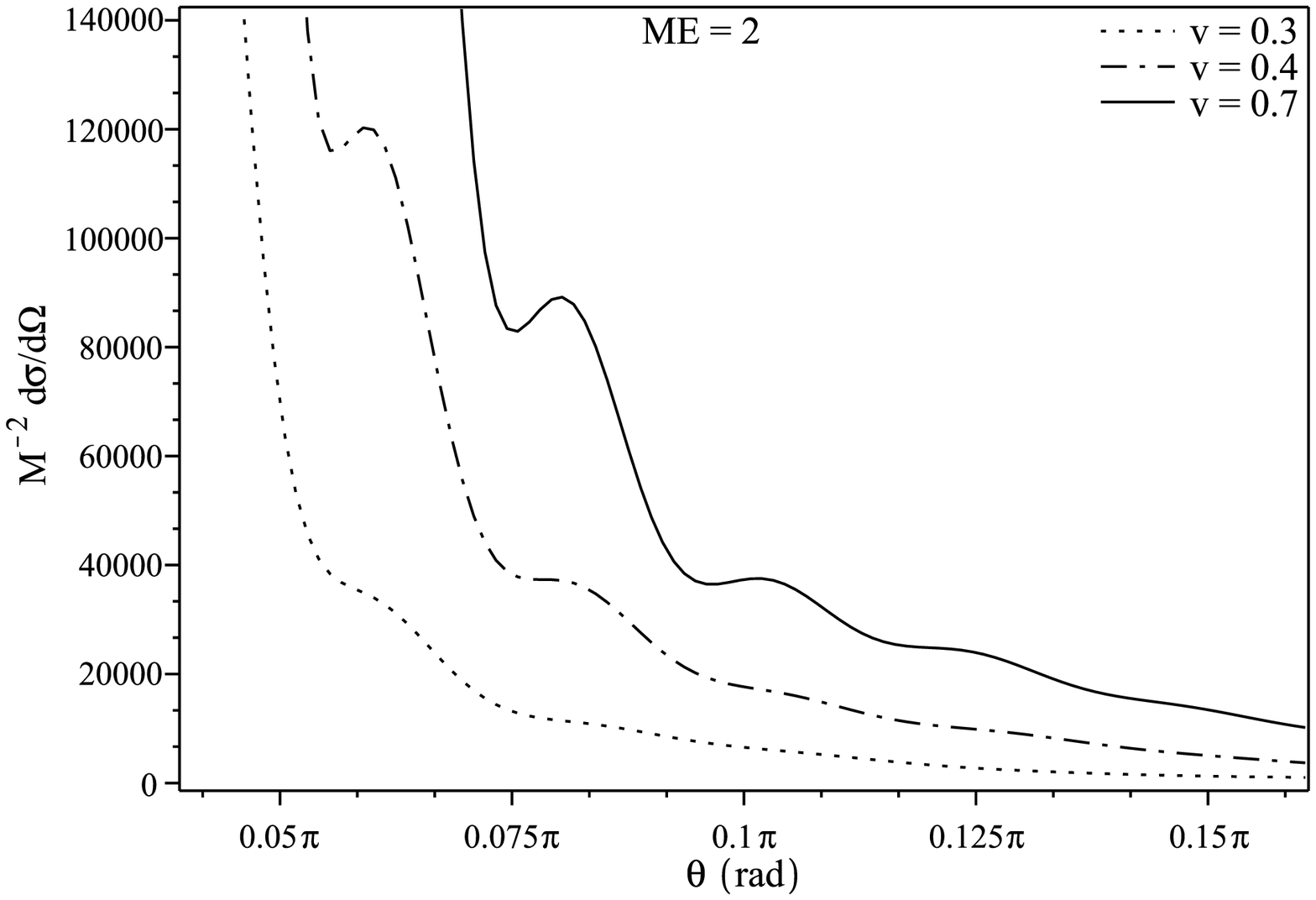}
\quad
\includegraphics[scale=0.4]{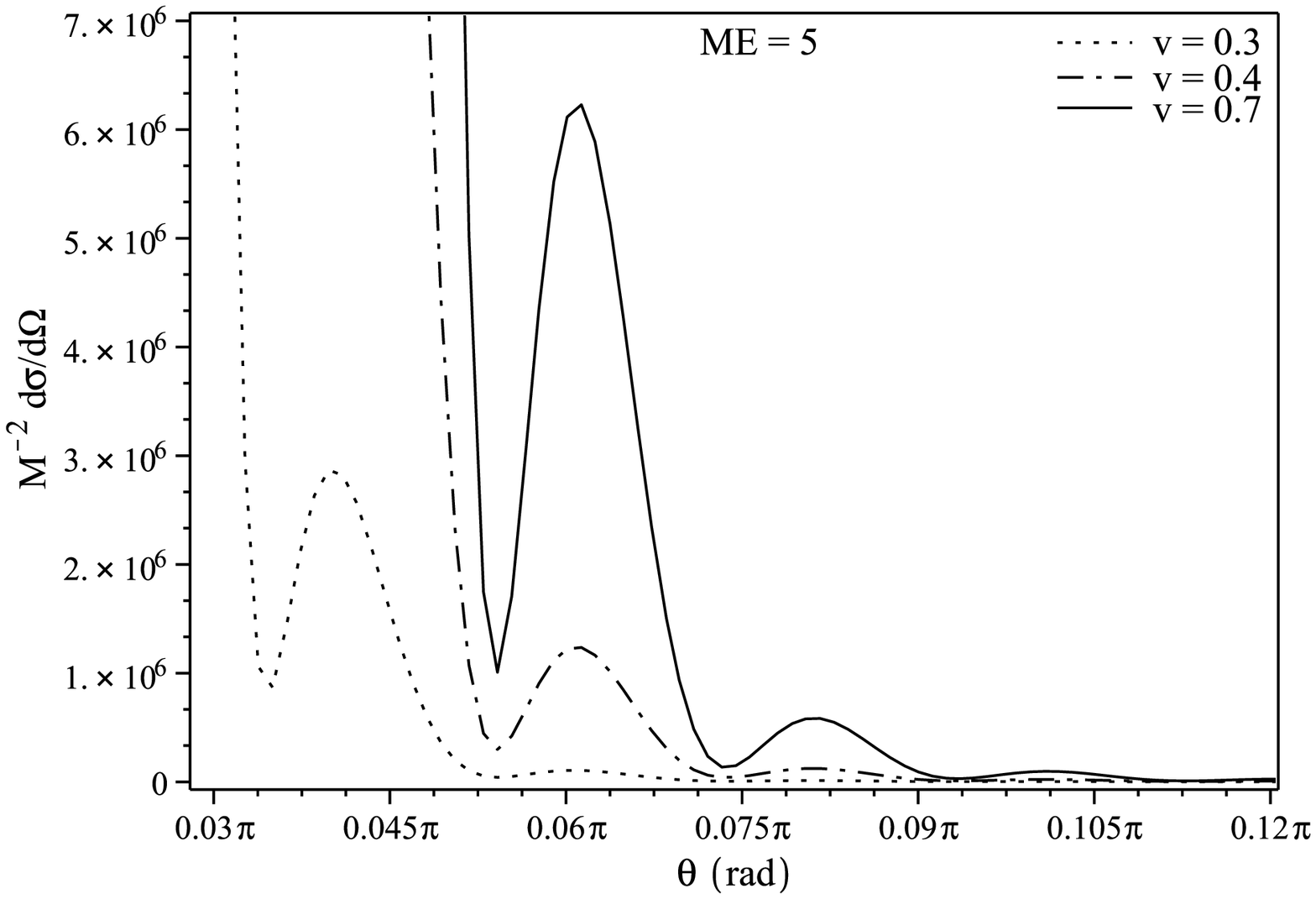}
\quad
\includegraphics[scale=0.4]{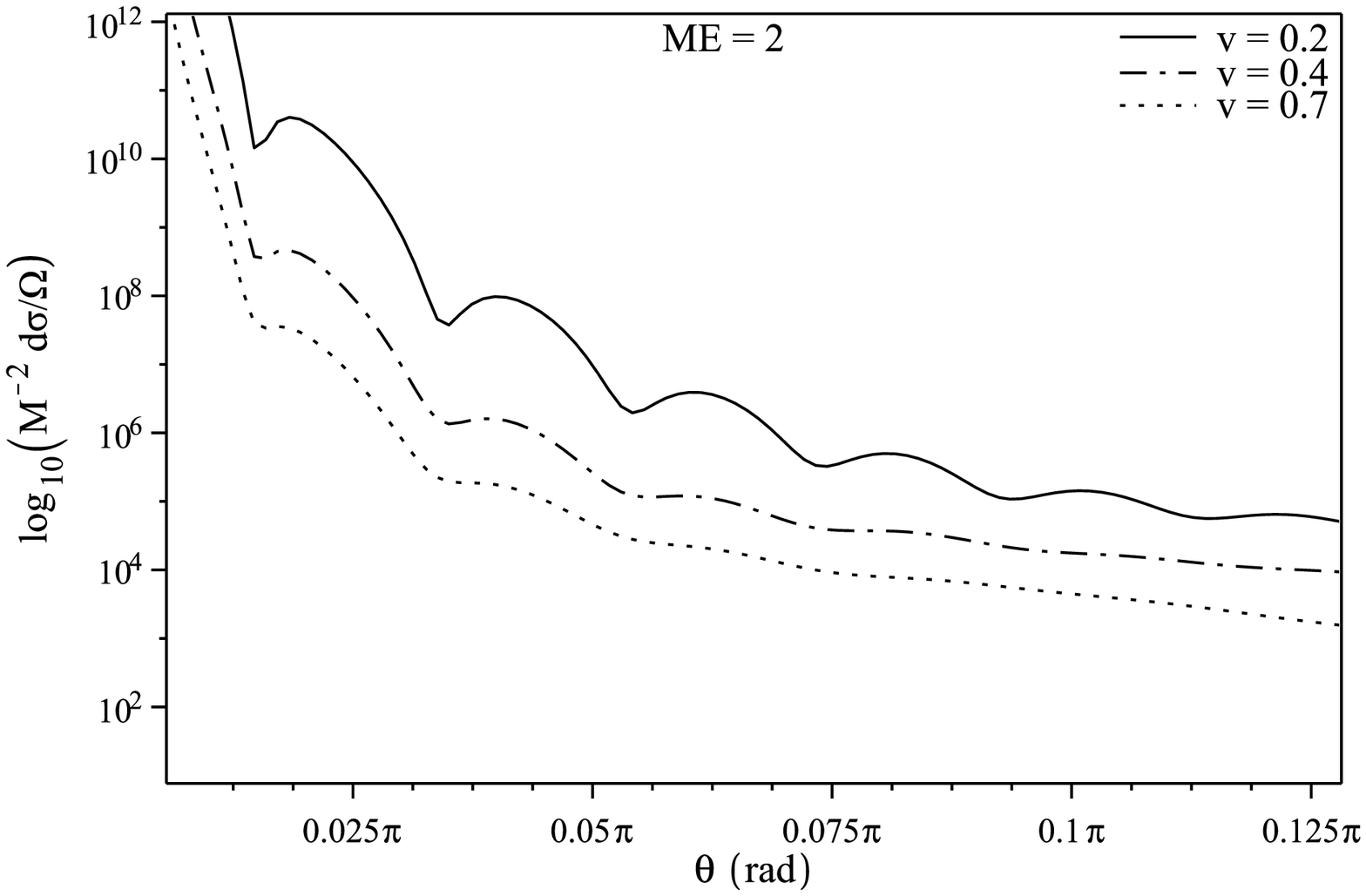}
\quad
\includegraphics[scale=0.4]{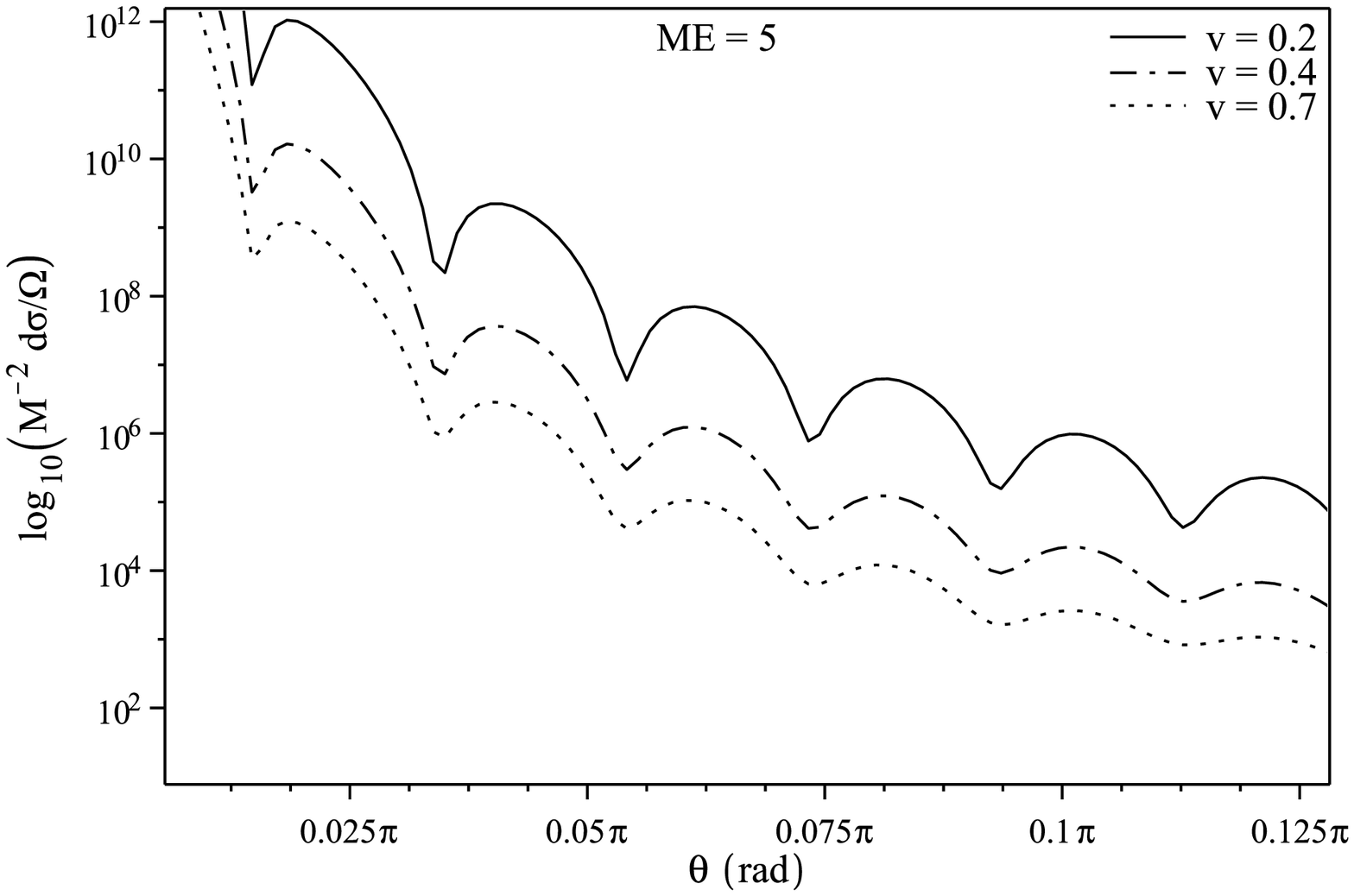}
\caption{The differential cross at small $\theta$,
 for different values of the fermion speed.}
\label{f2}
\end{figure}

\begin{figure}[h!t]
\includegraphics[scale=0.4]{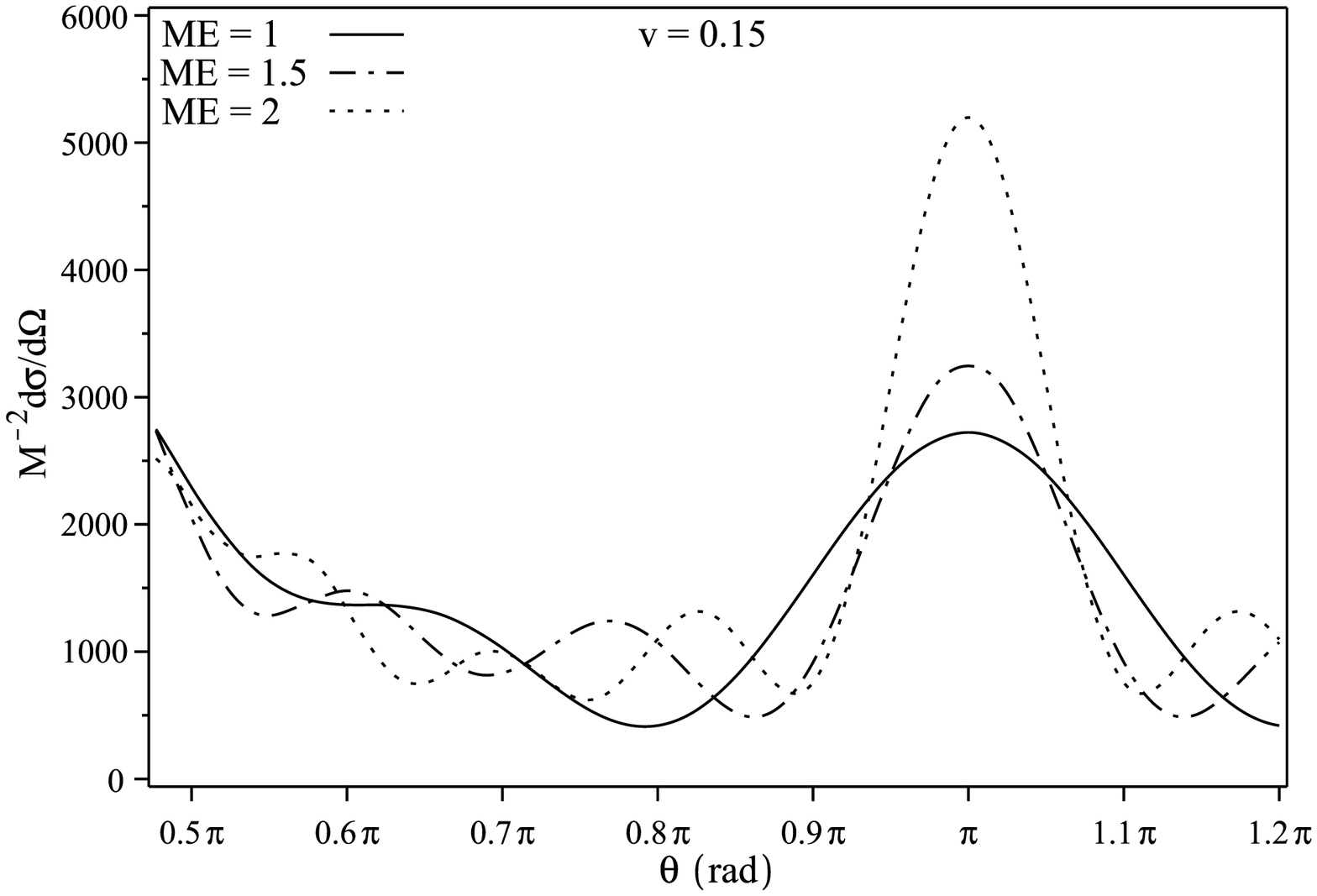}
\quad
\includegraphics[scale=0.4]{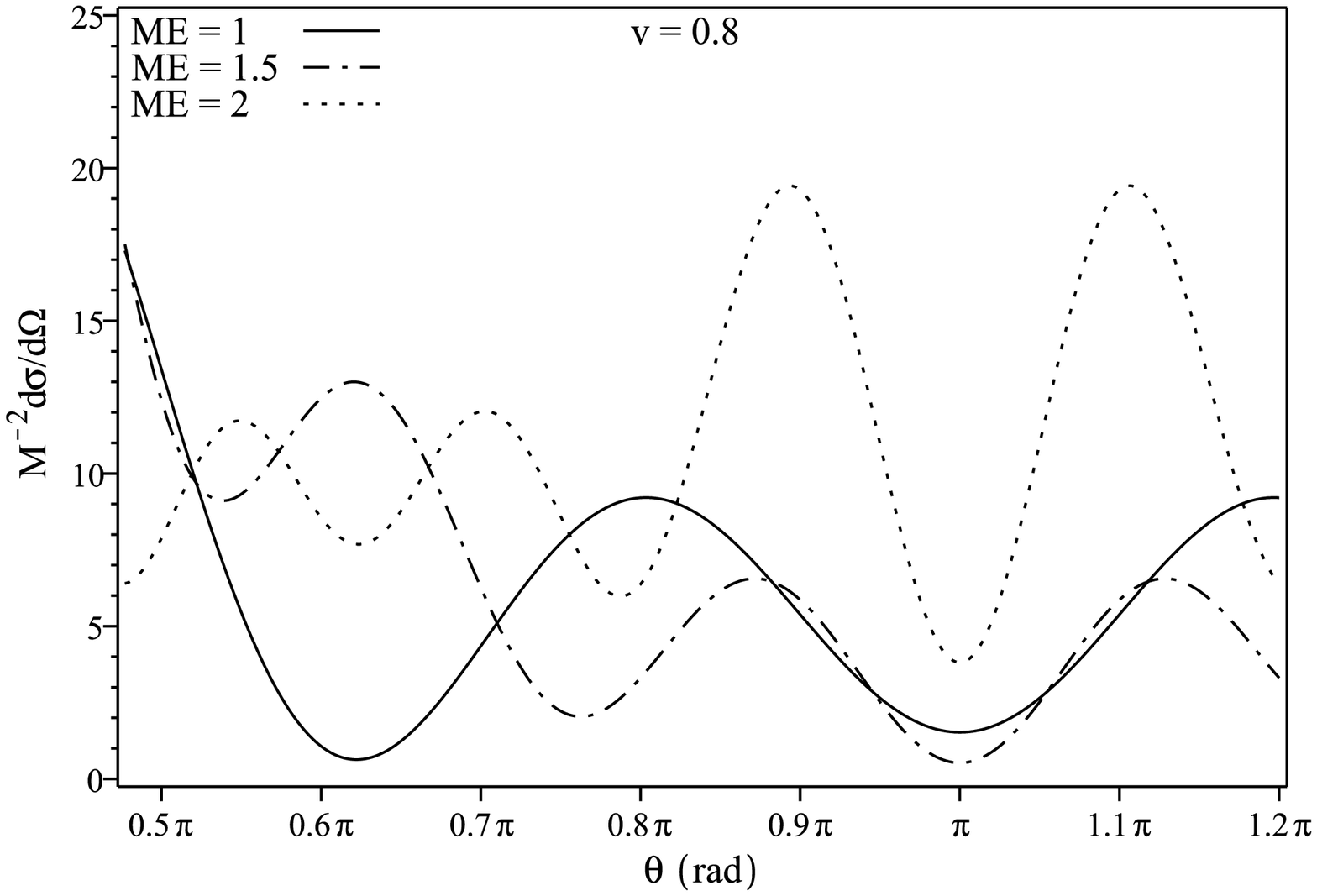}
\quad
\includegraphics[scale=0.4]{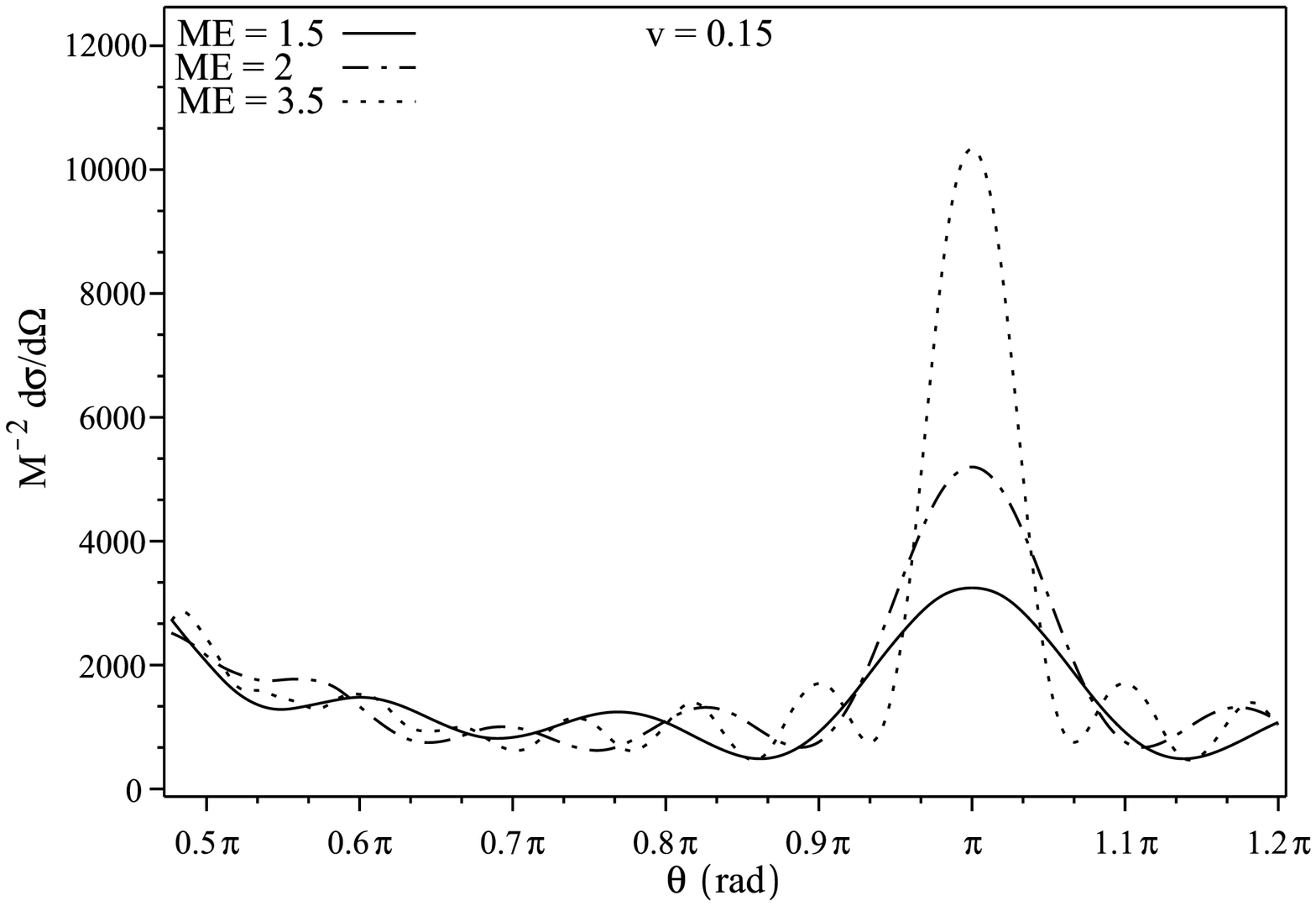}
\quad
\includegraphics[scale=0.4]{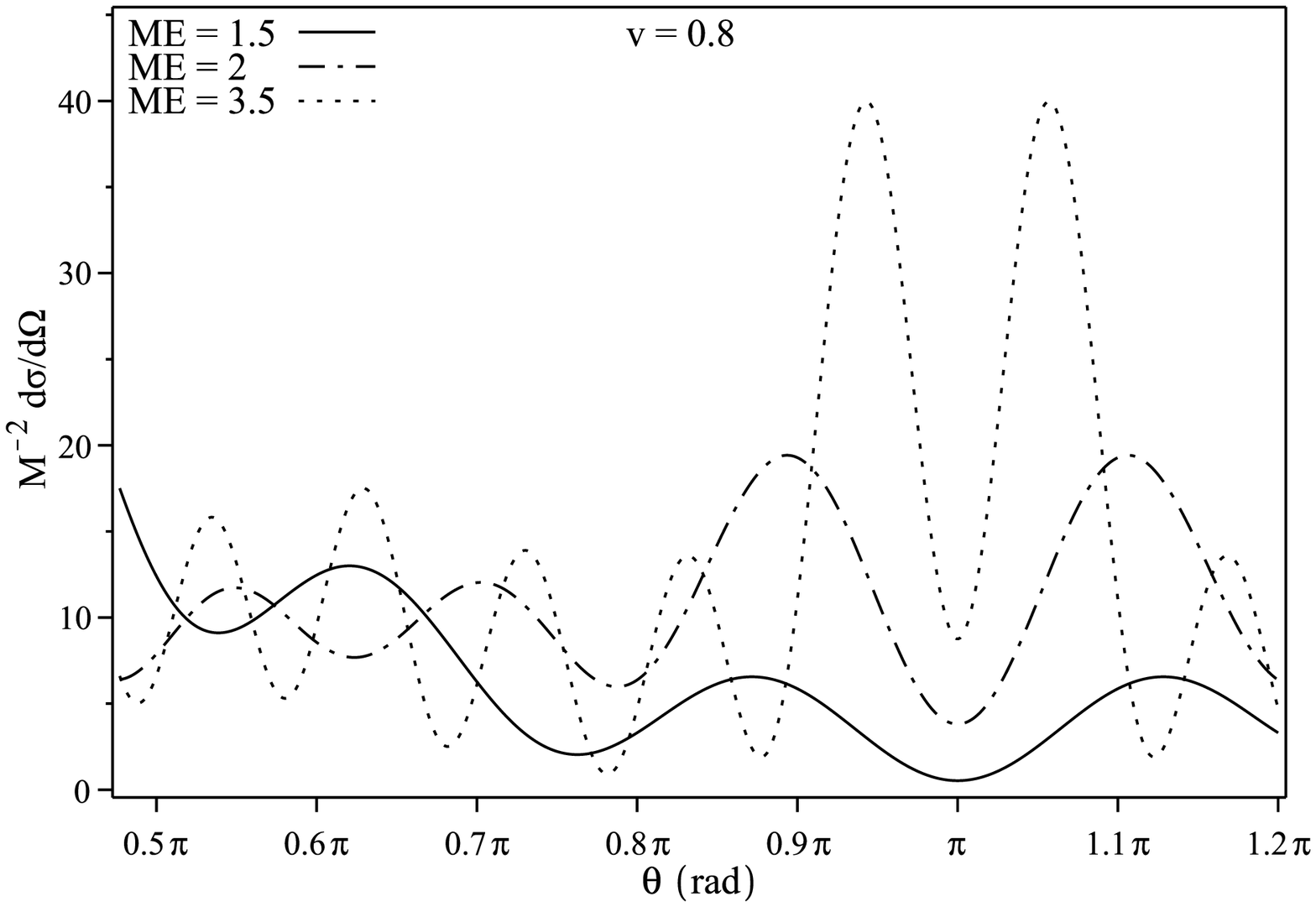}
\caption{The differential cross section as function of $\theta$  for
different $ME$ at fixed values of the fermion speed.} \label{f2b}
\end{figure}

Our graphs in Figs. (\ref{f2}) show that the scattering intensity
becomes divergent as $\theta\rightarrow 0$ (forward scattering) and
they also show the presence of oscillations around small values of
$\theta$, i.eE. the orbiting scattering \cite{WF}. We also observe
that the scattering intensity in the forward direction increases
with the parameter $ME$. The forward scattering is in fact a
diffraction on the black hole horizon. This could be possible only
in the case when the wavelength of the incident particle is
comparable with the size of the event horizon. From Figs. (\ref{f2})
we observe that the oscillatory behavior of scattering intensity in
the forward direction is more pronounced in the case of small
fermion velocities comparatively with the case of relativistic ones.
It is also worth to mention that the oscillatory behavior of the
scattering intensity is increasing with the parameter $ME$ (see
Figs. (\ref{f2b})). The conclusion is that the orbiting scattering
is negligible for relativistic particles while in the case of
particles with small velocities this phenomenon becomes important.

Furthermore, we address the problem of variation of the cross
section  with the black-hole mass given in Figs. (\ref{f2b}). We
observe that the scattering intensity in the backward direction is
increasing with the black hole mass but without changing its general
profile. This suggests that the positions of the relative extremes
are independent on the black hole mass.

Let us comment now the results of the scattering intensities
obtained  in graphs Figs. (\ref{f1}-\ref{f2b}). One knows that the
differential cross section represents the area that the incident
particle must cross in the target zone in order to be detected in
the solid angle $d\Omega$. In the case of a scattering process
between two quantum objects this quantity is very small. So in our
case it is not surprising that the differential cross section
becomes very large, since the target is of the size of the black
hole event horizon. This can be better understood if we recall the
result from classical physics according to which the scattering
intensity for a classical particle moving on spiral trajectory is
larger that the scattering intensity for a particle moving on a
straight line. A classical particle which has nonzero angular
momentum will always across a larger area in the target zone. The
situation is the same in the case of a quantum particle scattered by
a black hole but with the observation that in this case the notion
of trajectory is not well-defined. Taking into consideration that
the minimum area of the target is of the size of the event horizon
we see that the area crossed by the fermion to be detected in a
solid angle could be very large. As a final remark we can observe
that the scattering intensity in the forward/backward direction
increase with the black hole mass. This result is expected since the
area of the event horizon also increases with the black hole mass.
Our results are compatible with those obtained in the literature
using analytical-numerical methods \cite{S3}.

\newpage

\subsection{Dependence on energy}

We study now the behavior of the differential cross section in terms
of energy by plotting Eq. (\ref{int}) as function of ratio $E/m$ for
different scattering angles. Since for $\theta=0$ the differential
cross section is divergent, our analyze is done for
$\theta=\pi/3,\,\pi/4$.

\begin{figure}[h!t]
\includegraphics[scale=0.4]{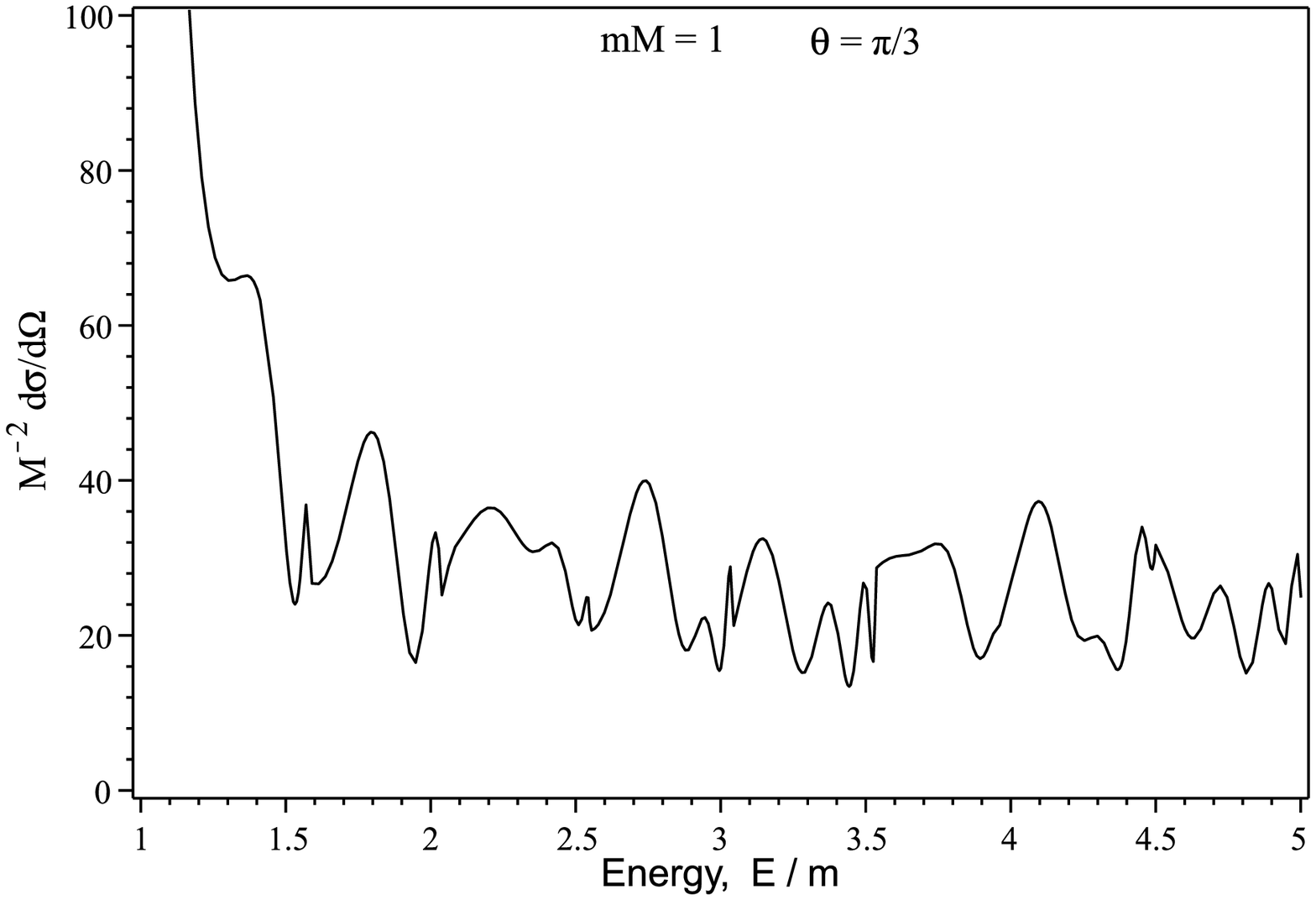}
\quad
\includegraphics[scale=0.4]{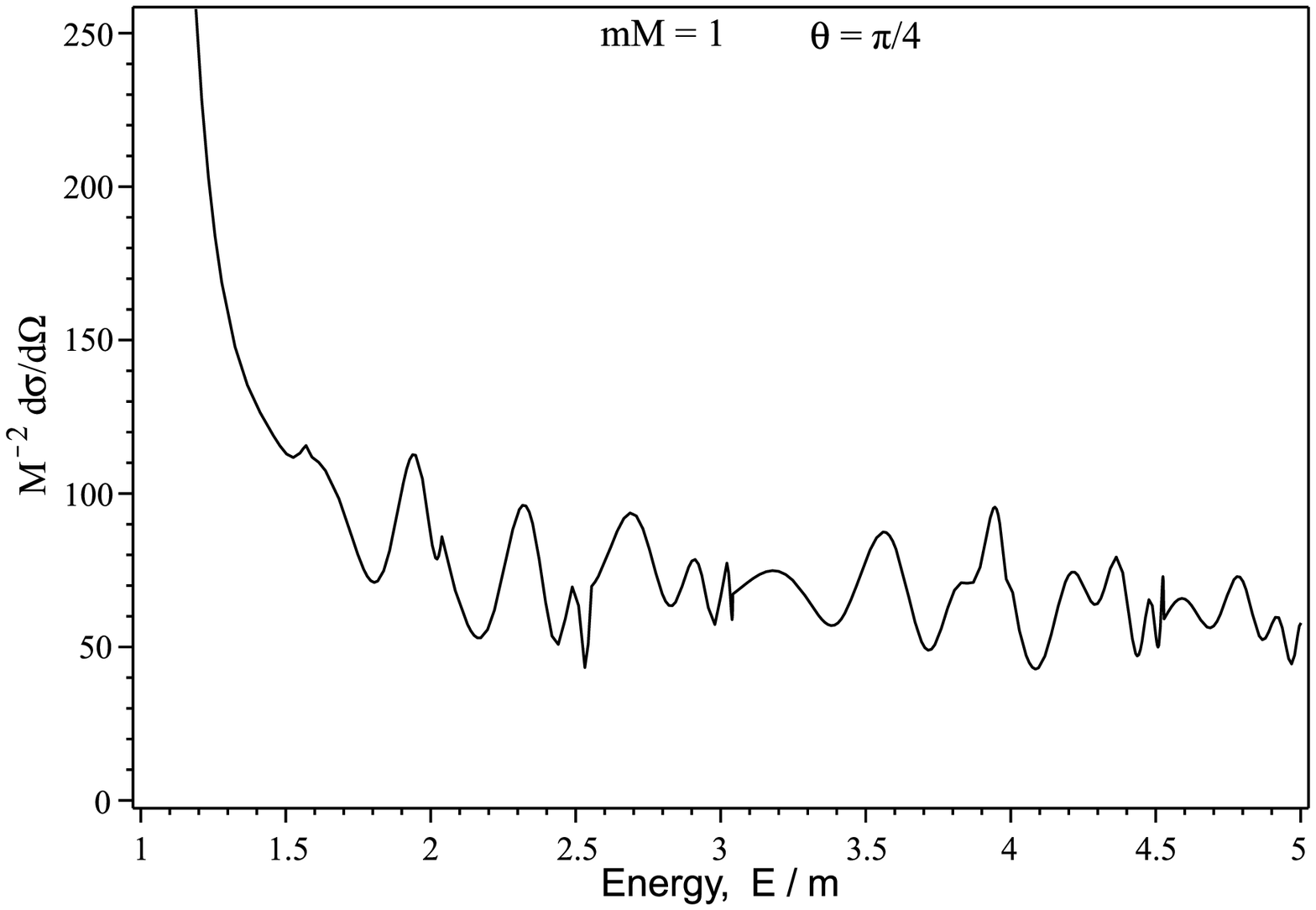}
\caption{Differential cross sections dependence of energy for different scattering angles and $mM=1$.}
\label{f4}
\end{figure}

\begin{figure}[h!t]
\includegraphics[scale=0.4]{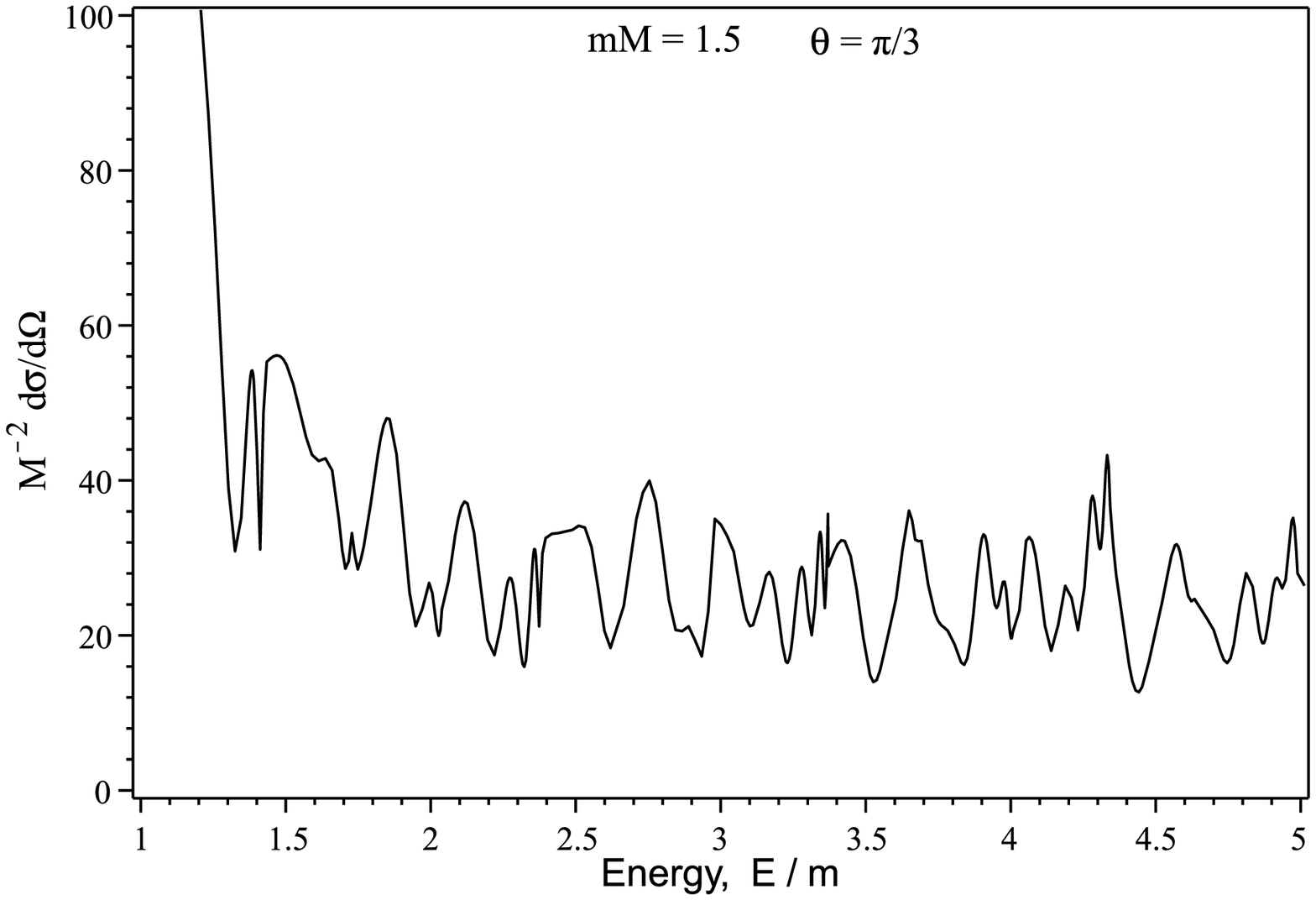}
\quad
\includegraphics[scale=0.4]{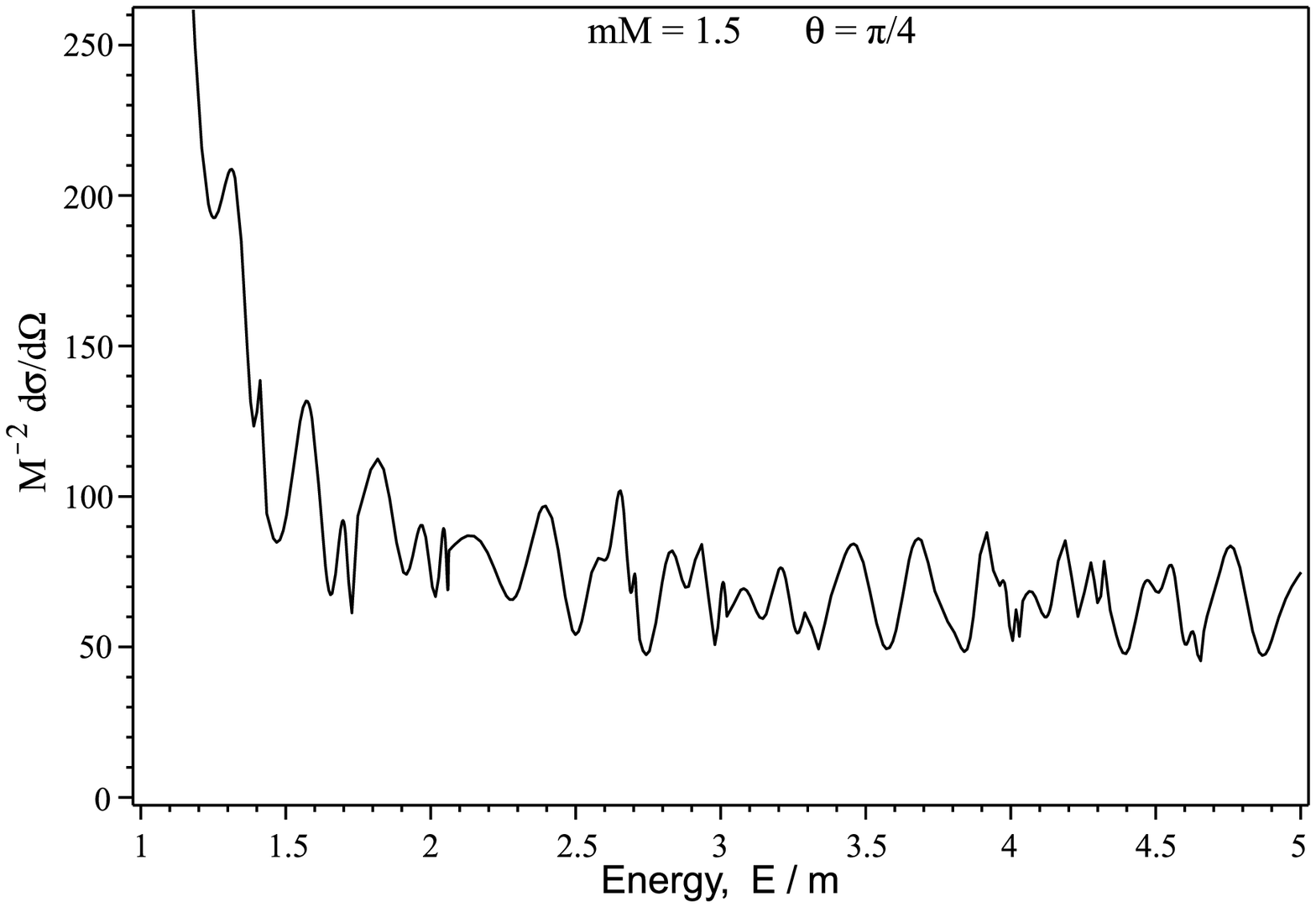}
\caption{Differential cross sections dependence of energy for different scattering angles and $mM=1.5$.}
\label{f4b}
\end{figure}

Another observation that emerge from Figs. (\ref{f4}-\ref{f4b}), is
that the energy dependence of  scattering intensity has a more
pronounced oscillatory behavior for small scattering angles and
large values of $mM$. On the other hand, we also observe from Figs.
(\ref{f4}-\ref{f4b}), that the scattering intensity is decreasing
when the energy increases and becomes divergent in the limit of
small energies. The shape of these graphs are the result of the fact
that our scattering intensity is proportional with the usual factor
$1/E^2$ and the oscillatory effect is given by the more complicated
dependence of energy from the phase shift given by Eq.
(\ref{final}).

\subsection{Polarization degree}

\begin{figure}[h!t]
\includegraphics[scale=0.4]{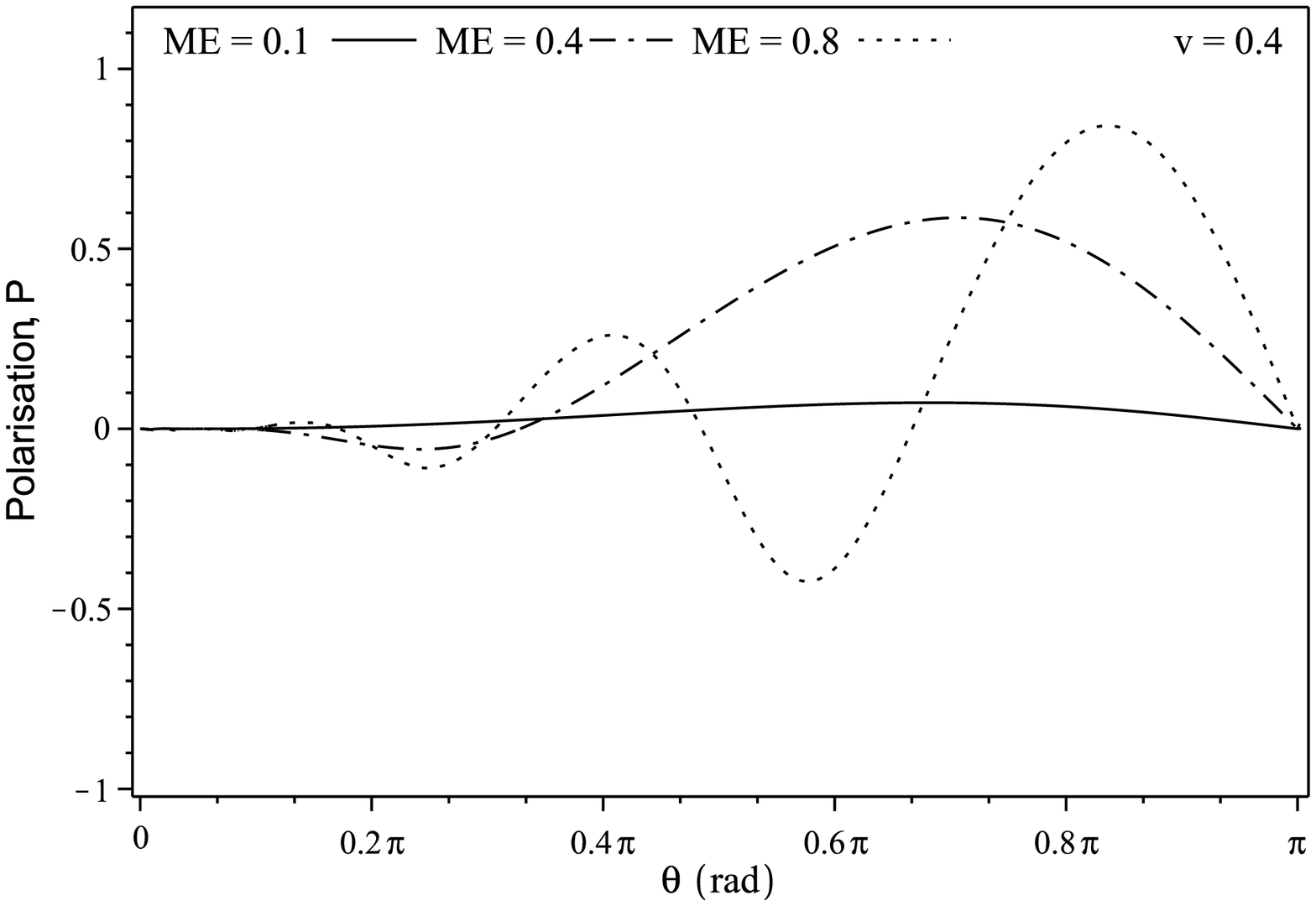}
\quad
\includegraphics[scale=0.4]{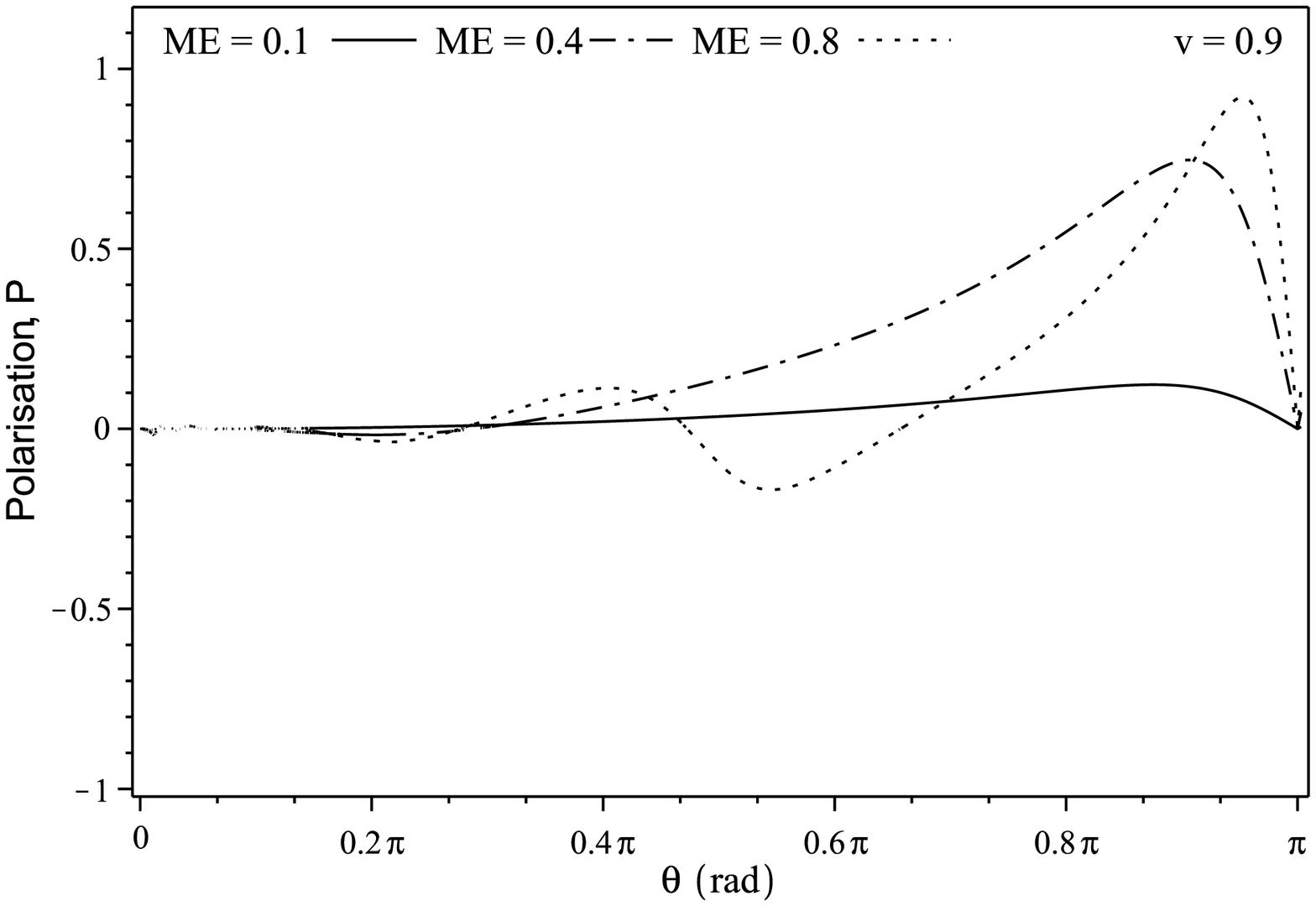}
\quad
\includegraphics[scale=0.4]{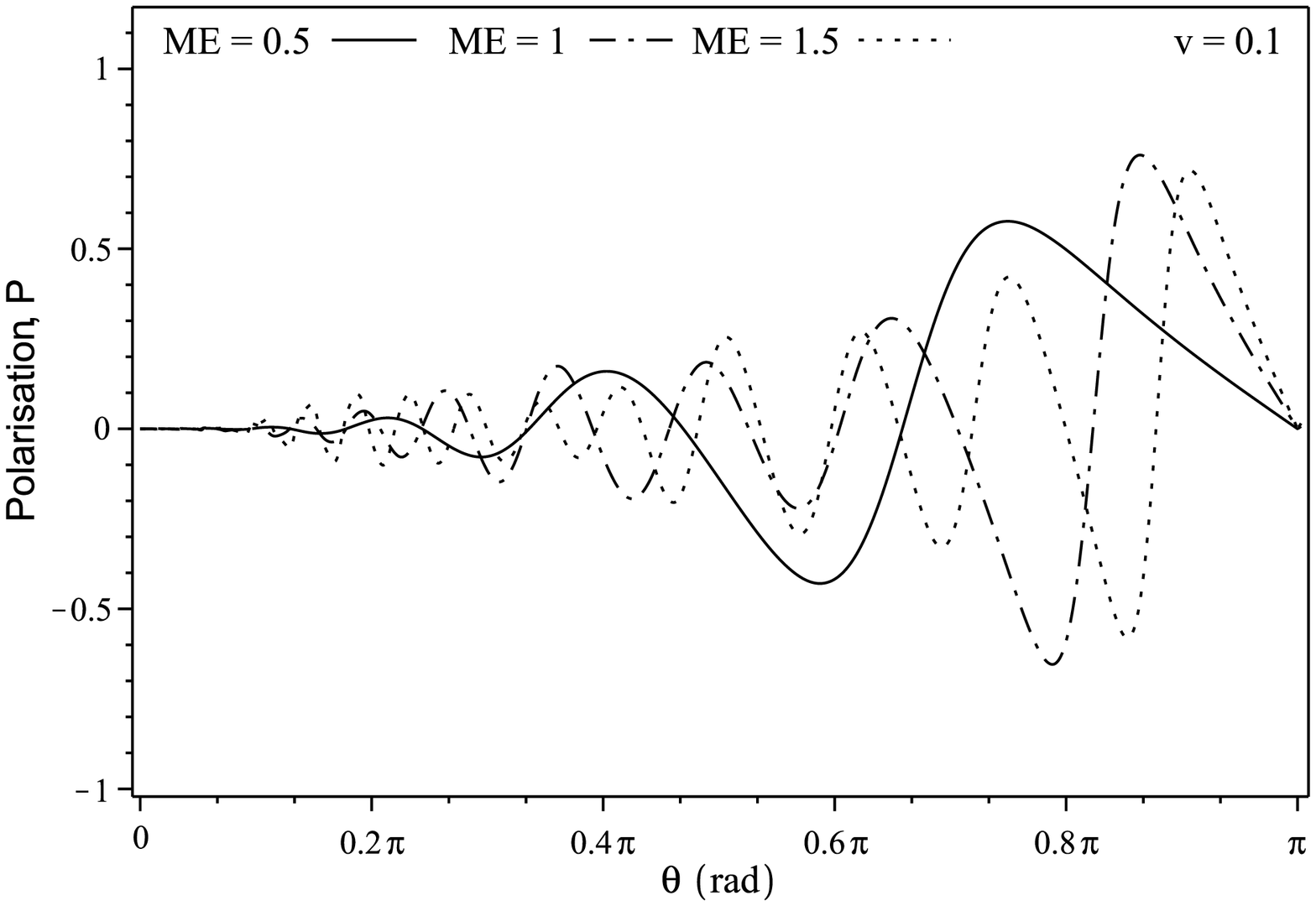}
\quad
\includegraphics[scale=0.4]{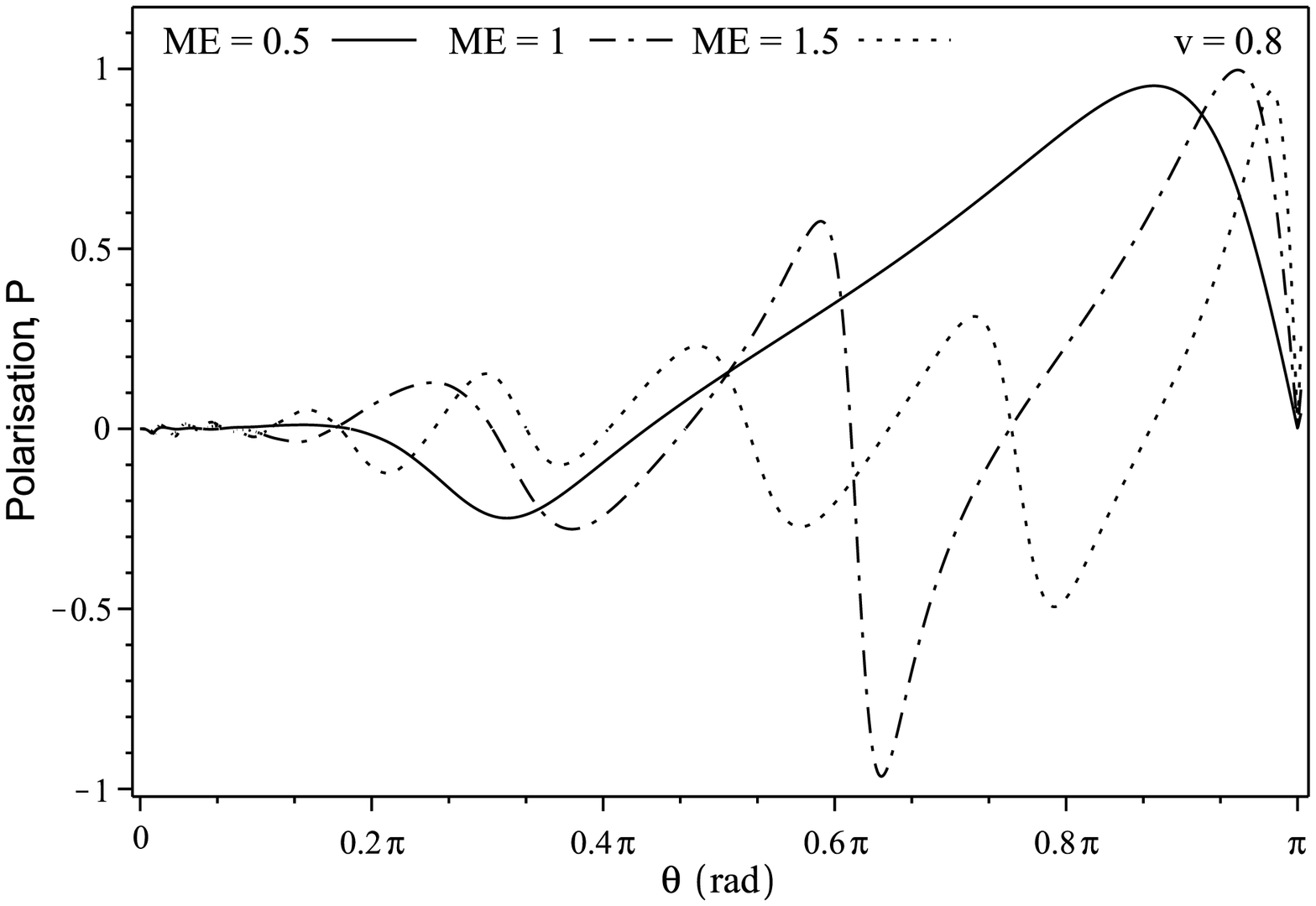}
\caption{The polarization degree dependence of $\theta$ for different  fermion velocities.}
\label{f5}
\end{figure}

\begin{figure}[h!t]
\includegraphics[scale=0.4]{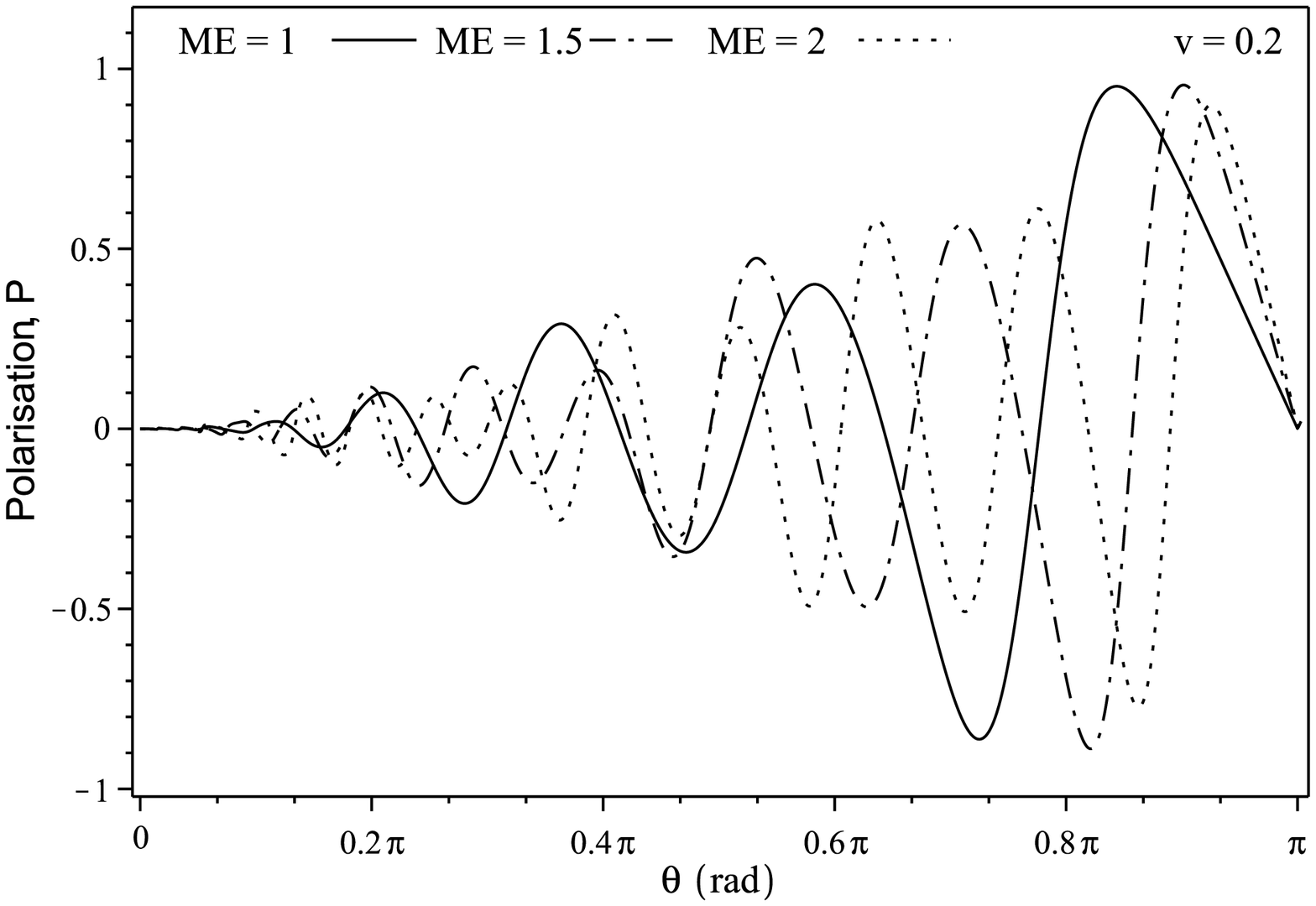}
\quad
\includegraphics[scale=0.4]{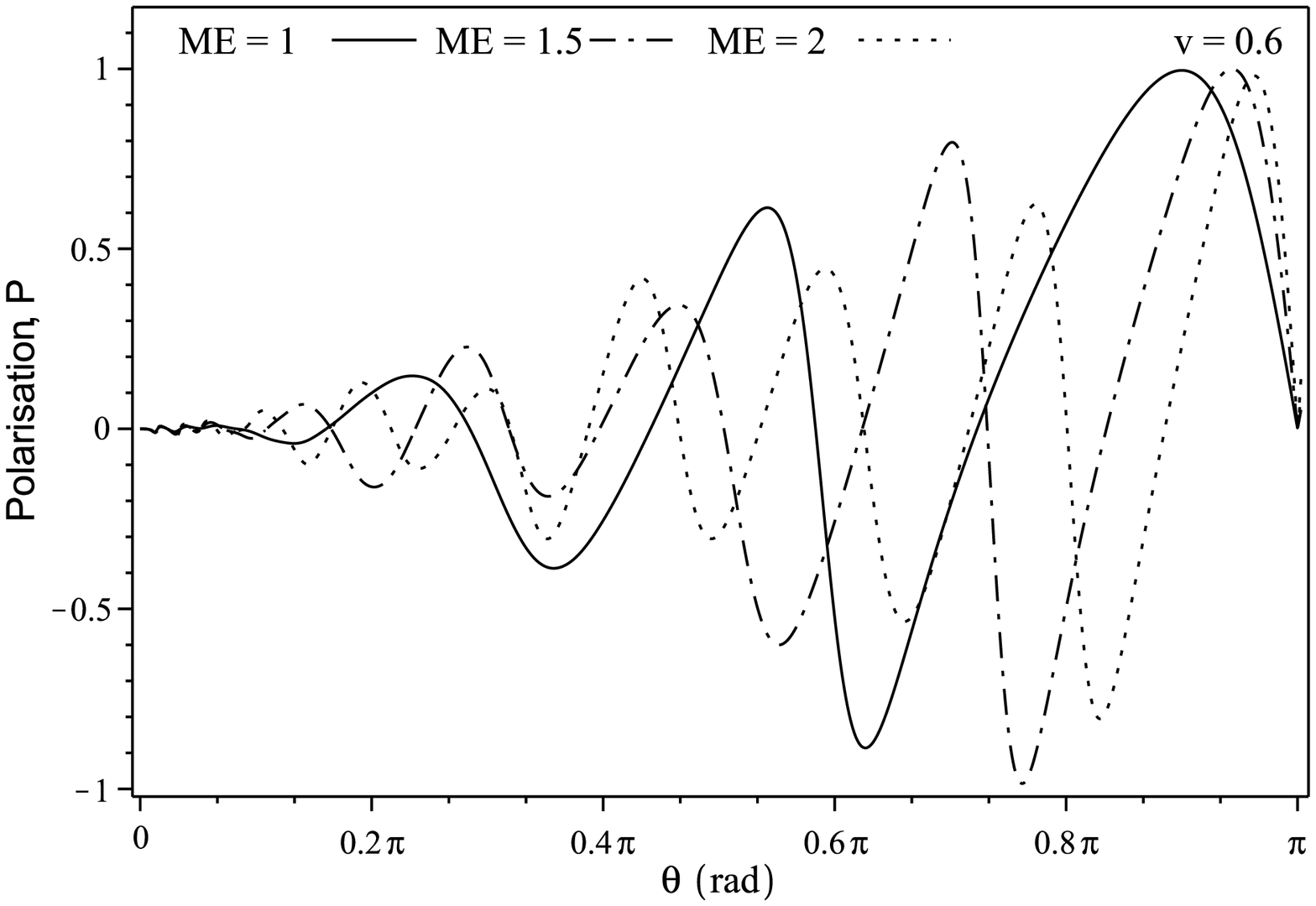}
\quad
\includegraphics[scale=0.4]{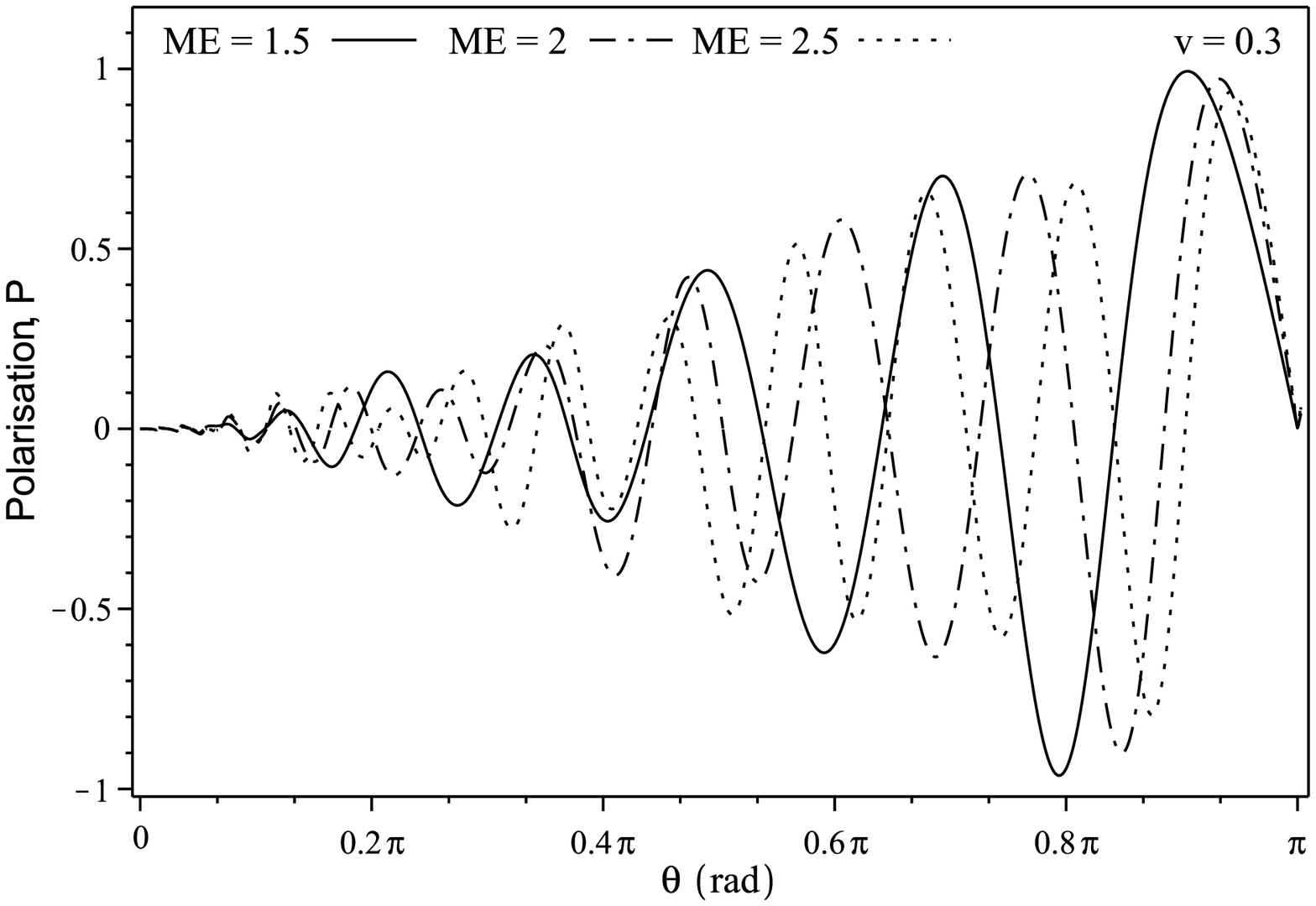}
\quad
\includegraphics[scale=0.4]{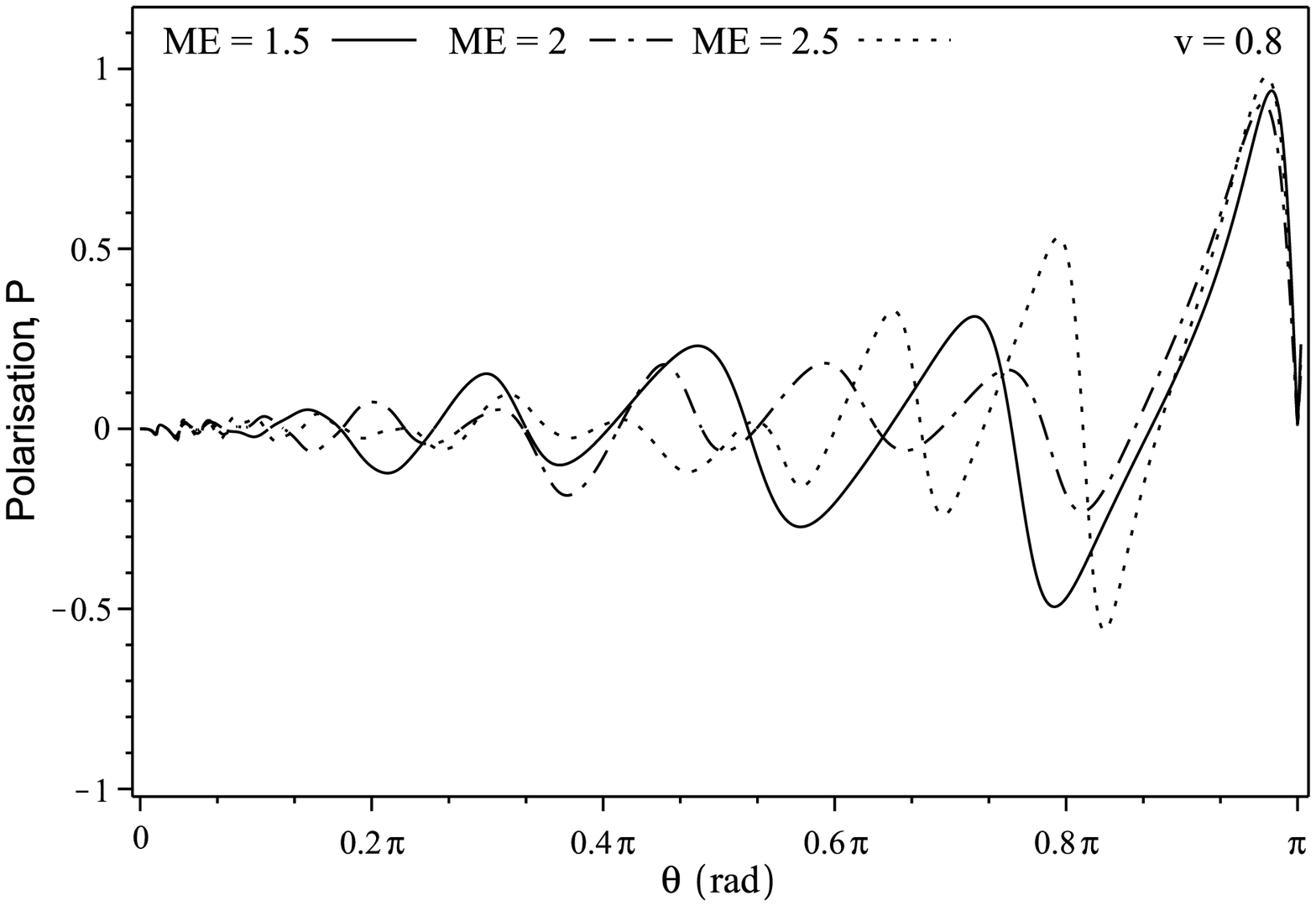}
\caption{The polarization degree dependence of $\theta$ for
different fermion velocities.} \label{f5a}
\end{figure}

If we consider that the fermions from the incident beam are not
polarized,  then after the interaction with the black hole, the
scattered beam could become partially polarized. It is interesting
to study this effect by plotting the degree of polarization
(\ref{pol}) as a function of scattering angle for given values of
$ME$ and different fermion velocities. Plotting the polarization as
function of scattering angle we obtain the results given in Figs.
(\ref{f5}-\ref{f5a}).

As we may observe from Figs. (\ref{f5}-\ref{f5a}), the polarization
is a very oscillatory function of the scattering angle. These
oscillations are the result of the forward and backward scattering
as well as the orbiting scattering. These three types of scattering
induce the oscillatory behavior of polarization, since the
scattering intensity also oscillates with the scattering angle.
Another result that is worth mentioning is  the oscillatory behavior
of the polarization, Figs. (\ref{f5}-\ref{f5a}), which is modified
as we change the parameter $ME$. If the fermion energy is fixed,
then we can draw the conclusion that the oscillatory behavior of
polarization depends on the black hole mass $M$, becoming more
pronounced as we increase the black-hole mass.

\begin{figure}[h!t]
\includegraphics[scale=0.35]{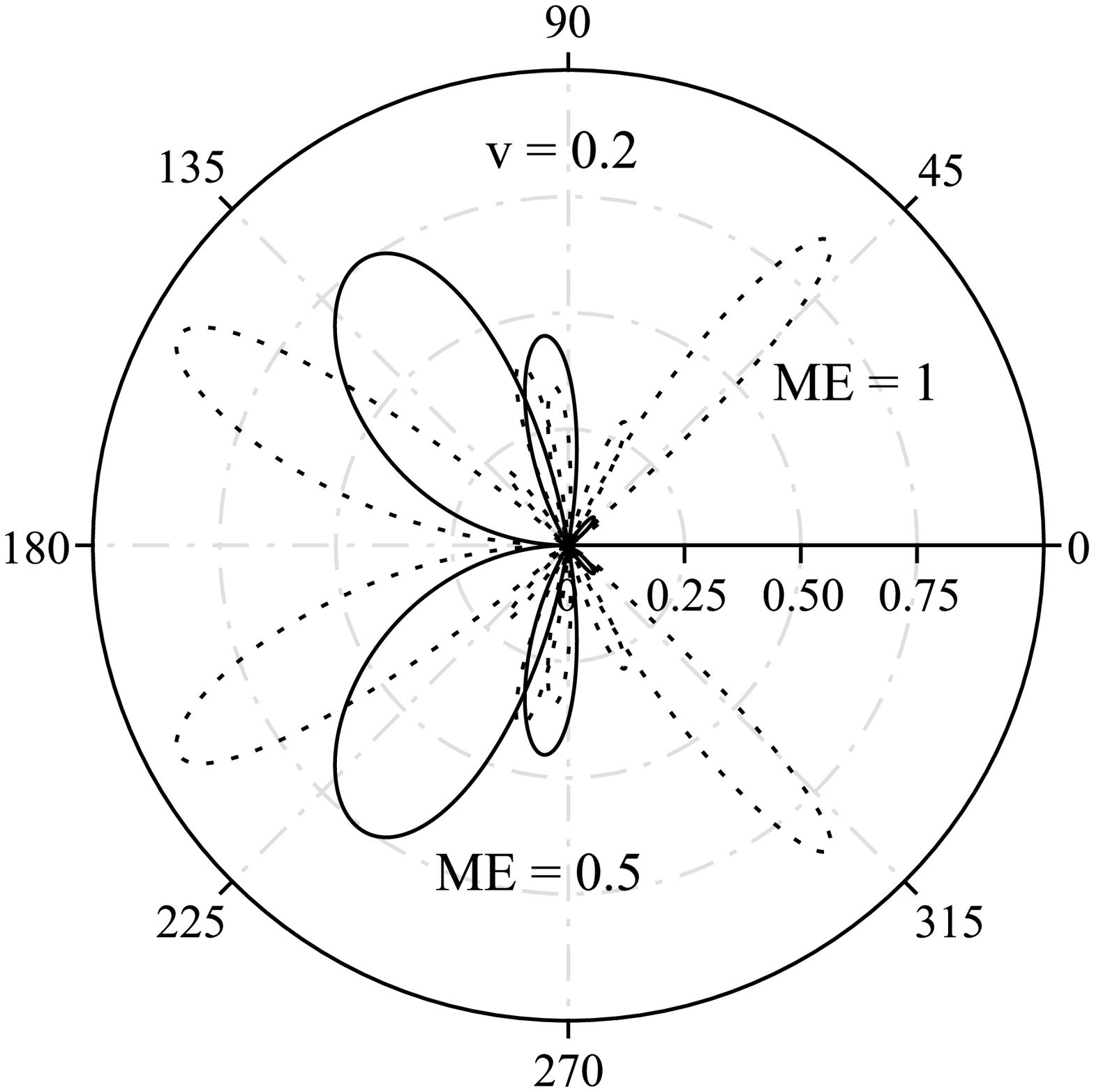}
\quad
\includegraphics[scale=0.35]{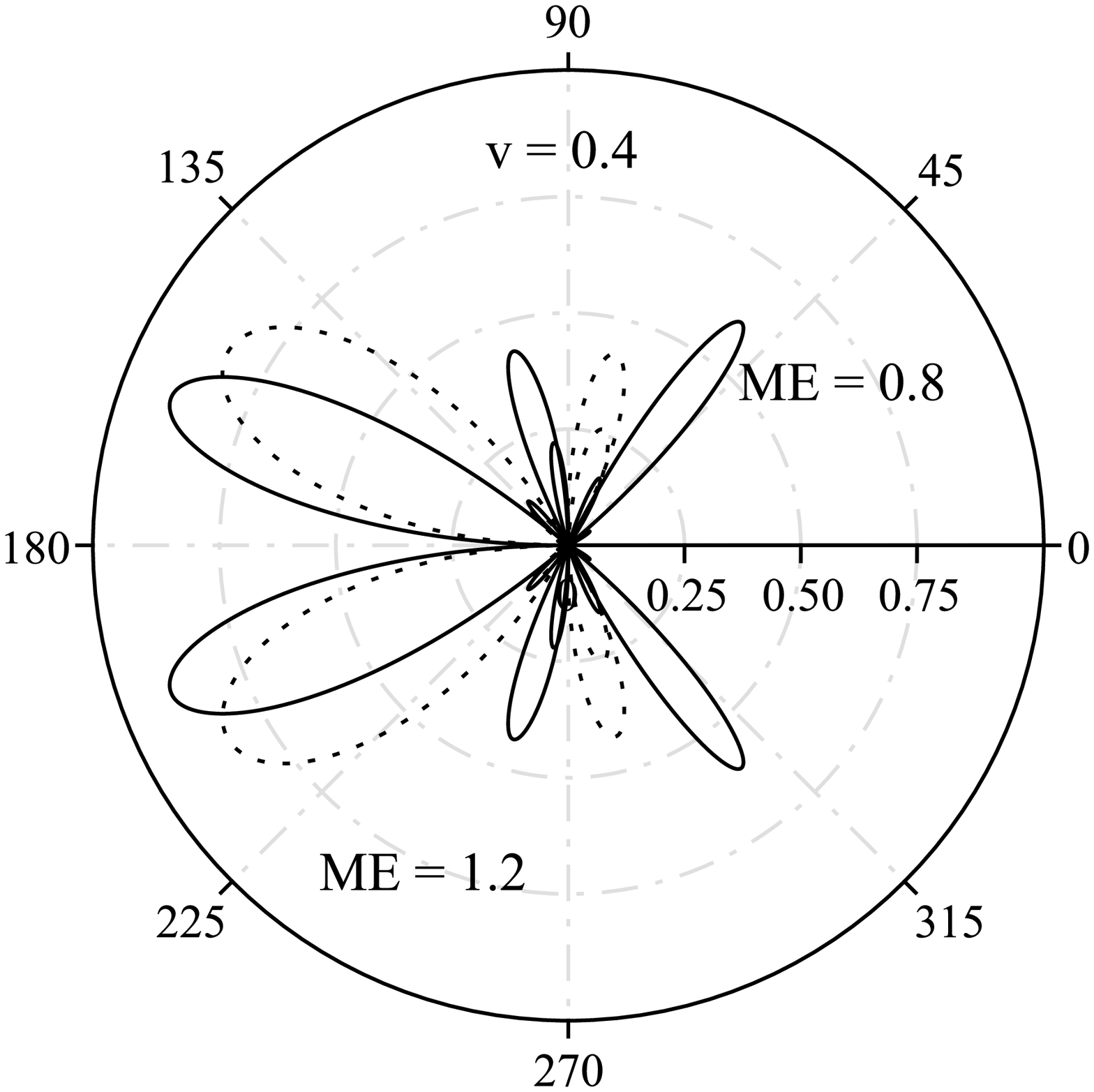}
\quad
\includegraphics[scale=0.35]{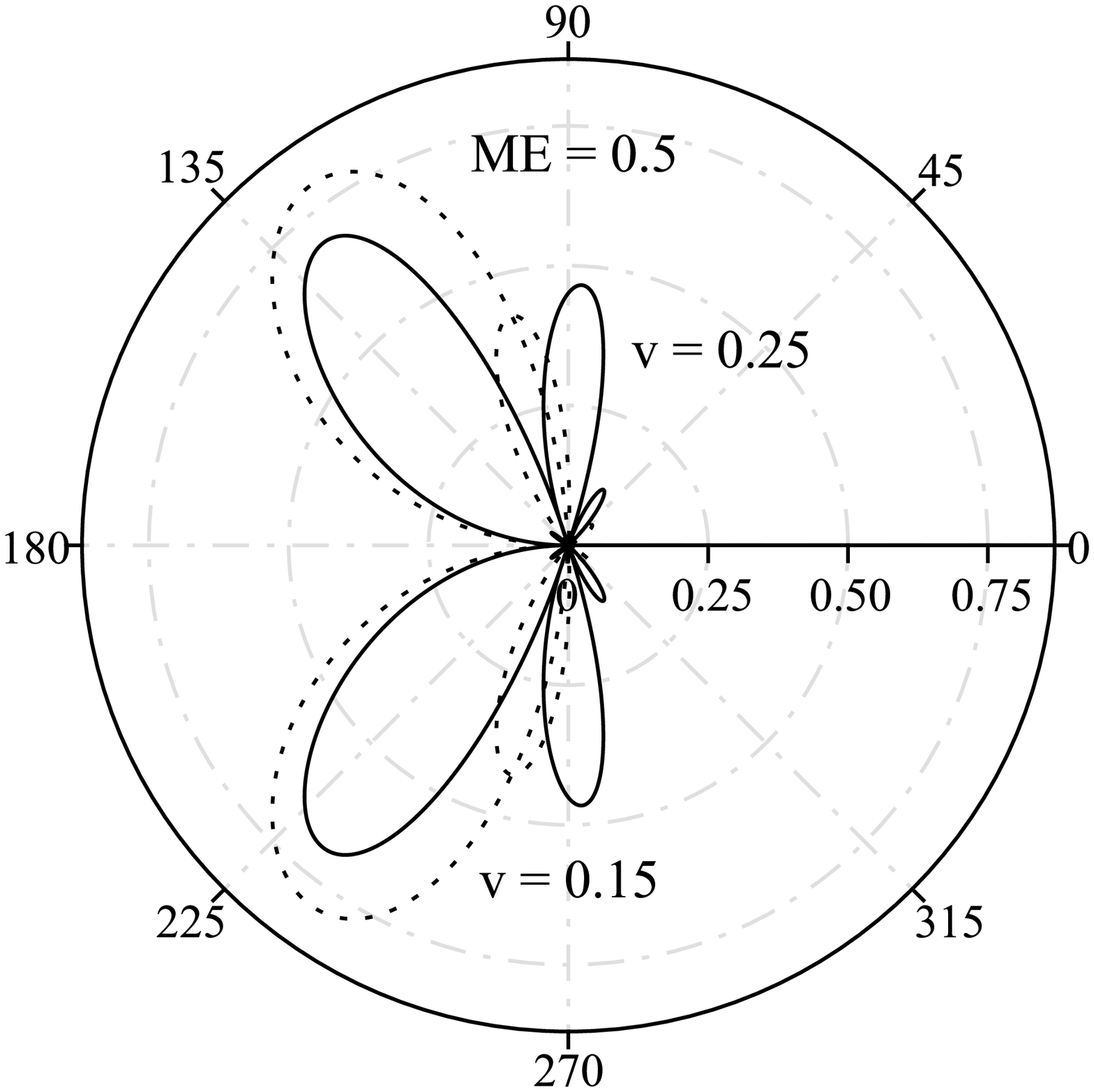}
\quad
\includegraphics[scale=0.35]{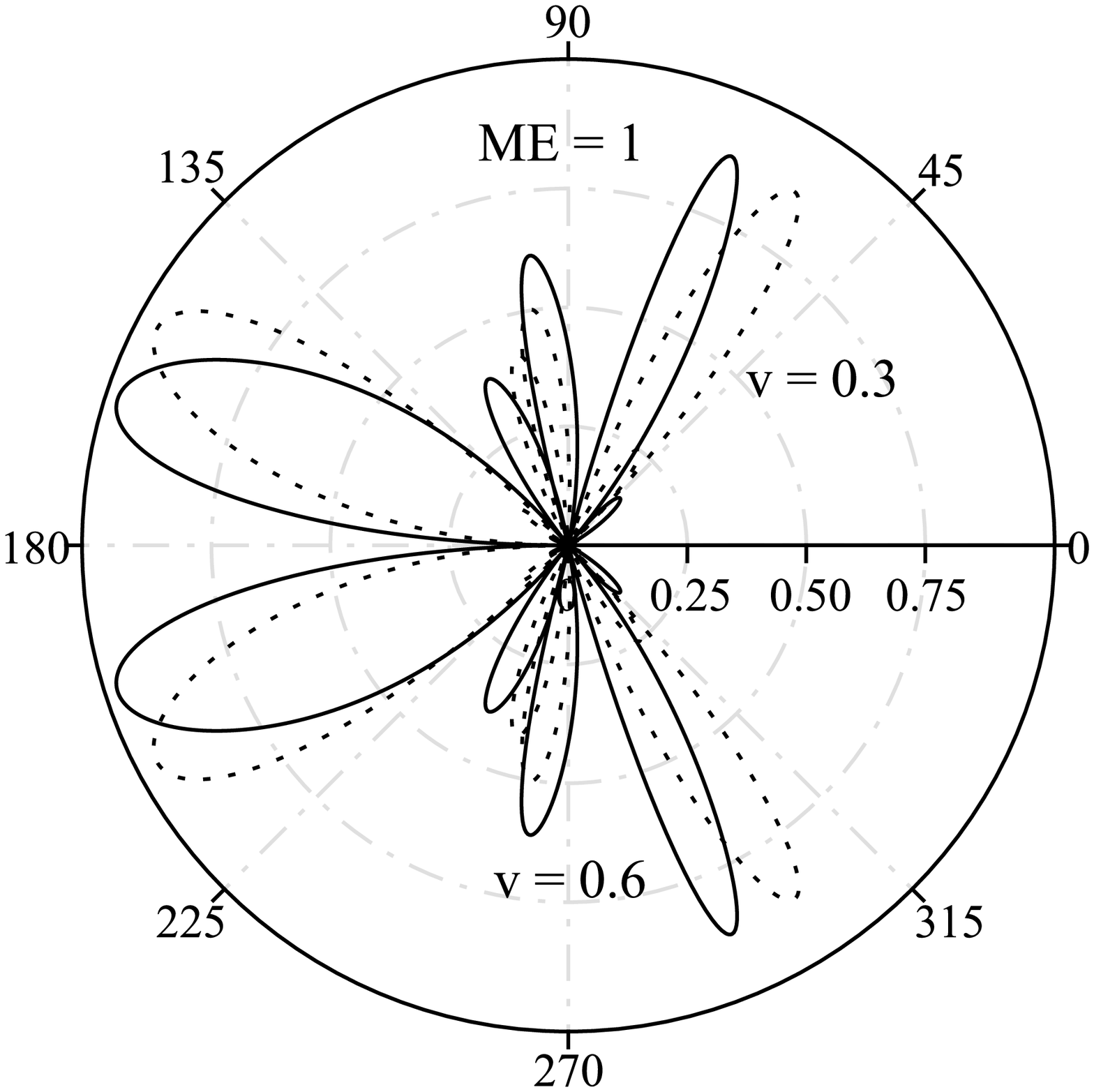}
\caption{Polar plot of ${\cal P}(\theta)$  for different values of
$ME$ and fermion velocities.} \label{f6}
\end{figure}

\begin{figure}[h!t]
\includegraphics[scale=0.35]{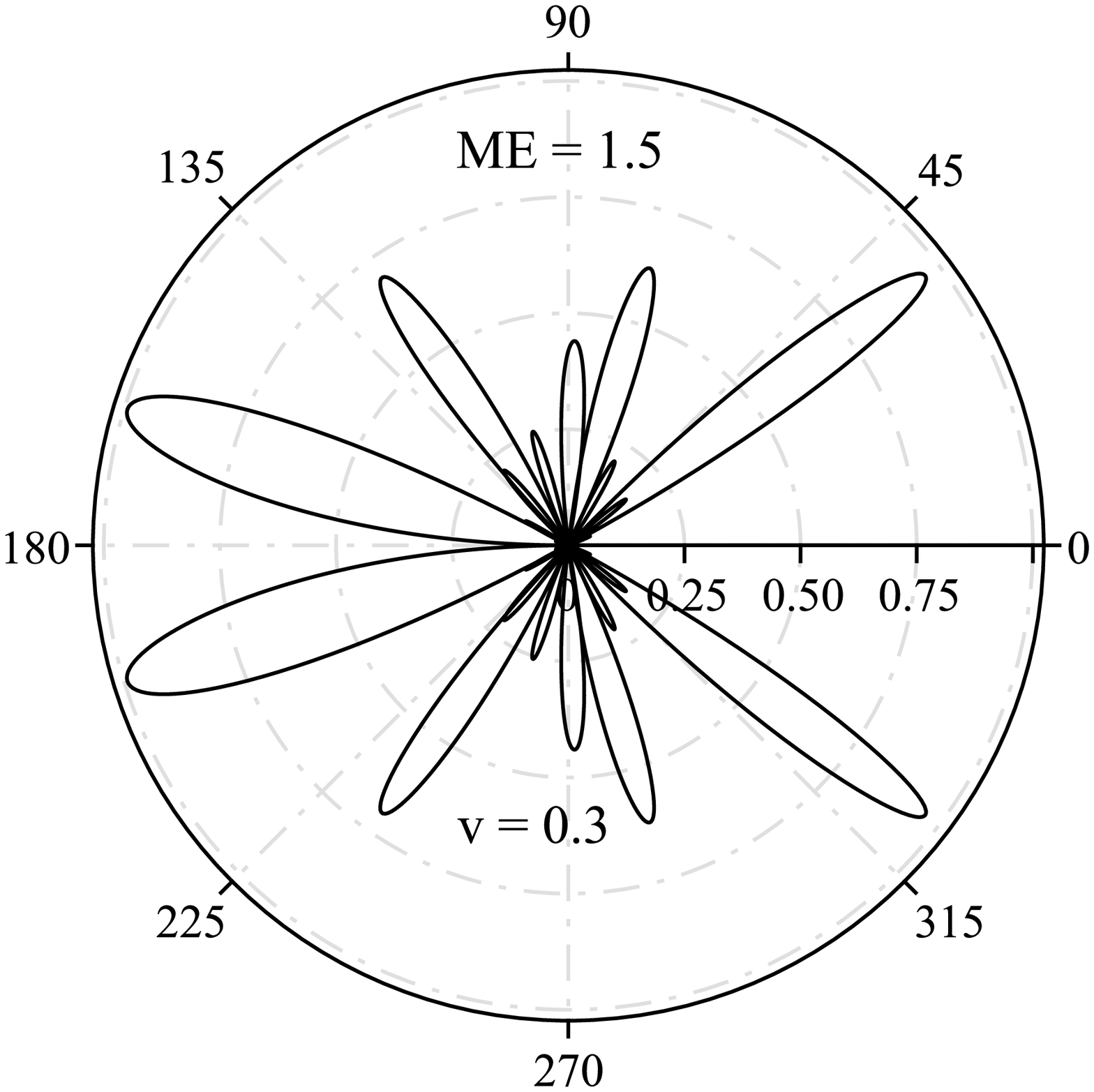}
\quad
\includegraphics[scale=0.35]{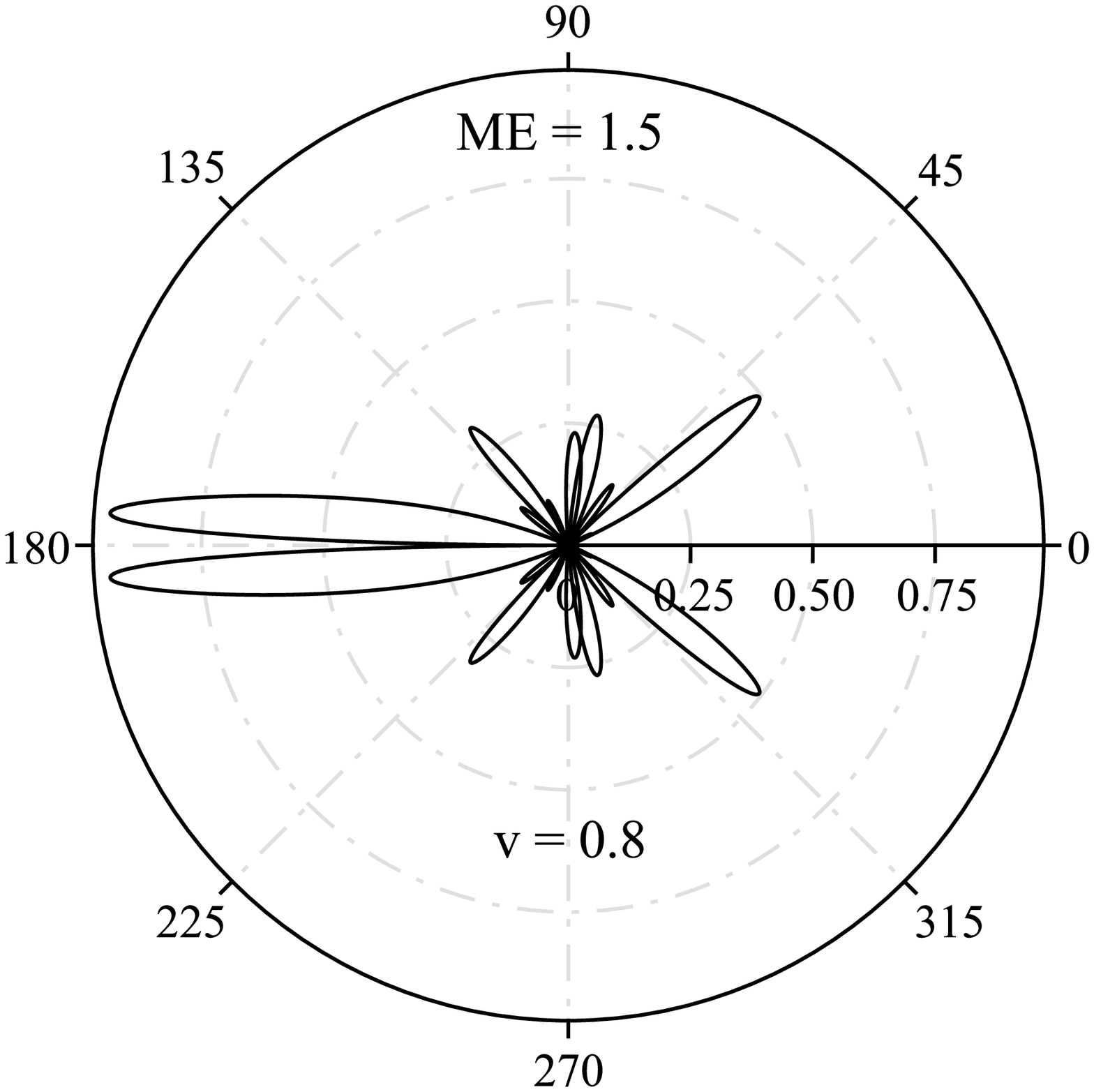}
\quad
\includegraphics[scale=0.35]{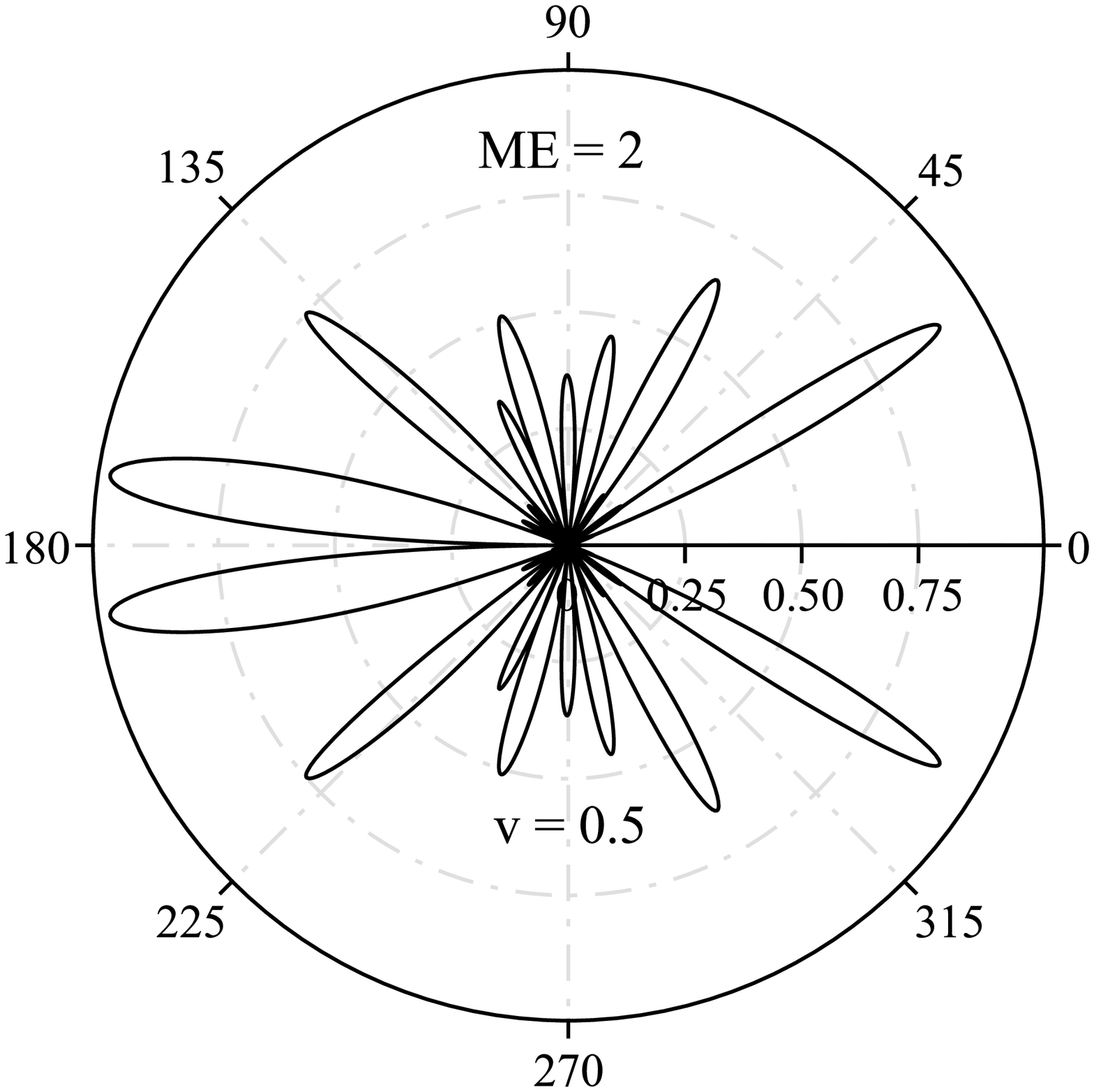}
\quad
\includegraphics[scale=0.35]{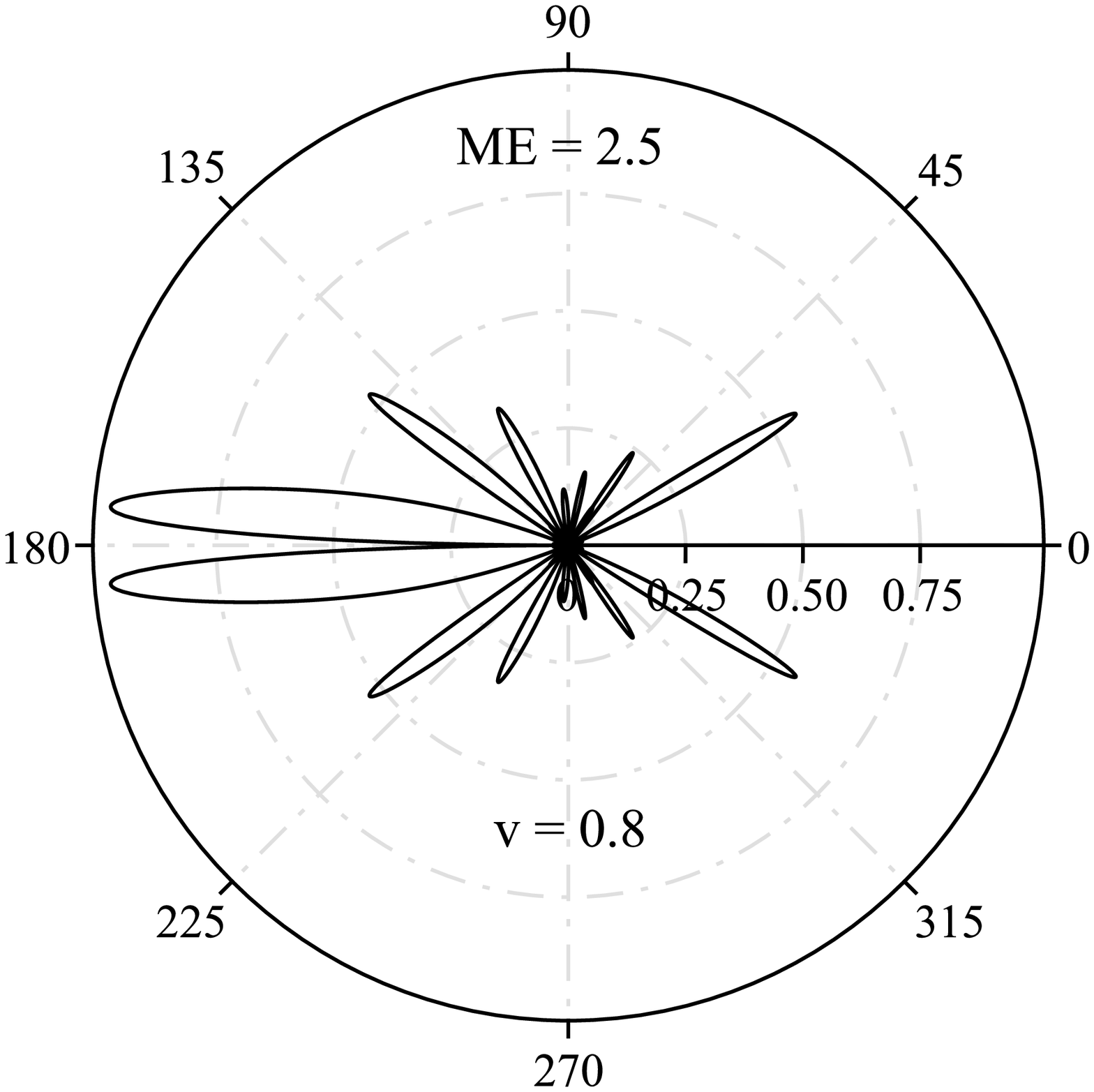}
\caption{Polar plot of ${\cal P}(\theta)$  for different values of
$ME$ and fermion velocities.} \label{f7}
\end{figure}

\newpage
To see how the spin of the fermion is aligned with a given direction
after scattering on a black hole, we present the polar plots for the
degree of polarization in Figs.(\ref{f6}-\ref{f7}). These graphs are
plotted for different values of $ME$ and different velocities.

\newpage

From our Figs. (\ref{f6}-\ref{f7}) we observe that the scattered
wave can  be partially polarized in the direction orthogonal to the
scattering plane. This phenomenon is similar with the Mott
polarization \cite{MT}, which appear in the electromagnetic
scattering. This conclusion was also underlined in  Ref. \cite{S3}.

\newpage
\subsection{Absorption cross section}

The above examples show that our analytical approach reproduces
correctly all  the results concerning the elastic scattering
obtained by using analytical-numerical methods  \cite{bh1,S3}.
However, there are significant differences in what concerns the
partial absorption cross sections (\ref{sigmaa}) or the total one
(\ref{abss}).

\begin{figure}[h!t]
\includegraphics[scale=0.4]{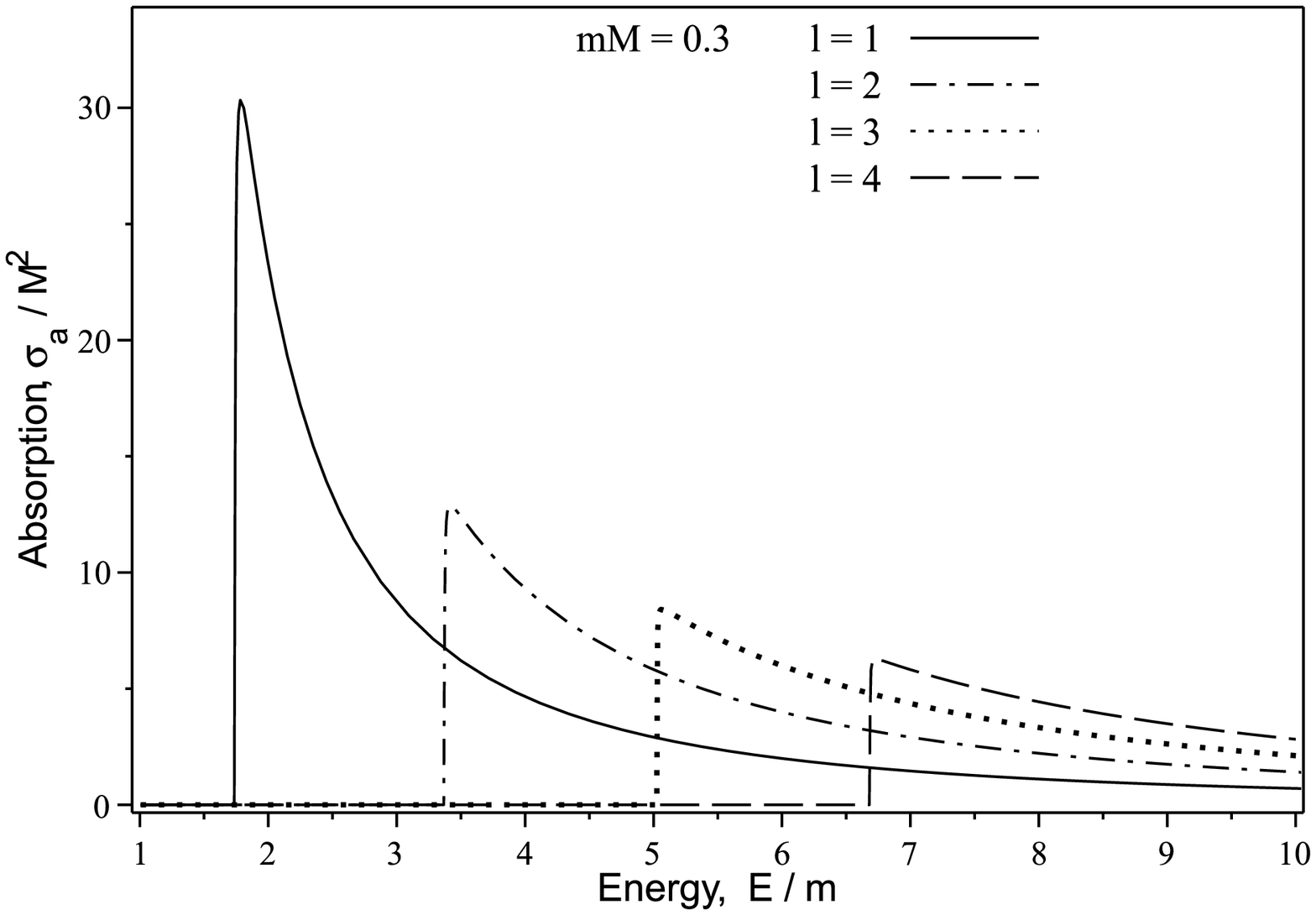}
\quad
\includegraphics[scale=0.4]{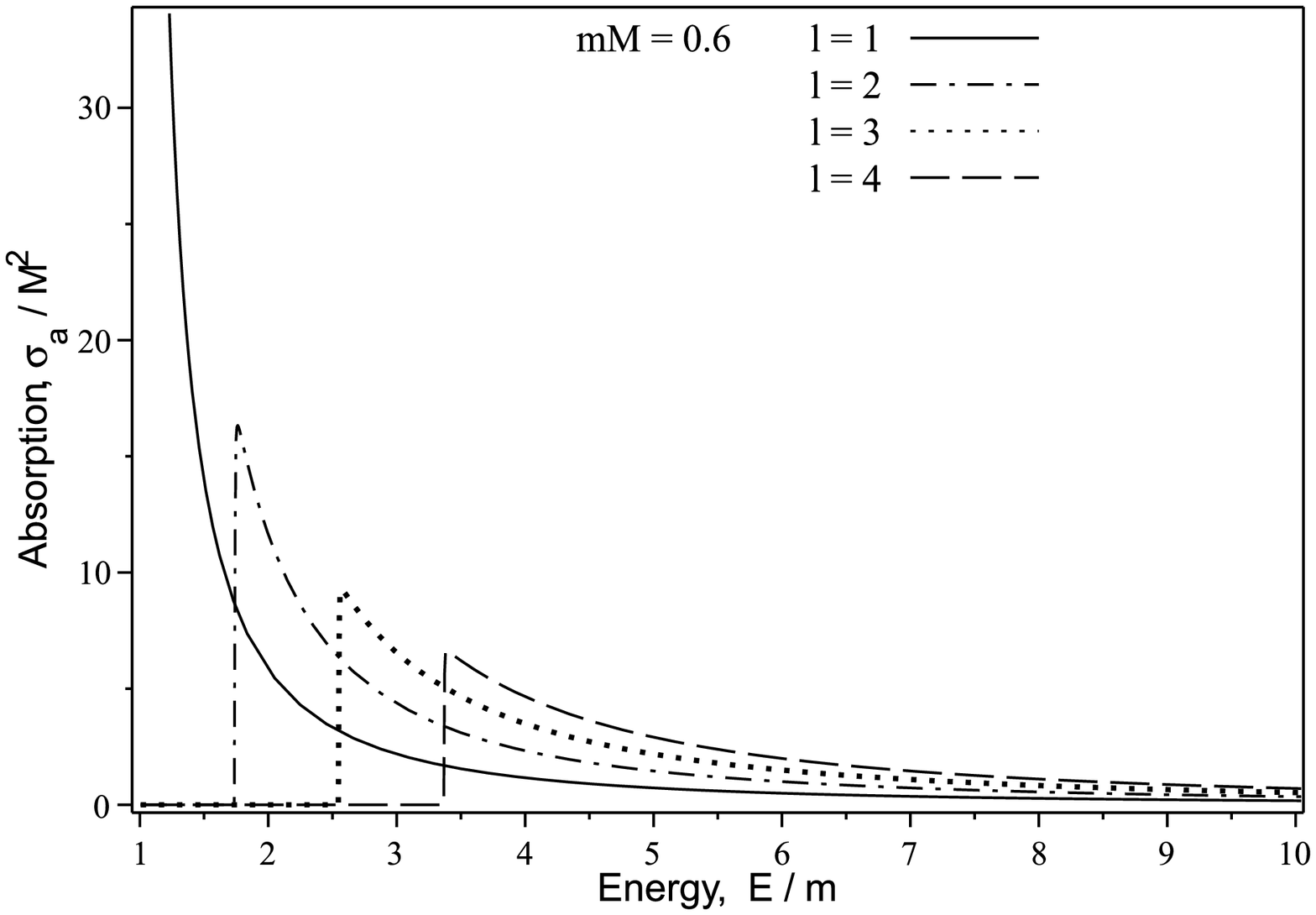}

\caption{Partial absorbtion cross sections as functions  of energy
for different values of $mM$.} \label{fx}
\end{figure}

\begin{figure}[h!t]
\includegraphics[scale=0.4]{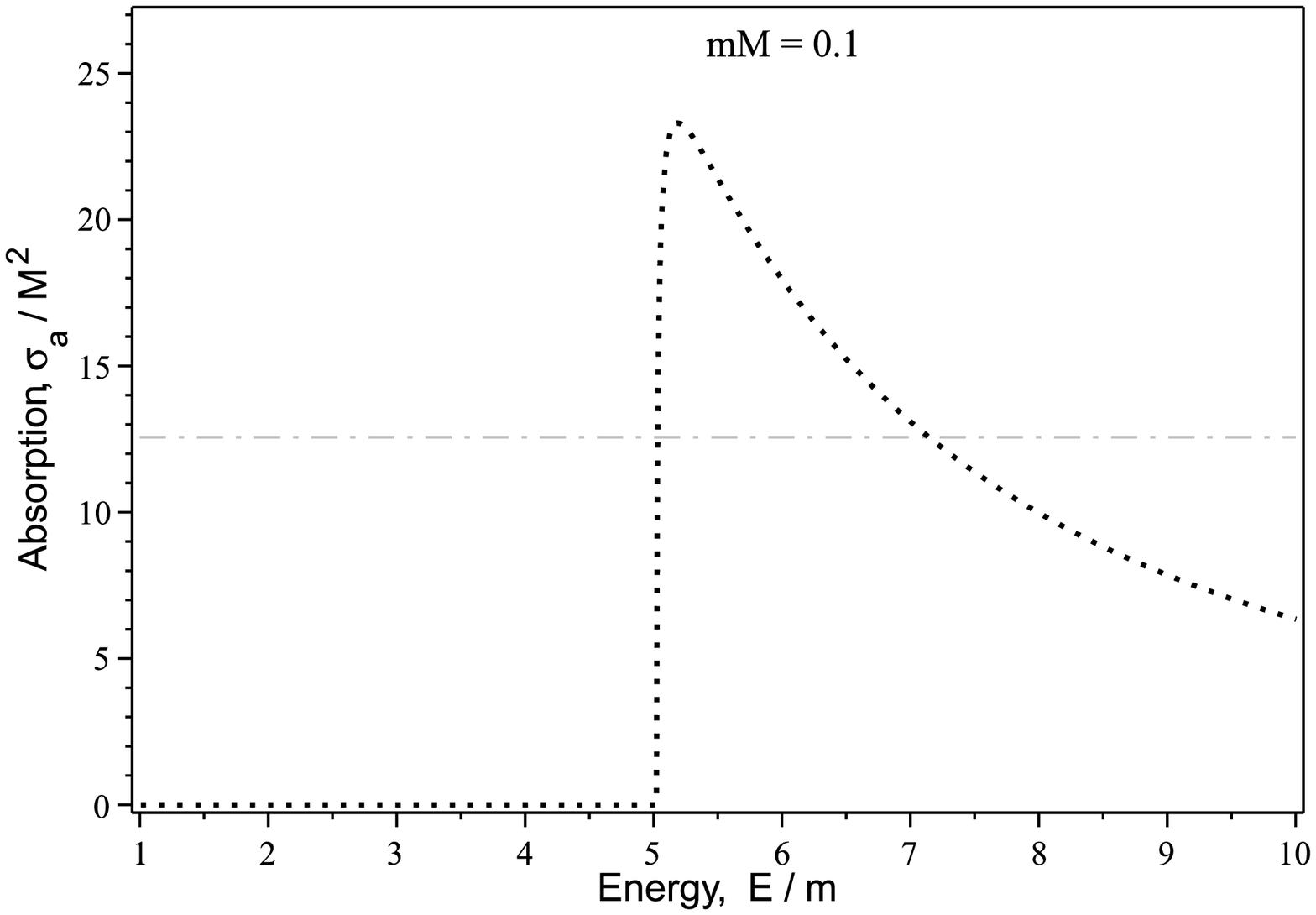}
\quad
\includegraphics[scale=0.4]{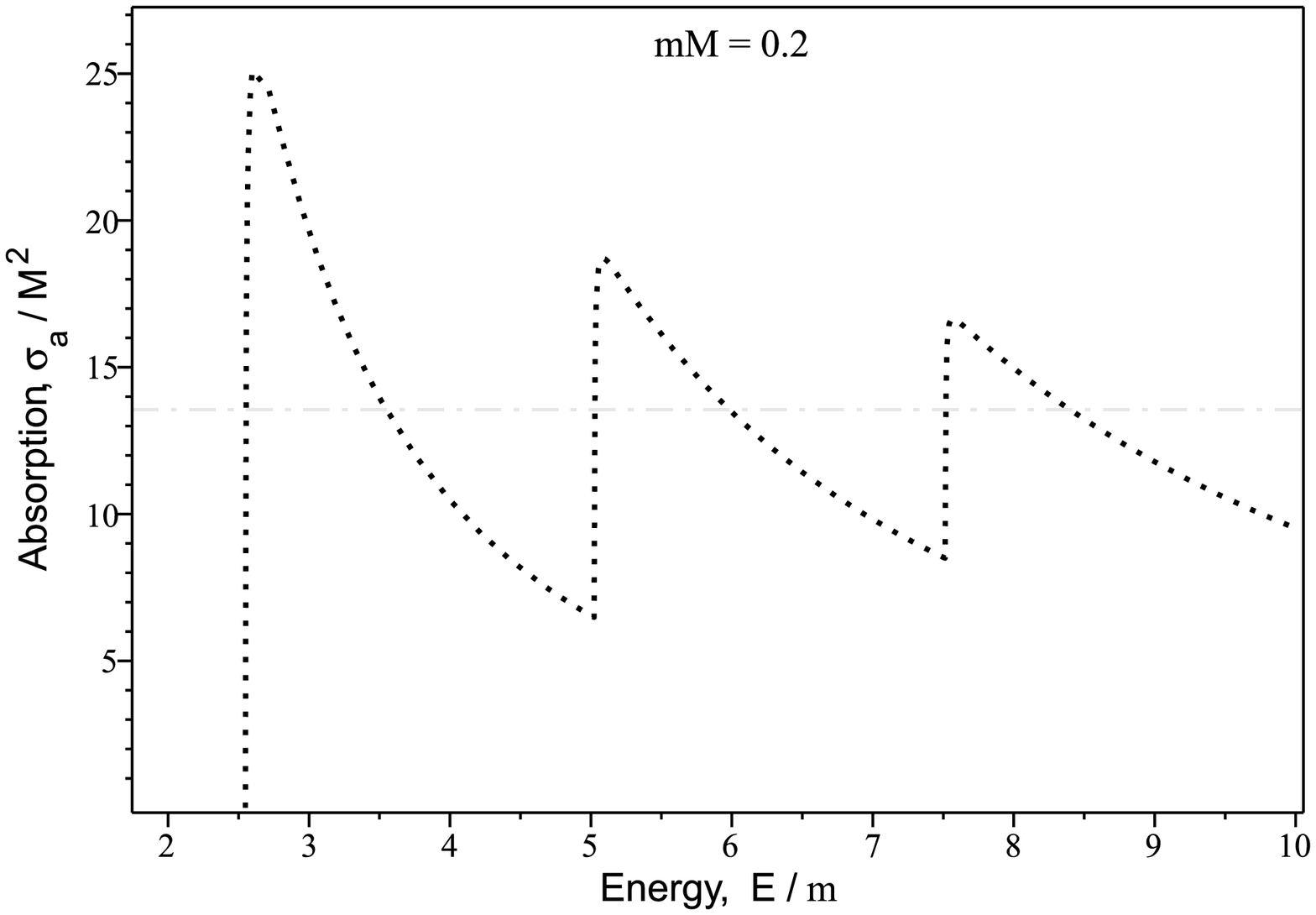}
\quad
\includegraphics[scale=0.4]{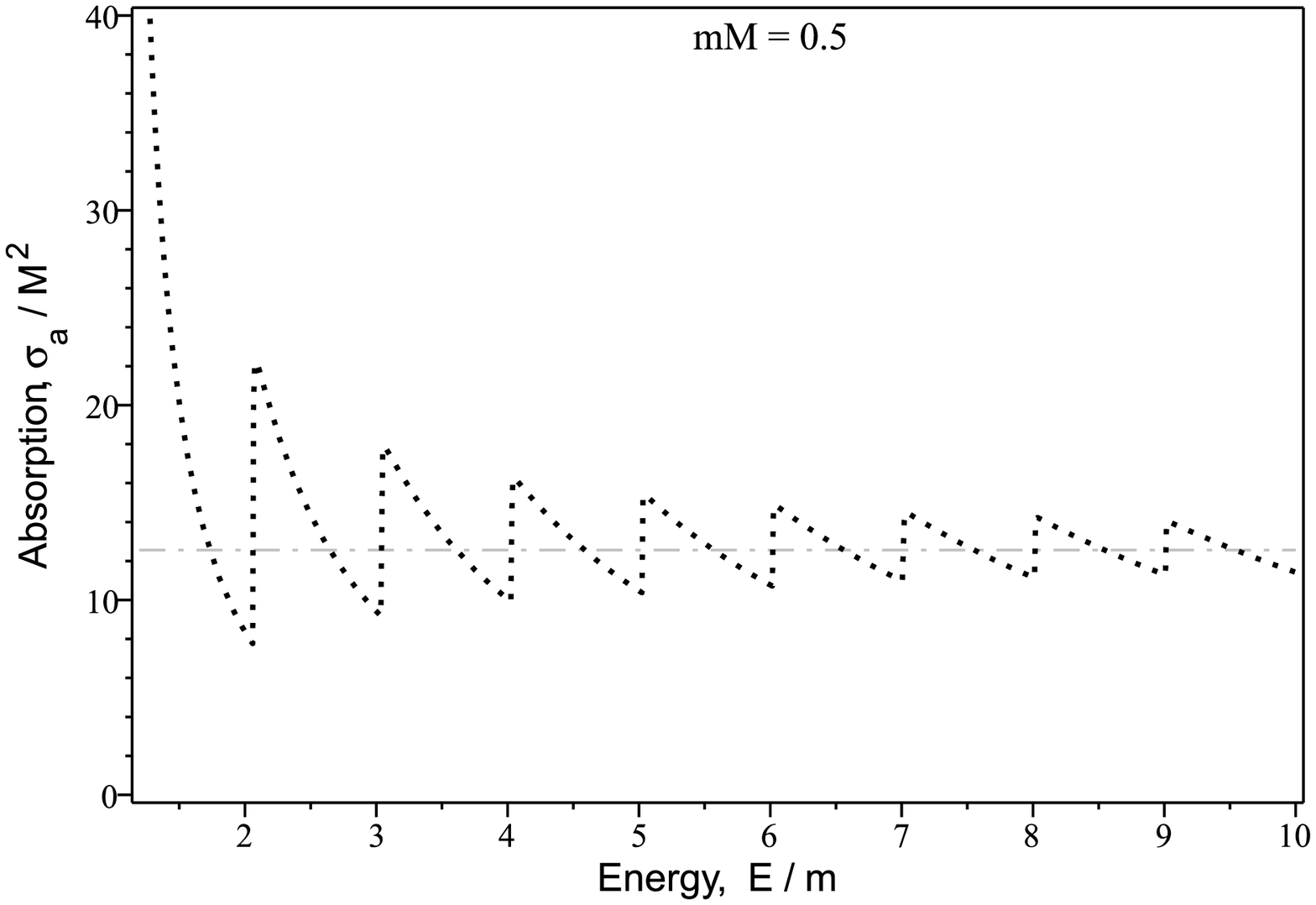}
\quad
\includegraphics[scale=0.4]{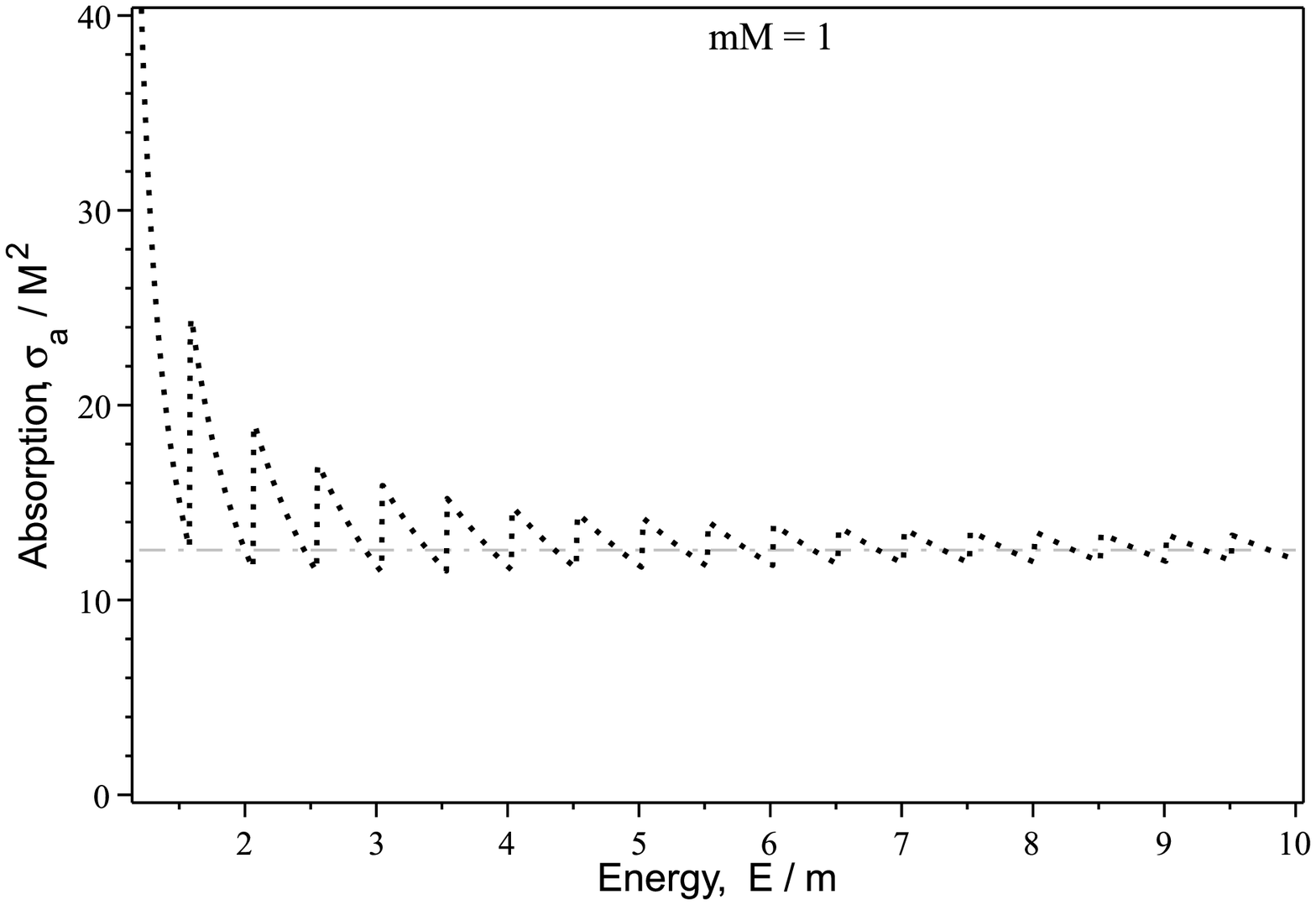}
\caption{Absorbtion cross sections as functions  of energy for
different values of $mM$.} \label{fy}
\end{figure}

We specify first that in our case the partial waves
$|S_{-\kappa}|=|S_{\kappa}|$ give  the {\em same} contribution to
the partial absorption cross sections. As observed in section
III.B., when $\sqrt{3}mM<1$ all these partial sections remain finite
in $E=m$ as we can see in Fig. (\ref{fx}a). However, if this
quantity becomes larger than 1 then the partial sections
$\sigma_a^l(p)$ with $1<l<\sqrt{3}mM$ reach the singularity point
$E=m$ where these are divergent, as in Fig. (\ref{fx}b) where only
$\sigma_a^1(p)$ becomes singular. Thus the profile of $\sigma_a(p)$
is strongly dependent on the parameter $mM$ as shown in Fig.
(\ref{fy}). In the first panel (\ref{fy}a) we plot the cross section
$\sigma_a(p)=\sigma_a^1(p)$ given by $|S_{-1}|=|S_{1}|$ that are the
only contributions in the energy range under consideration. The next
panels show the profiles of the absorption cross sections for
different values of $mM$, pointing out the mentioned effect of the
singularity in $E=m$ and the asymptotic behavior resulted from Eq.
(\ref{asy}).

We note that the analytical-numerical results obtained so far
\cite{bh1,S3} are somewhat different from our analytical ones we
present here. The first difference is that one obtains numerically
$|S_{-\kappa}|\not=|S_{\kappa}|$ such that the quantity
$\sigma_a^{-\kappa}(p)-\sigma_a^{\kappa}(p)$ is maximal for
$|\kappa|=1$ - when $\sigma_a^{-1}$ (with $l=0$) diverges for $E\to
m$ while $\sigma_a^1(p)$ remains finite - vanishing then rapidly as
increasing $|\kappa|$ \cite{bh1}. The second difference is the
numerical asymptote that corresponds to the geometrical photon value
$27\pi M^2$ while in our case this asymptote (\ref{asy})  is much
smaller. On the other hand, despite of these two discrepancies,
there are notable similarities as the general profile of the cross
sections and the values of the thresholds of the partial waves that
become equidistant for $m=0$.

Under such circumstances it seems that something is missing in our
analytic approach. In our opinion this is just a possible resonant
scattering produced by the bound states which were ignored here. The
argument is that the resonant scattering is dominant in the $s$ wave
(with $l=0$) and increases significantly the absorption cross
section such that this could explain satisfactory  the above
discussed differences.

\newpage

\section{Concluding remarks}

In this paper we performed the partial wave analysis of Dirac
fermions scattered by black-holes  by using the approximative
analytical scattering modes of the Dirac equation in Schwarzschild's
charts with Cartesian gauge.  Selecting a suitable asymptotic
condition we obtained the analytical expression of the phase shifts
that allow us to derive the scattering amplitudes, differential
cross sections and degrees of polarization. These depend on a
parameter ($s$) that takes real values in elastic collisions,
becoming pure imaginary when the fermion is absorbed by black hole.
Thus by using the same formalism we can study the elastic scattering
and absorption deriving the closed form of the absorption cross
section produced by the scattering modes (while the influence of the
bound states is neglected).

From our analytical and graphical results we established that the
spinor wave can  be scattered in forward and backward directions and
that the spiral scattering (orbital scattering) is present in both
these cases. The scattering intensity in both forward and backward
directions is increasing with the black hole mass. Also the
oscillations in scattering intensity around $\theta=\pi$ become more
pronounced as the black hole mass is increasing. The polarization
degree has an oscillatory behavior in terms of scattering angle and
depends on the black hole mass too. Our polar plots for
polarizations show what are the directions in which the spin is
aligned after the interaction. We note that thanks to our analytical
formulas the graphical analysis can be performed for any values of
the parameters $mM$, $ME$ and $v$ leading to similar conclusions.

The general conclusion is that our formalism is suitable for
describing the elastic fermion-black hole collisions for which we
recover the entire phenomenology pointed out by analytical-
numerical methods \cite{S1}-\cite{S3}. The principal problem that
remains open is the role of the bound states \cite{C4} in this
scattering. The challenge is to show how these states may complete
the present analytical approach in order to recover, in addition,
the well-known results concerning the fermion absorbtion by black
hole as obtained so far \cite{bh1}.

\appendix

\section{Wittaker functions}

The Wittaker functions $M$ \cite{NIST} have the  property
\begin{equation}\label{A2}
M_{-\kappa,\mu}(-z)=e^{i\pi(\mu+\frac{1}{2})}M_{\kappa,\mu}(z)\,,
\end{equation}
and the asymptotic representation for large $|z|$,
\begin{eqnarray}\label{A3}
M_{\kappa,\mu}(z)&\sim& \frac{\Gamma(1+2\mu)}{\Gamma(\frac{1}{2}
+\mu-\kappa)}e^{\frac{1}{2}z}z^{-\kappa}\left(1+O(z^{-1})\right) \nonumber\\
&+&e^{i(\frac{1}{2}+\mu-\kappa)\pi}\frac{\Gamma(1+2\mu)}{\Gamma(\frac{1}{2}+\mu+\kappa)}
e^{-\frac{1}{2}z}z^{\kappa}\left(1+O(z^{-1})\right)
\end{eqnarray}
that holds the case of our radial functions where  $-\frac{1}{2}\pi<{\rm ph \,}
 z=\frac{\pi}{2}<\frac{3}{2}\pi$.

\section{Newtonian limit}

In the case of scalar particles scattered from black holes,   the
phase shifts $\delta_l^N$  in the large-$l$ limit  satisfy
\cite{FHM}
\begin{equation}\label{shN}
e^{2i\delta^N_l}=\frac{\Gamma(1+l-iq)}{\Gamma(1+l+iq)}\,.
\end{equation}
where $q$ is given by Eq. (\ref{q}). Then the partial amplitudes
\begin{equation}
 f_l^N=\frac{1}{2ip}\left(e^{2i\delta_l^N}-1\right)\,,
\end{equation}
can be expanded as
\begin{equation}\label{fN}
f_l^N=-M\frac{\beta}{p^2}+iM^2\frac{\beta^2}{p^3}+O(M^3)\,, \quad
\beta=(2p^2+m^2)\psi(l+1)\,.
\end{equation}

\section{Condition of elastic scattering}

Let us consider the problem of the appropriate asymptotic
conditions determining the integration constants of the solutions
(\ref{E11}) and (\ref{E22}) that  satisfy Eqs. (\ref{C1C1}). It is
convenient to introduce the new notation  (up to a real arbitrary
common factor),
\begin{equation}
C_1^+=e^{i\theta_1}\,,\quad C_2^+=C e^{i\theta_2}\,,\quad C_1^-=e^{\theta_1}\frac{s-iq}{\kappa-i\lambda}\,,\quad C_2^-=-\frac{C}{\kappa-i\lambda}e^{i\theta_2}\,,
\end{equation}
where $C$,  $\theta_1$ and $\theta_2$ are real-valued parameters.
Then, according to Eq. ({\ref{A2}) we obtain  the general asymptotic
representation for large $|z|$,
 of our solutions,
 \begin{eqnarray}
\hat f^+&\sim& e^{i\nu^2 x^2}(2\nu)^{\frac{1}{2}+iq}x^{2iq}e^{-\frac{1}{2}\pi q}e^{i\theta_1}
\frac{\Gamma(1+2s)}{\Gamma(1+s+iq)}\,,\nonumber\\
\hat f^-&\sim & e^{i\nu^2 x^2}(2\nu)^{\frac{1}{2}-iq} x^{2iq}\left[e^{\frac{1}{2}\pi q}e^{i\pi s}e^{-\pi q}e^{i\theta_1}\frac{s-iq}{\kappa-i\lambda}\frac{\Gamma(1+2s)}{\Gamma(1+s-iq)}-C\frac{e^{i\theta_2}e^{\frac{1}{2}\pi q}}{\kappa-i\lambda}\right]\nonumber\,.
\end{eqnarray}
These solutions can be put in the trigonometric form (\ref{AsiF})
where  the argument is given by $\frac{1}{2}{\rm arg}
\frac{f^+}{f^-}$, as mentioned before in Sec. III A.  Then by using
the method indicated therein we deduce the general expression of
the phase shifts
\begin{equation}
e^{2i\delta_{\kappa}}=\frac{\kappa-i\lambda}{s+iq}\,\frac{\frac{\Gamma(1+2s)}{\Gamma(s+iq)}
e^{i\pi(l-s)}}{\frac{\Gamma(1+2s)}{\Gamma(s-iq)}-Ce^{i\theta}e^{-i\pi(s+iq)}}\,,
\end{equation}
with arbitrary integration constants, that holds for any value of
the parameter $s$ which can take either real values, $s=|s|$, or
pure imaginary ones, $s=\pm i|s|$. We observe that the phase shifts
depend now only on two real integration constants $C$ and the
relative phase $\theta=\theta_1-\theta_2$.

The elastic scattering can arise only when  we have
\begin{equation}\label{eqeq}
\left |e^{2i\delta_{\kappa}}\right |=1\,.
\end{equation}
There are two cases. In the first one, when $s=|s|$,  the equation
(\ref{eqeq}) has two real solutions, $C=0$ and
\begin{equation}\label{Cbad}
C=e^{-\pi q}\frac{\Gamma(1+2s)}{|\Gamma(s+iq)|^2}\left[e^{i(\pi s-\theta)}\Gamma(s+iq)+e^{-i(\pi s-\theta)}\Gamma(s-iq)\right] \,.
\end{equation}
The second case is of pure imaginary  $s=\pm i|s|$ when the above
equation  has no real solutions.

Furthermore, we observe that the phase shifts have correct Newtonian
limits  (\ref{shN}) for large $l$ only if we chose the asymptotic
condition $C=0$ (i. e.  $C_2^+=C_2^-=0$) when the phase shifts are
completely determined being given by Eq. (\ref{final}). Otherwise,
if we consider the solution  (\ref{Cbad}) we obtain  non-determinate
phase shifts
\begin{equation}
e^{2i\delta_{\kappa}}=-\frac{\kappa-i\lambda}{s+iq}e^{i(\pi l+\pi s-2\theta)}
\end{equation}
that are still depending on the arbitrary phase $\theta$. Obviously,
in this  case we cannot speak about the Newtonian limit.

The conclusion is that we must consider the asymptotic condition
$C=0$ giving  the correct phase shifts in elastic collisions for
$s=|s|$.  However, it is natural to keep the same condition for
$s=\pm i|s|$ when the collision is no longer elastic because of the
absorption of the fermions by black hole.

\section*{Acknowledgements}

 I.I. Cotaescu and C. Crucean were supported by a grant of the Romanian National
 Authority for Scientific Research,
Programme for research-Space Technology and Advanced Research-STAR, project nr. 72/29.11.2013 between
Romanian Space Agency and West University of Timisoara.

C.A. Sporea was supported by the strategic grant
POSDRU/159/1.5/S/137750, Project “Doctoral and Postdoctoral
programs support  for increased competitiveness in Exact Sciences
research” cofinanced by the European Social Found within the
Sectorial Operational Program Human Resources Development 2007-2013.


\begin{thebibliography}{20}

\bibitem{sc1}
 R. A. Matzner, {\em J. Math. Phys.} (N.Y.) {\bf9}, 163 (1968).

\bibitem{sc2}
R. Fabbri, {\em Phys. Rev. D} {\bf 12}, 933 (1975).

\bibitem{sc3}
P. C. Peters, {\em Phys. Rev. D} {\bf 13}, 775 (1976).

\bibitem{sc4}
W. K. de Logi and S. J. Kova´cs, {\em Phys. Rev. D} {\bf 16}, 237
(1977).

\bibitem{sc5}
N. G. Sa´nchez, {\em J. Math. Phys.} (N.Y.) {\bf 17}, 688 (1976).

\bibitem{sc5}
N. G. Sa´nchez, {\em Phys. Rev. D} {\bf 16}, 937 (1977).

\bibitem{sc7}
N. G. Sa´nchez, {\em Phys. Rev. D} {\bf 18}, 1030 (1978).

\bibitem{sc8}
N. G. Sa´nchez, {\em Phys. Rev. D} {\bf 18}, 1798 (1978).

\bibitem{sc9}
T. R. Zhang and C. DeWitt-Morette, {\em Phys. Rev. Lett.} {\bf 52},
2313 (1984).

\bibitem{sc10}
R. A. Matzner, C. DeWitt-Morette, B. Nelson, and T. R.
Zhang, {\em Phys. Rev. D} {\bf 31}, 1869 (1985).

\bibitem{sc11}
P. Anninos, C. DeWitt-Morette, R. A. Matzner, P. Yioutas,
and T. R. Zhang, {\em Phys. Rev. D} {\bf 46}, 4477 (1992).

\bibitem{sc12}
N. Andersson, {\em Phys. Rev. D} {\bf 52}, 1808 (1995).

\bibitem{sc13}
N. Andersson and B. P. Jensen, gr-qc/0011025.

\bibitem{sc14}
C. J. L. Doran and A. N. Lasenby, {\em Phys. Rev. D} {\bf 66}, 024006
(2002).

\bibitem{ch}
S. Chandrasekhar, {\em The Mathematical Theory of Black
Holes} (Oxford University Press, New York, 1983).

\bibitem{no}
V. P. Frolov and I. D. Novikov, {\em Black hole physics: Basic
concepts and new developments} (Kluwer Academic Publishers, Dordrecht, 1998).

\bibitem{n1}
J. Chen, H. Liao, and Y. Wang, {\em Eur. Phys. J. C} {\bf 73}, 2395 (2013).

\bibitem{n2}
D. Batic, N. G. Kelkar and M. Nowakowski, {\em Eur. Phys. J. C} {\bf 71}, 1831 (2011).

\bibitem{n3}
D. Batic, N. G. Kelkar and M. Nowakowski, {\em Phys. Rev. D} {\bf 86}, 104060 (2012).

\bibitem{FHM}
J. A. H. Futterman, F. A. Handler, and R. A. Matzner,
{\em Scattering from Black Holes} (Cambridge University
Press, Cambridge, England, 1988).

\bibitem{S1}
J. Jing, {\em Phys. Rev. D} {\bf 70}, 065004 (2004); {\em Phys. Rev.
D} {\bf 71}, 124006 (2005).

\bibitem{S2}
K. H. C. Castello-Branco, R. A. Konoplya and A. Zhidenko, {\em Phys.
Rev. D} {\bf 71}, 047502 (2005).

\bibitem{bh1}
C. Doran, A. Lasenby, S. Dolan and I. Hinder, {\em Phys. Rev. D} {\bf 71}, 124020 (2005).

\bibitem{S3}
S. Dolan, C. Doran and A. Lasenby, {\em Phys. Rev. D} {\bf 74}, 064005 (2006).

\bibitem{C4}
I. I. Cot\u aescu, {\em Mod. Phys. Lett. A} {\bf 22}, 2493  (2007).

\bibitem{ES}
I. I. Cot\u aescu, {\em J. Phys. A: Math. Gen.} {\bf 33}, 1977
(2000).

\bibitem{TH}
B. Thaller,  {\it The Dirac Equation} (Springer Verlag, Berlin
Heidelberg, 1992).


\bibitem{C1}
I. I. Cot\u aescu, {\em Mod. Phys. Lett. A} {\bf 13}, 2923   (1998).

\bibitem{C1A}
I. I. Cot\u aescu, {\em Mod. Phys. Lett. A} {\bf 13}, 2991   (1998).

\bibitem{C1B}
I. I. Cot\u aescu, {\em Phys. Rev. D} {\bf 60}, 124006-010 (1999).

\bibitem{C1C}
I. I. Cot\u aescu, {\em Int. J. Mod. Phys. A} {\bf 19}, 2217 (2004).



\bibitem{Nov}
I. D. Novikov, {\em doctoral disertation}, Sthernberg Astronomical Institute (1963).

\bibitem{GRAV}
C. W. Misner, K. S. Thorne and J. A. Wheeler, {\em Gravitation} (Freeman \& Co., San Francisco, 1971).

\bibitem{LL}
V. B. Berestetski, E. M. Lifshitz and L. P. Pitaevski, {\em Quantum Electrodynamics} (Pergamon Press, Oxford 1982).

\bibitem{Un}
W. G. Unruh, {\em Phys. Rev. D}  {\bf 14}, 3251 (1976).

\bibitem{C3}
I. I. Cot\u aescu, {\em Phys. Rev. D} {\bf 60}, 124006 (1999).

\bibitem{NIST}
F. W. J. Olver, D. W. Lozier, R. F. Boisvert and C. W. Clark, {\em NIST Handbook of Mathematical Functions} (Cambridge University Press, 2010).



\bibitem{Yeni}
D. R. Yennie, D. G. Ravenhall, and R. N. Wilson, {\em Phys. Rev.} {\bf 95}, 500 (1954).

\bibitem{bh2}
J. Jing, {\em Phys. Rev. D} {\bf 72}, 027501 (2005).

\bibitem{bh3}
R. A. Konoplya and A. Zhidenko, {\em Phys. Rev. D} {\bf 76}, 084018 (2007).

\bibitem{bh4}
C. L. Benone, E. S. Oliveira, S. R. Dolan and L. C. B. Crispino, {\em Phys. Rev. D} {\bf 89}, 104053.
\bibitem{bh5}
L. C. B. Crispino, S. R. Dolan, E. S. Oliveira, {\em Phys. Rev. Lett.} {\bf 102}, 231103 (2009).
\bibitem{WF}
K. W. Ford, G. A. Wheeler, {\em Ann.Phys.} {\bf 7}, 259 (1959).
\bibitem{MT}
N. F. Mott and H. S. W. Massey, {\em The Theory of Atomic Collisions} (Oxford University Press, London, 1965).
\end{thebibliography}
\end{document}